\documentclass[preprint,3p,12pt]{elsarticle_preprint}
\usepackage{amssymb,amsmath} 
\usepackage{mathtools}

\usepackage{overpic,rotating,contour,footnote}                        
\usepackage{graphics,subfigure,enumerate}
\usepackage{graphicx,psfrag}
\usepackage{epsfig,wasysym}  
\usepackage[all]{xy}
\usepackage[color=aliceblue]{todonotes}
\usepackage{xcolor}
\usepackage{hyperref} 

\definecolor{aliceblue}{rgb}{0.94, 0.97, 1.0}

\usepackage{mathptmx}       % selects Times Roman as basic font
\usepackage{helvet,overpic}         % selects Helvetica as sans-serif font
\usepackage{courier}        % selects Courier as typewriter font
\usepackage{graphicx,subfigure}        % standard LaTeX graphics tool 
\usepackage[bottom]{footmisc} % places footnotes at page bottom    
      
\def\Beweisende{\square}            % Beweisende-Zeichen
\def\BewEnde{\hfill{\Beweisende}}

\def\phm{{\hphantom{-}}} % Laesst genausoviel Platz wie ein Minus!

\def\phi{\varphi}

\def\CC{{\mathbb C}}
\def\RR{{\mathbb R}}

\def\NN{{\mathbb N}}

\def\Vkt#1{{\mathbf #1}} 

\def\Area{\mbox{A}}
\def\Vol{\mbox{Vol}}

\newcommand{\go}[1]{{\sf #1}}

\usepackage[T1]{fontenc}% http://ctan.org/pkg/fontenc
\usepackage{xhfill}% http://ctan.org/pkg/xhfill

\newtheorem{thm}{Theorem}

\newtheorem{rmk}{Remark} 
\newtheorem{example}{Example}

\newtheorem{definition}{Definition}

\begin{document}

\begin{frontmatter}

\title{Multi-stable design of triangulated origami structures \\ on cones of revolution}

\author{Georg Nawratil\corref{mycorrespondingauthor}}
\address{
	Institute of Discrete Mathematics and Geometry \& Center for Geometry and Computational Design, 
	TU Wien, \\
	Wiedner Hauptstrasse 8-10/104, Vienna 1040, Austria
    }	  
\cortext[mycorrespondingauthor]{Corresponding author}
\ead{nawratil@geometrie.tuwien.ac.at}

%%%%%%%%%%%%%%%%%%%%%%%%%%%%%%%%%%%%%%%%%%%%%%%%%%%%%%%%%%%%%%%%%%%%%%
\begin{abstract}
It is well-known that the Kresling pattern of congruent triangles can be arranged either circularly on a cylinder of revolution or in 
a helical way. In both cases the resulting cylindrical structures are multi-stable. We generalize these arrangements with respect to cones of revolution, 
where our approach allows to construct structures, which snap between conical realizations whose apex angles serve as design parameters. 
In this context we also figure out shaky realizations, intervals for self-intersection free realizations and 
an interesting property related to the cross sectional area. 
Finally, we analyze these origami structures with respect to their capability to snap by means of the so-called snappability index.  
\end{abstract}
%%%%%%%%%%%%%%%%%%%%%%%%%%%%%%%%%%%%%%%%%%%%%%%%%%%%%%%%%%%%%%%%%%%%%%

\begin{keyword}
Origami, flat-foldability, triangulated cone structure, multi-stability, Kresling pattern, snapping, shakiness
\end{keyword}

\end{frontmatter}

\section{Introduction}\label{sec:intro}

\begin{figure}
\phm\hfill
\begin{minipage}{77mm} 
\begin{overpic}
    [width=70mm]{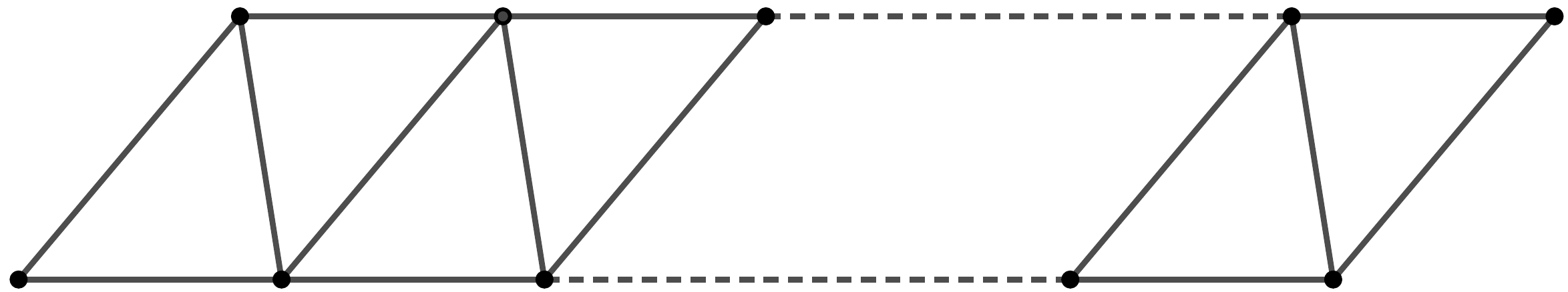}
\begin{small}
\put(-3,15){a)}
\put(0,-4){$A_0$}
\put(17,-4){$A_1$}
\put(34,-4){$A_2$}
\put(66,-4){$A_{n-1}$}
\put(83.5,-4){$A_n$}
\put(17,13){$B_0$}
\put(33.5,13){$B_1$}
\put(49.5,13){$B_2$}
\put(83.5,13){$B_{n-1}$}
\put(99,13){$B_n$}
\end{small}     
  \end{overpic} 
$\phm$ \medskip \newline \\	
\begin{overpic}
    [width=77mm]{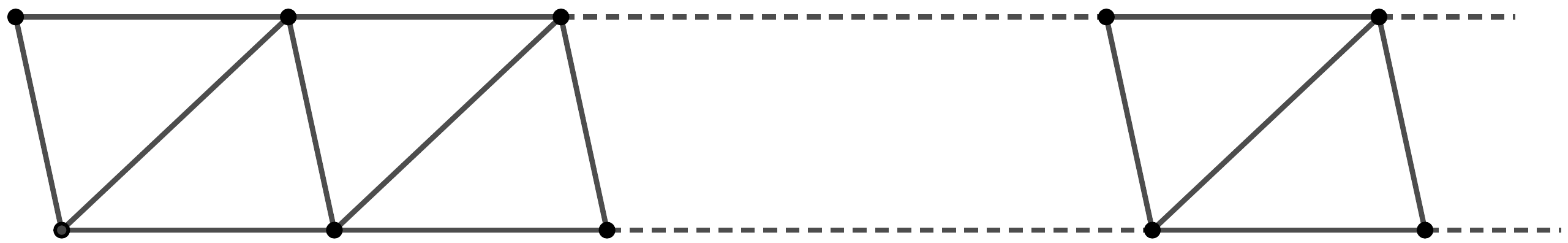}
\begin{small}
\put(-3,0){b)}
\put(3,-4){$V_0$}
\put(20,-4){$V_1$}
\put(37.5,-4){$V_2$}
\put(72,-4){$V_{n-1}$}
\put(90,-4){$V_n$}
\put(2.4,10){$V_{n-1}$}
\put(20,10){$V_n$}
\put(37.5,10){$V_{n+1}$}
\put(89.5,10){$V_{2n-1}$}
\put(71.5,10){$V_{2n-2}$}
\end{small}         
  \end{overpic} 	
$\phm$ \vspace{-2mm}\newline	
\end{minipage}	
\hfill
\begin{minipage}{55mm} 
 \begin{overpic}
    [width=20mm]{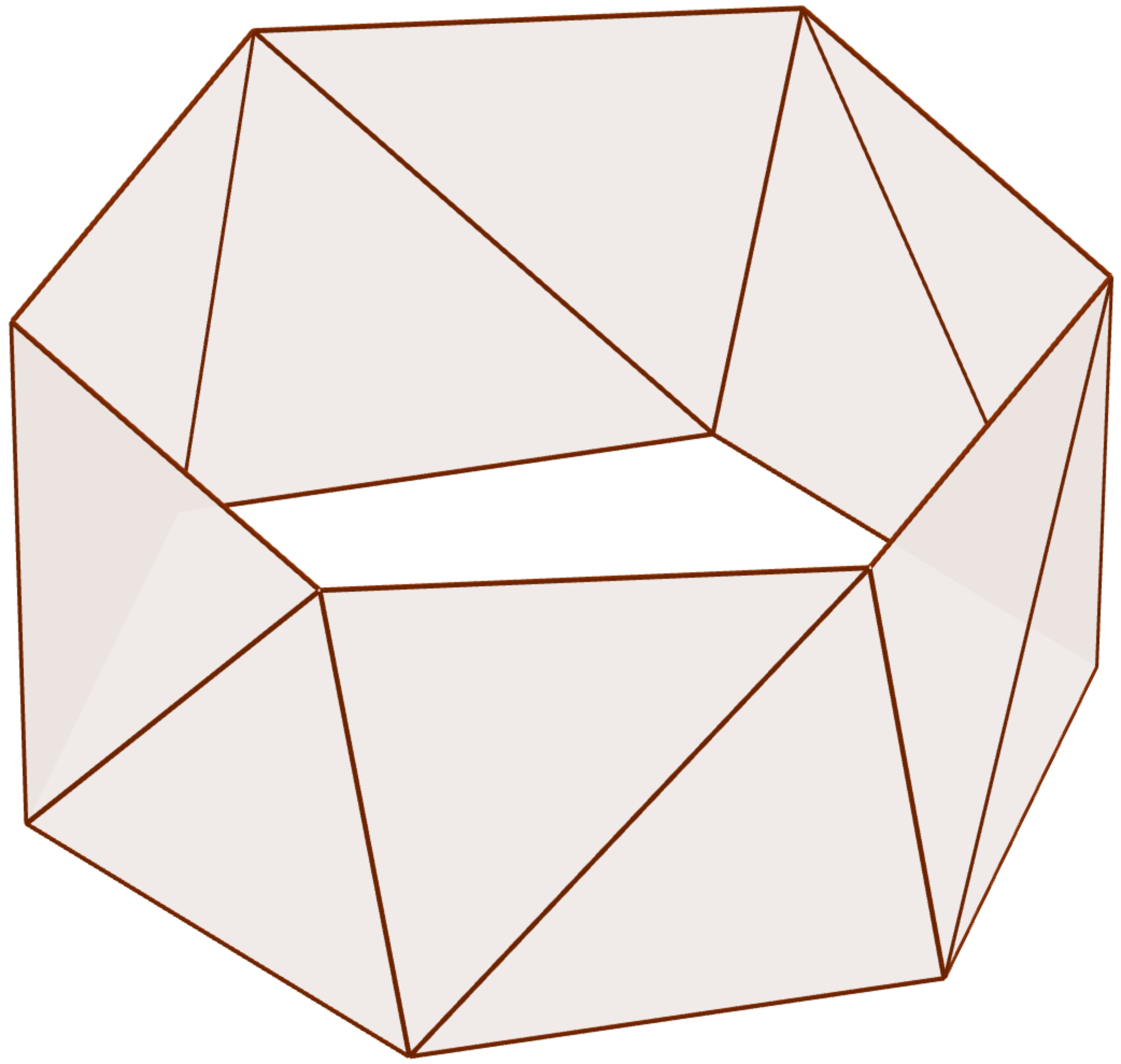}
\begin{small}
\put(0,0){c)}
\end{small}         
  \end{overpic} 
	\hfill
\begin{overpic}
    [width=20mm]{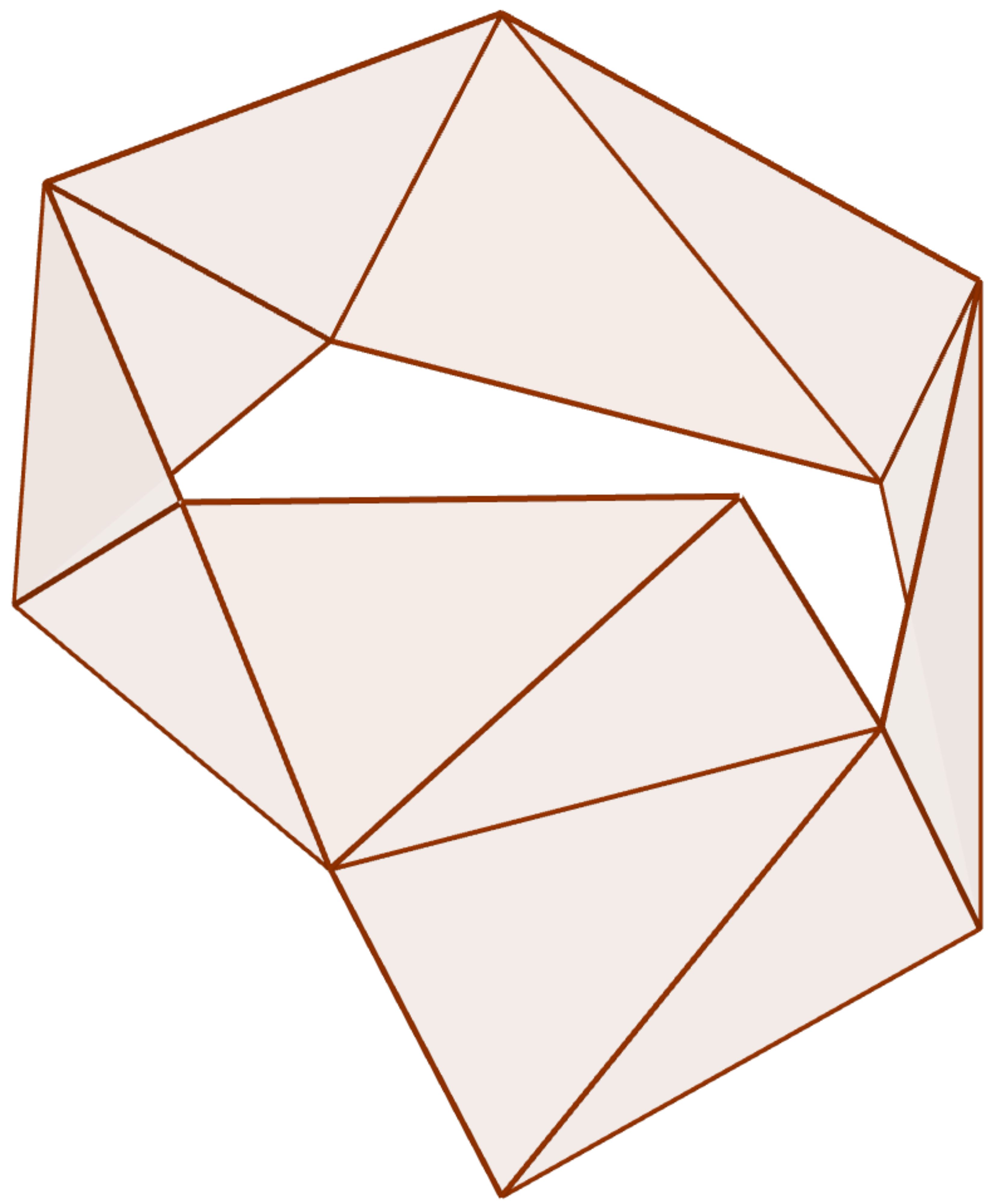}
\begin{small}
\put(0,0){d)}
\end{small}         
  \end{overpic} 	
 \newline	
$\phm$ \newline	
\end{minipage}	
\hfill $\phm$
\caption{
(a/b) Kresling pattern for a circular/helical arrangement.  
(c) Anti-prism for $n=6$. (d) Illustration of the first $2n$ triangles of the helical arrangement for $n=6$.
}
  \label{fig1}
\end{figure}

\begin{figure}
\begin{overpic}
    [width=25mm]{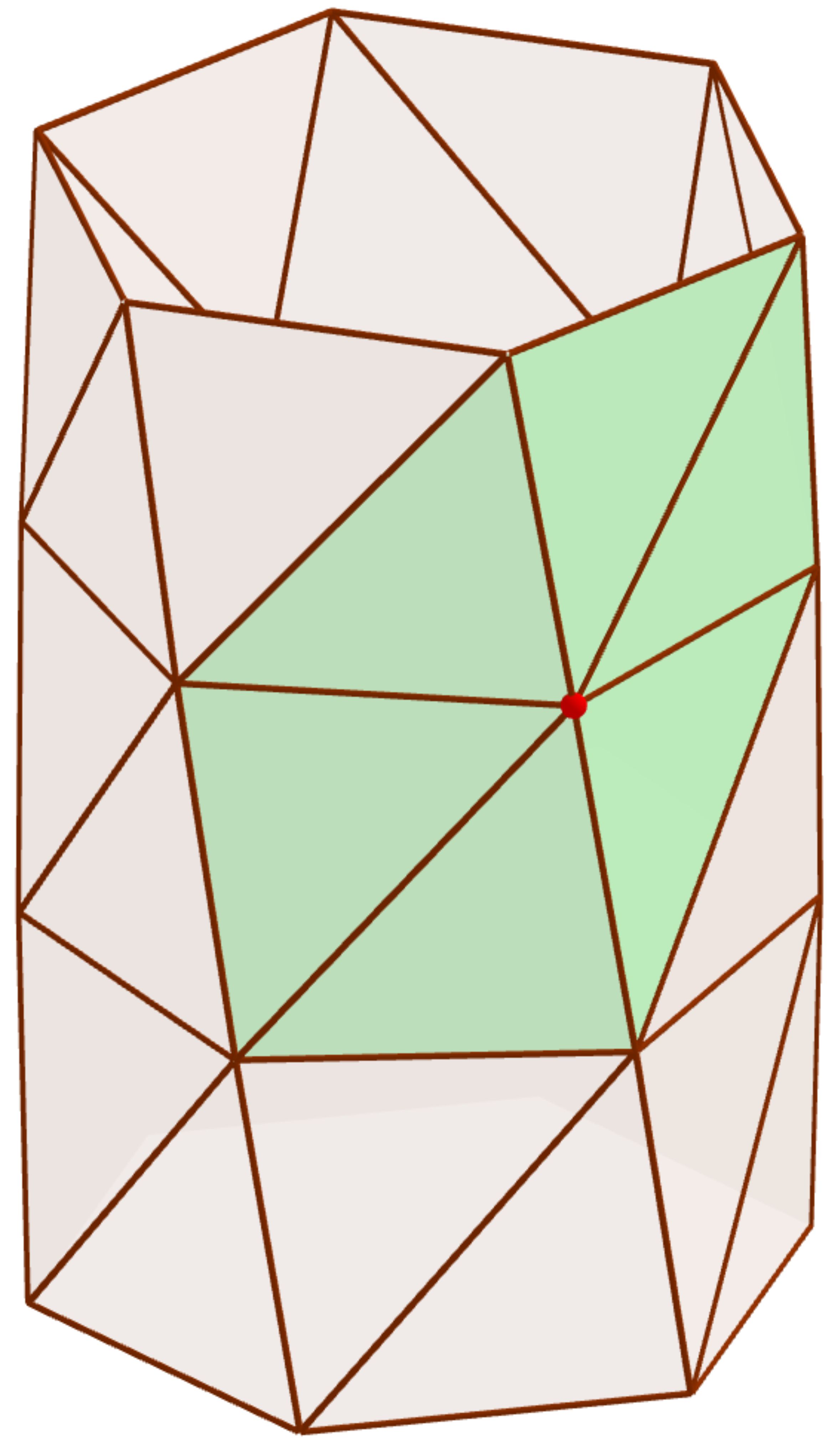}
\begin{small}
\put(42.5,46){$V$}
\put(0,0){a)}
\end{small}     
  \end{overpic} 
\hfill
 \begin{overpic}
    [width=25mm]{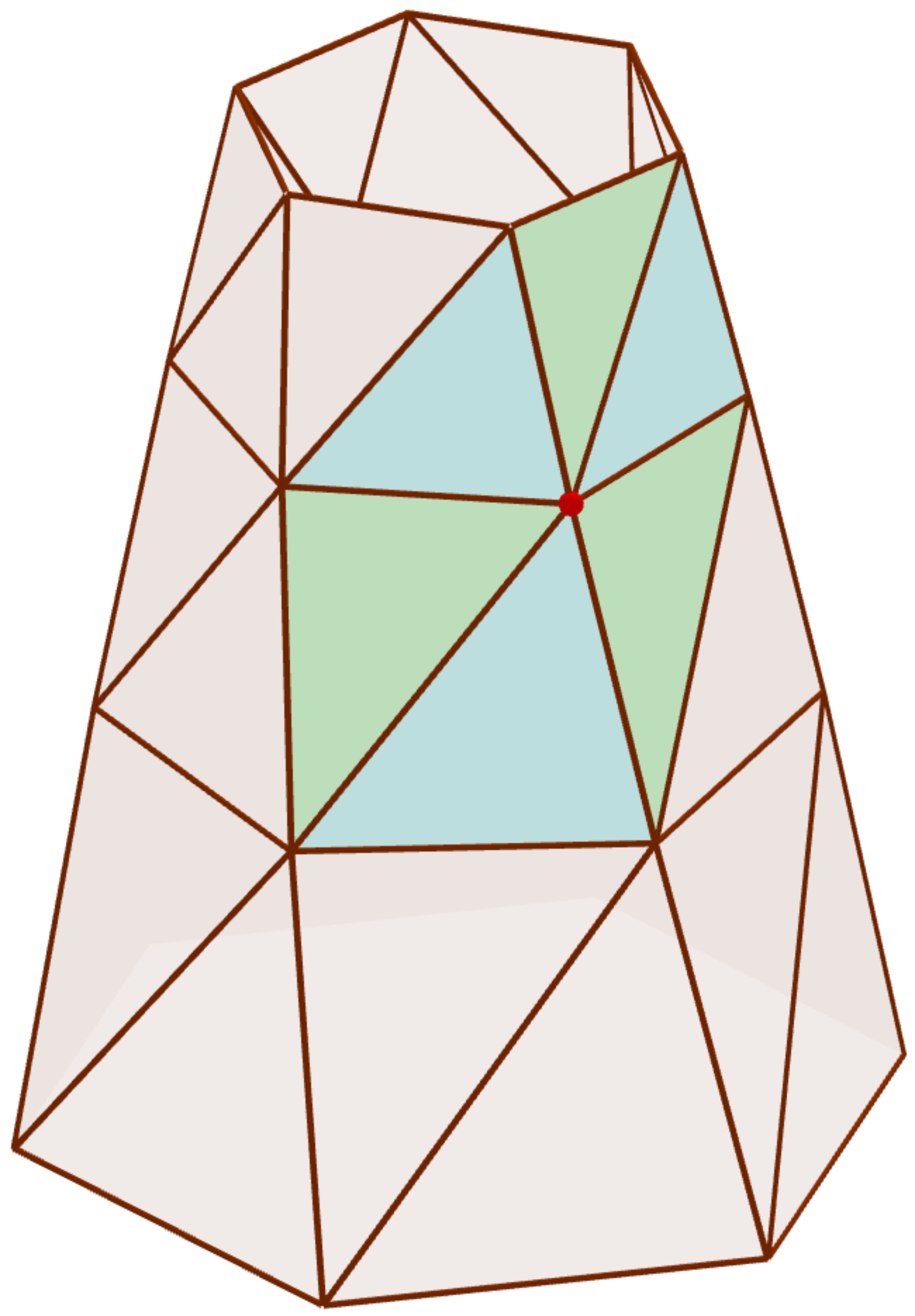}
\begin{small}
\put(46,55.5){$V$}
\put(0,0){b)}
\end{small}         
  \end{overpic} 
	\hfill
\begin{overpic}
    [width=25mm]{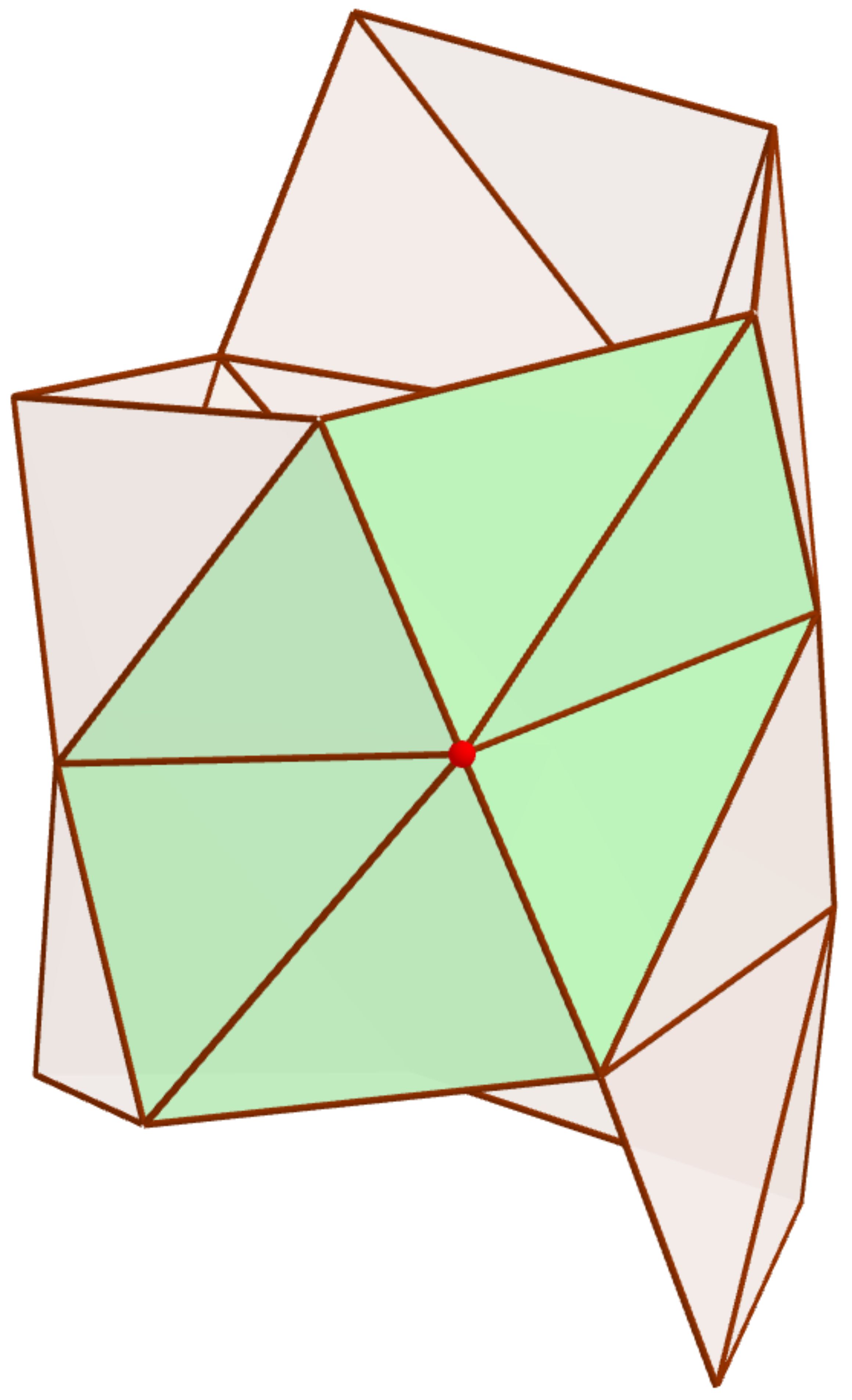}
\begin{small}
\put(37,41){$V$}
\put(0,0){c)}
\end{small}         
  \end{overpic} 	
		\hfill
\begin{overpic}
    [width=21mm]{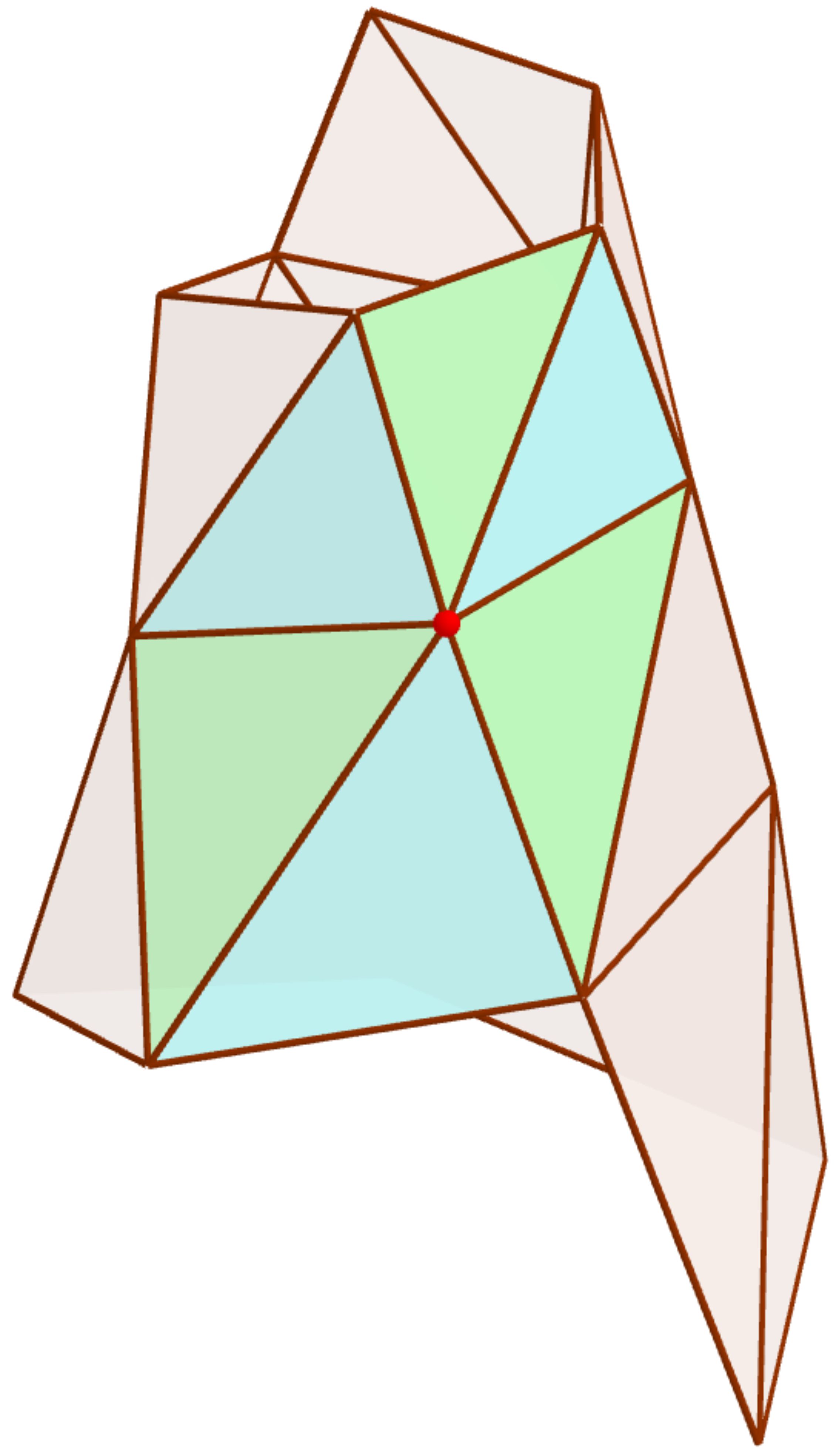}
\begin{small}
\put(35,51){$V$}
\put(0,0){d)}
\end{small}         
  \end{overpic} 	
\caption{
(a/b) Triangulated cylinder/cone built up by anti-prisms/frusta.  
(c/d) Polyhedral cylinder/cone based on the helical/spiral arrangement of triangles. 
All four illustrations are done for $n=6$. 
Moreover, congruent and similar triangles around a vertex $V$ are displayed in the 
same color in (a/c) and (b/d), respectively. 
}
  \label{fig2}
\end{figure}
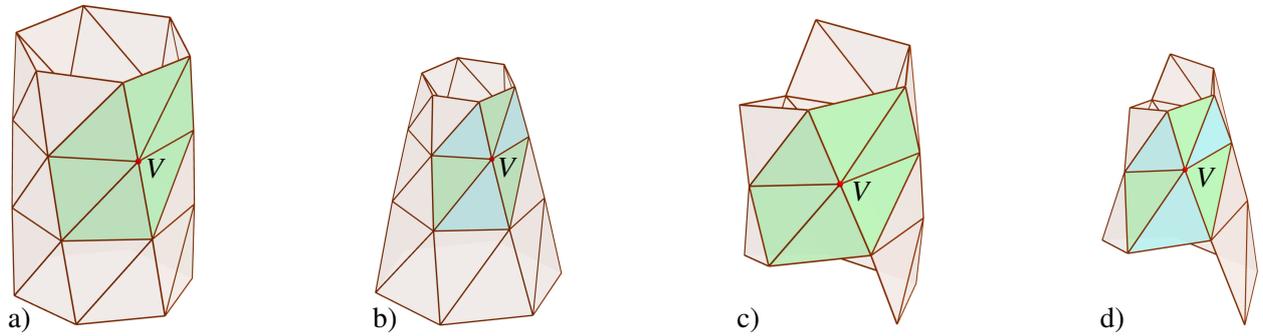

Let us start with a flat strip of congruent triangles as illustrated in Fig.\ \ref{fig1}, which is also known as 
Kresling pattern \cite{kresling} and can be arranged in two different ways:
\begin{enumerate}[1.]
\item
{\bf Circular arrangement to a closed strip:}  
In this case the strip has finite length (Fig.\ \ref{fig1}a) and is folded such that $A_0=A_n$ and $B_0=B_n$ holds ($n\geq 3$) and that
$A_1,\ldots,A_n$ and  $B_1,\ldots,B_n$ form regular $n$-gons with centers $A$ and $B$, respectively, located in parallel planes $\alpha$ and $\beta$ (Fig.\ \ref{fig1}c).
Moreover the line $AB$ is orthogonal to $\alpha$ and $\beta$. 
This discretized cylindrical strip has a bi-stable behavior, which was already known to  Wunderlich \cite{wunderlich_antiprism}, where these 
structures appear as special cases (regular ones) of snapping anti-prisms, which are a generalization of his snapping octahedra construction \cite{wunderlich_achtflach}. 
These snapping regular anti-prisms can be composed repetitively to cylindrical towers \cite{kresling,hunt,cai1,cai2,liu2017,kidambi,moshtaghzadeh} (Fig.\ \ref{fig2}a)
and find practical application as energy/shock absorbers \cite{wu,zhao}, vibration isolators \cite{ishida},  rigidizable inflatable booms/tubes \cite{barker,schenk}, 
modular fluidic actuators \cite{forte} 
or crawling mechanisms \cite{pagano}.
\item
{\bf Helical arrangement according to C.R.\ Calladine:}  
In this case the strip can be assumed of infinite length (Fig.\ \ref{fig1}b) where every label $V_i$ with $i\geq n-1$ ($n\geq 3$)
appears twice (once on the lower rim and once on the upper one). Based on this labeling one can fold up the strip in a way that 
points with the same labels match and are located on a helix, which results in a triangulated cylinder studied in \cite{guestI,guestII,guestIII,wittenburg} 
(Figs.\ \ref{fig1}d and \ref{fig2}c). 
This polyhedral structure is also multi-stable and from the formulation used in \cite{wittenburg}, it can be seen that there exists in general 
$n-2$ cylindrical realizations (without taking reality or self-intersections into account).
\end{enumerate}

Within this paper we want to generalize these two constructions for conical structures, i.e.\ vertices are located on a cone of revolution (Figs.\ \ref{fig2}b and  \ref{fig2}d), 
which can e.g.\ be used as reconfigurable antenna \cite{antenna}, foldable horn speakers \cite{speaker},  
functional folded light art \cite{wul} or 
building blocks of inflatable origami structures at the metre scale \cite{melancon} for designing deployable systems. 
For these conical structures much less literature is available, namely a journal article of Ishida et al.\ \cite{conformal}  and a 
presentation manuscript of Nojima \cite{nojima1} written in English language. 
Both works are based on the following papers written in Japanese \cite{conformal_j,nojima2,nojima3}.
Within the papers \cite{conformal,conformal_j}  the crease patterns of conical structures are generated from crease patterns of cylindrical structures by applying 
the following planar conformal map. The Cartesian coordinates $(x,y)$ of a node of the cylindrical crease pattern are composed to 
the complex number $z=x+yI$, which is mapped by $ke^z$ with $k\in\CC$ to another complex number $\zeta=\xi+\upsilon I$, whose real and imaginary part are the 
Cartesian coordinates $(\xi,\upsilon)$ of the corresponding node of the conical crease pattern. 

The drawback of this indeed interesting approach is that it does not allow direct access to its spatial conical shape, 
with exception of the flat foldable state, which can be added as an extra condition to the crease pattern (cf.\ \cite[Eq.\ (14)]{conformal}). 

In contrast to this approach we do not generate the planar crease pattern but construct directly the triangulation on the 3-dimensional shape (cone). 
Moreover one can construct a structure which snaps between conical realizations whose apex angles serve as design parameters. Note that structures 
with a flat-foldability are obtained as a special case within our framework (apex angle equals \medskip $\pi$).

{\bf Outline.}
In Section \ref{sec:geometry} we focus on the geometry of the cylindrical/conical triangulated structures 
concerning their construction and degrees of freedom.  
The Sections \ref{sec:frusta} and \ref{sec:spiral} deal with the design 
of conical structures snapping between (two or even three) different realizations, where the apex angles 
of the underlying cones serve as design parameters. In this context we also derive shaky realizations (cf.\ Subsection \ref{subsec:shaky}) 
and intervals, which are free of self-intersections (cf.\ Subsection \ref{subsec:prop}). 
A further interesting property related to the cross sectional area is studied in Subsection \ref{sec:cross}.
Finally in Section \ref{sec:snap}, we study these structures with respect to their capability to snap 
using the concept of the so-called {\it snappability}.

\section{Geometry of the cylindrical/conical triangulated structures}\label{sec:geometry}

In the following we give the basic idea for the kinematic construction of triangulated structures, which are 
the conical analogues of the cylindrical ones presented in Section \ref{sec:intro}.

\begin{enumerate}[1.]
\item
{\bf Anti-frustum based conical triangulation:} 
We start with a cone of revolution, which is sliced along parallel planes orthogonal to the rotation axis. 
Moreover, we discretize the circles by regular $n$-gons, where adjacent ones are connected in the combinatorics of an anti-prism. 
We call these geometric objects regular anti-frusta (as a frustum results from the truncation of a pyramid). 
In order to archive the periodicity of the structure we demand that each anti-frustum can be transformed into any other 
one by a so-called spiral displacement $\sigma$; i.e.\ composition of a rotation about the cone 
axis and a scaling with center in the apex, which is a subgroup of the equiform motion group. 
In this way we get the conical structure illustrated in Fig.\ \ref{fig2}b, which is arranged by 
scaled copies of an initial anit-frustum.
\item
{\bf Spiral-motion based conical triangulation:} 
The helical arrangement of Fig.\ \ref{fig2}c can also be generated as follows. One starts with a line-segment $V_0V_1$ with non-zero slope, whose end points 
are located on the cylinder. Now we apply a so-called  helical (or screw) displacement $\eta$; i.e.\ composition of a rotation about the cylinder axis and a translation along its axis, 
to this line segment, in such a way that $\eta(V_0)=V_1$ holds. By iterating the process, i.e.\ $\eta^i(V_0)=\eta^{i-1}(V_1)$, we get all vertices of the structure. 

By replacing cylinder by cone and helical displacements by spiral ones, we get all vertices of the conical structure, which 
are connected in the same combinatorics as the corresponding cylindrical one.

\begin{rmk}
Note that the set of vertices $\sigma^i(V_0)=\sigma^{i-1}(V_1)$ is located on a so-called concho-spiral, which is also known as conical helix. 
In the remainder of the paper we will name this curve a spiral for short.  \hfill $\diamond$
\end{rmk}
\end{enumerate}

It is well-known that a cylinder/cone is a developable surface, but also the proposed triangulations of the cylinder/cone are developable, which can be argued as follows:
The triangulated cylinder consists of congruent triangles and in each vertex all three angles of this triangle meet two times, which yield $2\pi$ (Fig.\ \ref{fig2}a,c). 
The triangulated cone consists of two types of triangles with respect to spiral displacements. In every vertex all six angles determined by these 
two triangles meet and sum up to $2\pi$ (Fig.\ \ref{fig2}b,d). As a consequence the triangulated cylinders/cones can be folded up (plus one glue-line) 
from a planar crease pattern, which can easily be generated, thus they can also be seen as {\it origami structures}.

\subsection{Degrees of freedom}\label{dof}

\begin{figure}[t]
\begin{overpic}
    [width=25mm]{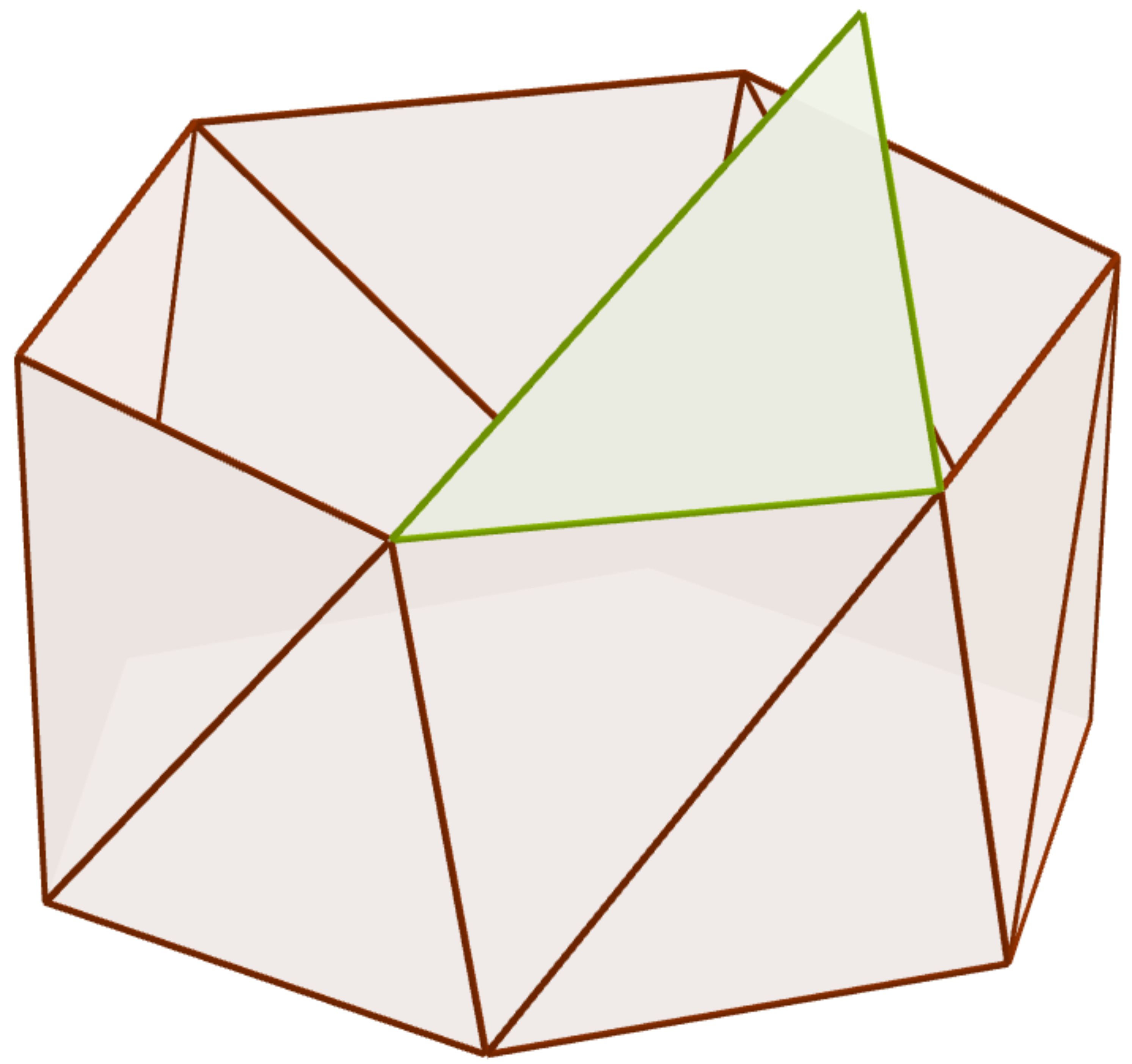}
\begin{small}
\put(0,0){a)}
\put(60,58){$F$}
\end{small}     
  \end{overpic} 
\hfill
 \begin{overpic}
    [width=25mm]{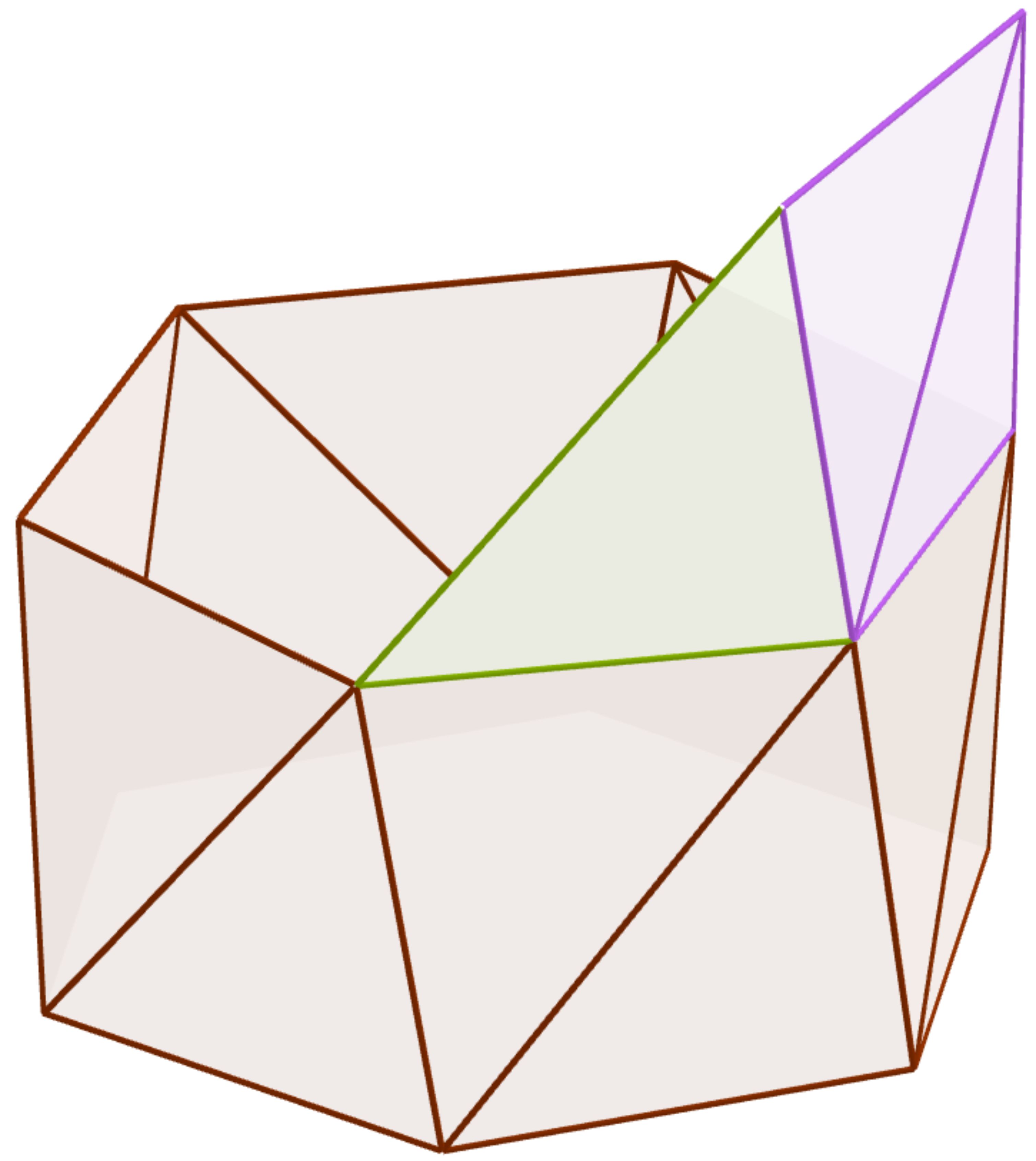}
\begin{small}
\put(0,0){b)}
\end{small}         
  \end{overpic} 
	\hfill
\begin{overpic}
    [width=25mm]{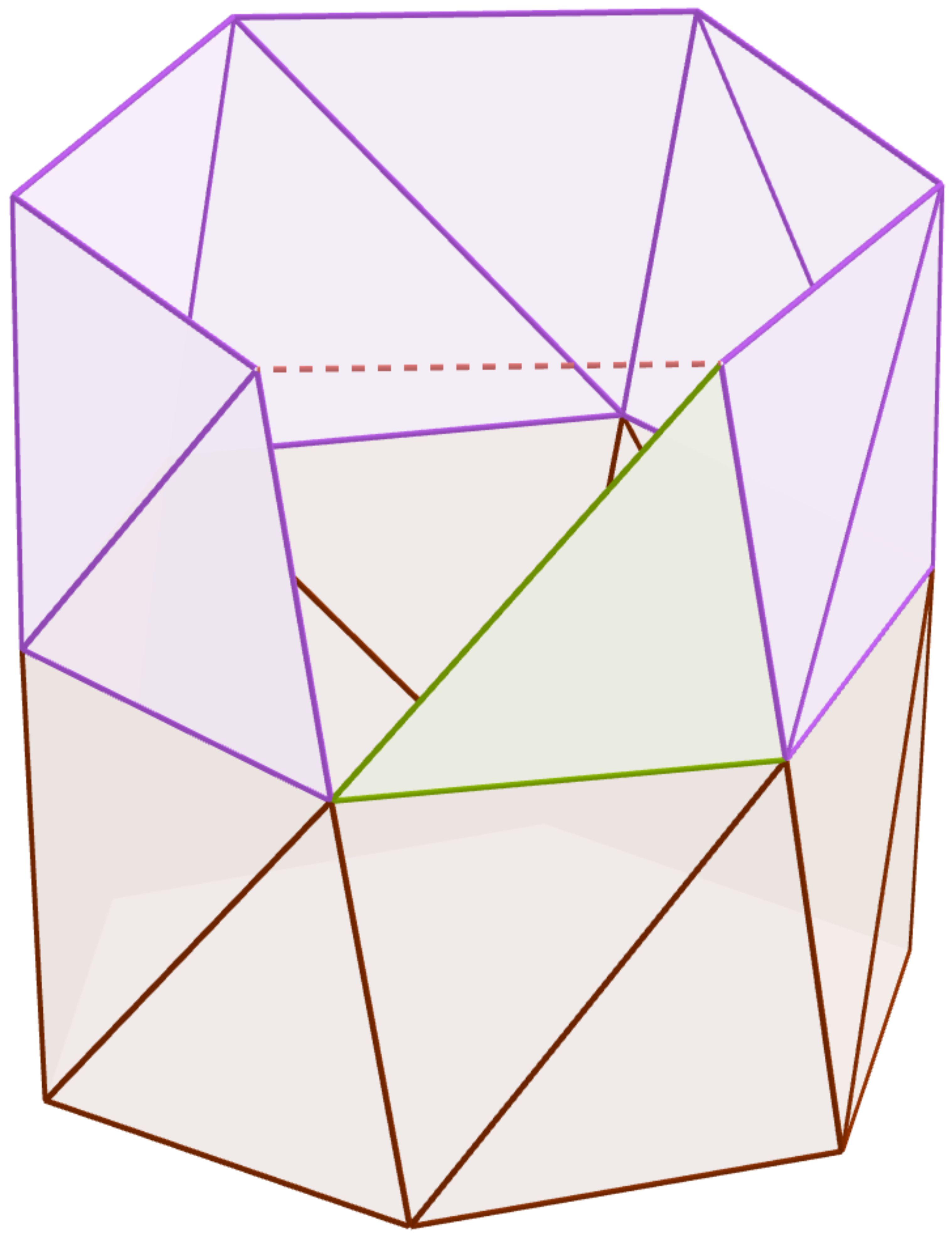}
\begin{small}
\put(0,0){c)}
\end{small}         
  \end{overpic} 	
		\hfill
\begin{overpic}
    [width=25mm]{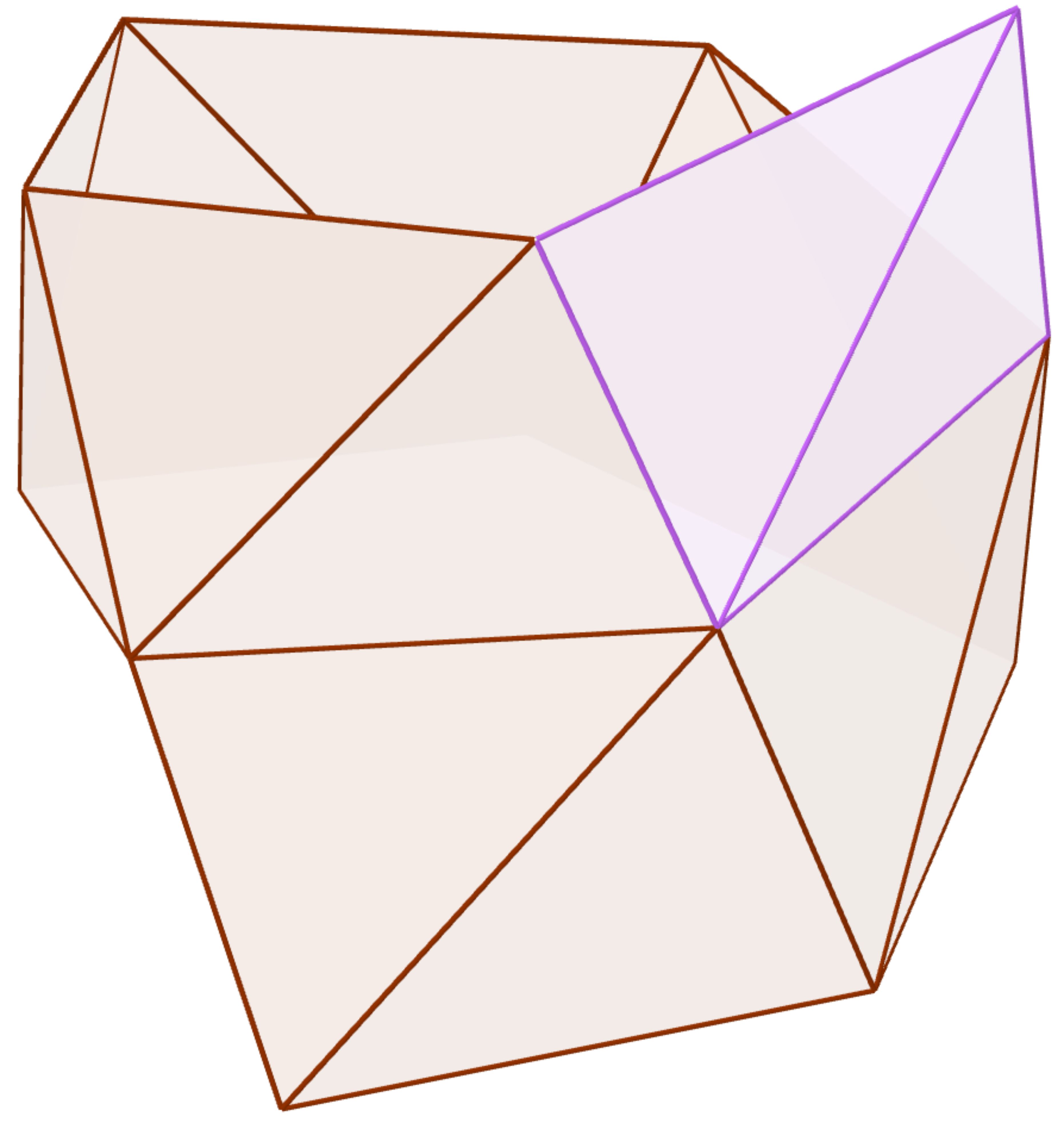}
\begin{small}
\put(0,0){d)}
\end{small}         
  \end{overpic} 	
\caption{Only the cylindrical case is illustrated as the conical is combinatorial equivalent.
(a) Attachment of the triangle $F$ in green.
(b,d) Attaching of further two triangles in purple. 
(c) Iterating the last step until $2n-1$ triangles of this level are assembled. The distance constraint is 
illustrated by the red dashed line. 
}
  \label{fig3}
\end{figure}

Now we want to clarify the degree of freedom (dof) of both kinds of triangulated cylinders/cones given above. 
\begin{enumerate}[{ad} 1.]
\item
Let us start with a cylindrical/conical tower composed of an arbitrary number of regular anti-prisms/frusta, where the top and the base are open; i.e.\ 
the structure is not bounded by two regular $n$-gonal shaped faces (Fig.\ \ref{fig2}a). Now the question arise if this combinatorial structure is rigid or 
continuous flexible? Therefore we built up the structure level by level. According to the formula of Gr\"ubler the closed strip at the bottom has $2n-6$ dofs as it can be seen 
as a closed serial chain with $2n$ rotational axes. The number of $2n-6$ gives already the mobility of the complete structure 
as each further level can only be assembled in finite many ways, which can be seen as follows: 
We assume now that we fix the base strip in one of its possible configurations. Then we attach one triangle $F$ of the next strip (Fig.\ \ref{fig3}a), 
whose loose vertex has one dof (rotation about the hinge by the angle $\chi$). Then the attachment of further two triangles, which corresponds to the adding 
of a further vertex (Fig.\ \ref{fig3}b), can only be done in two ways with respect to $\chi$. 
This argument can be iterated until $2n-1$ triangles of this level are assembled (Fig.\ \ref{fig3}c). 
Now the last triangle has to fit into the gap resulting in a distance constraint. This constraint can be formulated in dependence of the 
parameter $\chi$, which has in general a finite number of solutions. 
\item
Now we consider the cylindrical/conical structure with the helical/spiral arrangement and an arbitrary number of windings (Fig.\ \ref{fig2}c). 
To clarify if this combinatorial structure is rigid or continuous flexible we consider the first winding composed of the $2n$ 
triangles (Fig.\ \ref{fig1}d). It forms a closed serial chain with $2n$ rotational axes possessing $2n-6$ dofs according to the formula of Gr\"ubler, which is 
also the mobility of the complete structure. This can be seen as follows: 
We  fix the $2n$ rotational chain in one of its possible configurations. Then the attachment of further two triangles of the strip to 
the structure, which corresponds to the adding of a further vertex (Fig.\ \ref{fig3}d), can only be done in two ways. An iteration of this 
argument yields the result\footnote{These considerations also show that in general the triangles of the strip of Fig.\ \ref{fig1}b 
in the spatially assembled state are not necessarily the faces of a polyhedral cylinder as stated in \cite[page 46]{wittenburg}.}. 
\end{enumerate}

From the above considerations it follows that the combinatorial structures of the triangulated cylinders/cones 
are only rigid for $n=3$. In all other cases ($n>3$) the structures get rigid by 
stating additional conditions, which are identified next: 
Suppose a triangulated cone/cylinder is given and we ask for the conditions to impose on the structure to be minimally rigid\footnote{A structure is call minimally rigid (isostatic) 
if the removal of any constraint will make it continuous flexible.}. 
For that we consider the mentioned closed serial chain with $2n-6$ dofs, which is composed of $2n$ triangles and has $2n$ vertices. We pick out six of these vertices and determine 
the discrete set\footnote{It is well-known \cite{gfrerrer}  
that there exist twelve (over $\CC$) cones of revolution through six points in general position, where cylinders can be seen as cones with zero apex angles.} of cones of revolution through them. 
Then  the remaining $2n-6$ vertices also have to be located on these cones, where each vertex implies one condition. 
This number fits exactly with the number of dofs of the closed serial chain and we are done.

Therefore there exists in general a finite number of realizations of such a cylindrical/conical framework, where the word {\it realization} refers to the embedding of 
a framework with prescribed inner geometry (combinatorial structure plus lengths of the edges) into the Euclidean 3-space. Based on this notation a {\it snapping realization} 
can be defined as follows:

\begin{definition}
A realization is called a snapping realization if it is close enough to another incongruent realization such that the physical model can snap into this 
neighboring realization due to non-destructive elastic deformations of material.          
\end{definition}

The elastic deformation of the structure can be quantified by the total elastic strain energy. In general a configuration is called {\it stable} if it 
corresponds to a local minimum of that energy function, but within the body of literature on snapping structures the word {\it stable} is mostly used exclusively for
unstrained configurations, which are {\it realizations} within our nomenclature. We also use it in this way within the article at hand, thus e.g.\ 
the bi-stability of the anti-prism mentioned in item 1 of Section \ref{sec:intro} refers to the fact that it snaps between two realizations.

Moreover, it is well-known that infinitesimal flexibility can be seen as 
the limiting case where two realizations coincide. We stress this point of view for the determination/characterization of 
shaky ($=$ infinitesimal flexible) realizations.
 
\begin{rmk}\label{rmk:shaky}
As a regular anti-prism/frustum can be seen as a parallel robot from the structural point of view, its infinitesimal flexibility 
can also be characterized in a line-geometric way as follows \cite{merlet}: 
If the $2n$ edges, which  connect the vertices of the upper and lower $n$-gon, belong to a linear line-complex, 
then the regular anti-prism/frustum is shaky. As all the levels of the triangulated cylinder/cone of item 1 are obtained by 
a helical/spiral displacement of the base anti-prism/frustum, the complete structure is shaky if this condition holds. 
\hfill $\diamond$
\end{rmk}

\section{Anti-frustum based conical triangulation}\label{sec:frusta}

In the following we give the geometric construction of a regular anti-frustum, which can snap between two realizations 
$\mathcal{R}_+$ and $\mathcal{R}_-$ and corresponds to a special case in the study \cite{wunderlich_antiprism} done by Wunderlich.
The half apex angles of the underlying cones $\Lambda_{\pm}$ with apex $S_{\pm}$ are denoted by 
$\lambda_{\pm}\in]0;\tfrac{\pi}{2}]$ and their tangents values by $l_{\pm}:=\tan{\lambda_{\pm}}$. Without loss of generality we can assume that 
$\lambda_->\lambda_+$ holds.

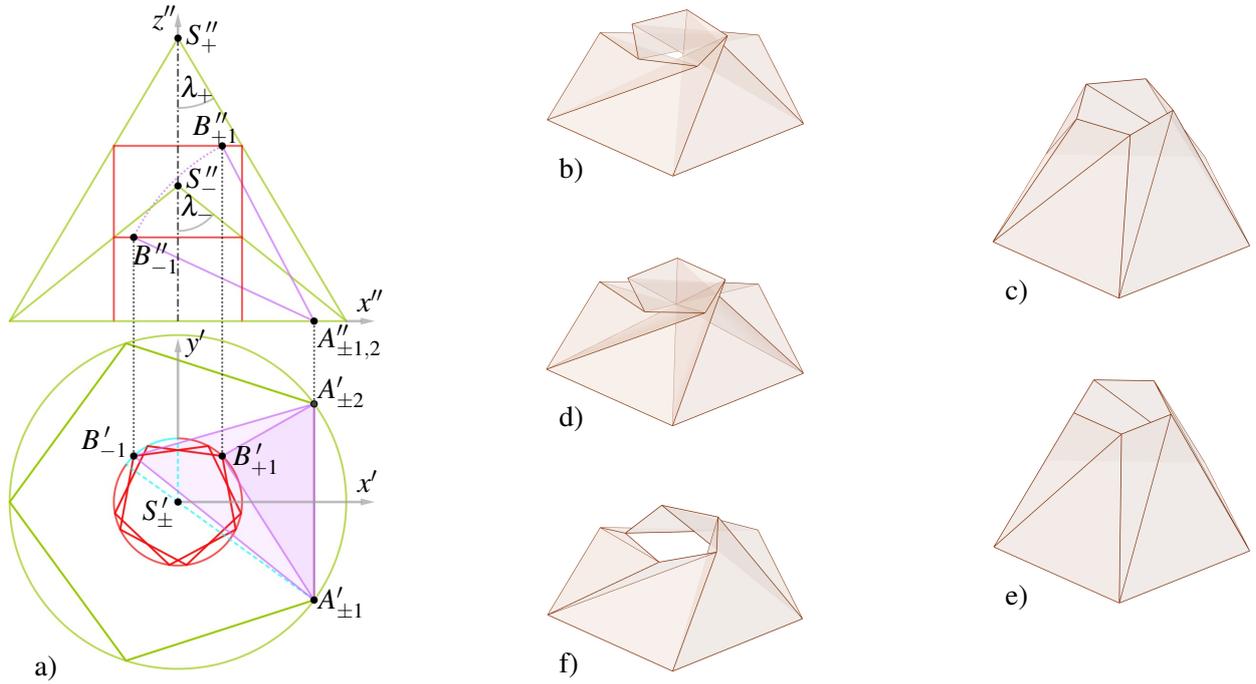
\begin{figure}[t]
\begin{minipage}{47mm} 
\begin{overpic}
    [width=50mm]{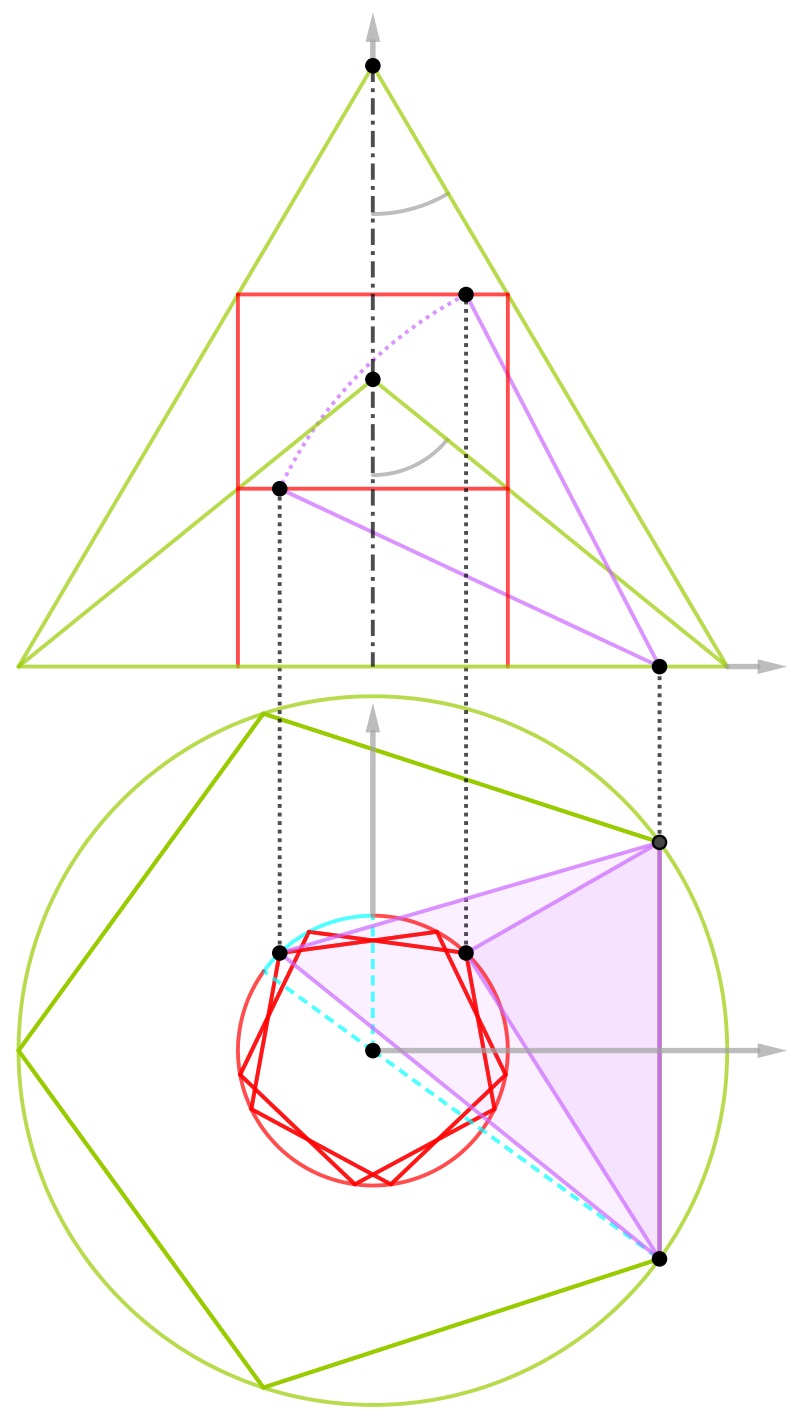}
\begin{small}
\put(5,0){a)}
\put(47.2,49.2){$A_{\pm 1,2}''$}
\put(47.2,10){$A_{\pm 1}'$}
\put(47.2,41.5){$A_{\pm 2}'$}
\put(53,54.5){$x''$}
\put(53,27.5){$x'$}
\put(27.5,48.5){$y'$}
\put(22.5,97){$z''$}
\put(12,34){$B_{-1}^{\,\,\prime}$}
\put(34.5,31.5){$B_{+1}^{\,\,\prime}$}
\put(21,23.4){$S_{\pm}^{\,\,\prime}$}
\put(19.6,61.5){$B_{-1}^{\,\,\prime\prime}$}
\put(28.5,81){$B_{+1}^{\,\,\prime\prime}$}
\put(27.5,73){$S_{-}^{\,\,\prime\prime}$}
\put(27.5,94.6){$S_{+}^{\,\,\prime\prime}$}
\put(26.7,86.7){$\lambda_{+}$}
\put(26.7,68.8){$\lambda_{-}$}
\end{small}     
  \end{overpic} 
\end{minipage}	
\hfill
\begin{minipage}{34mm} 
 \begin{overpic}
    [width=34mm]{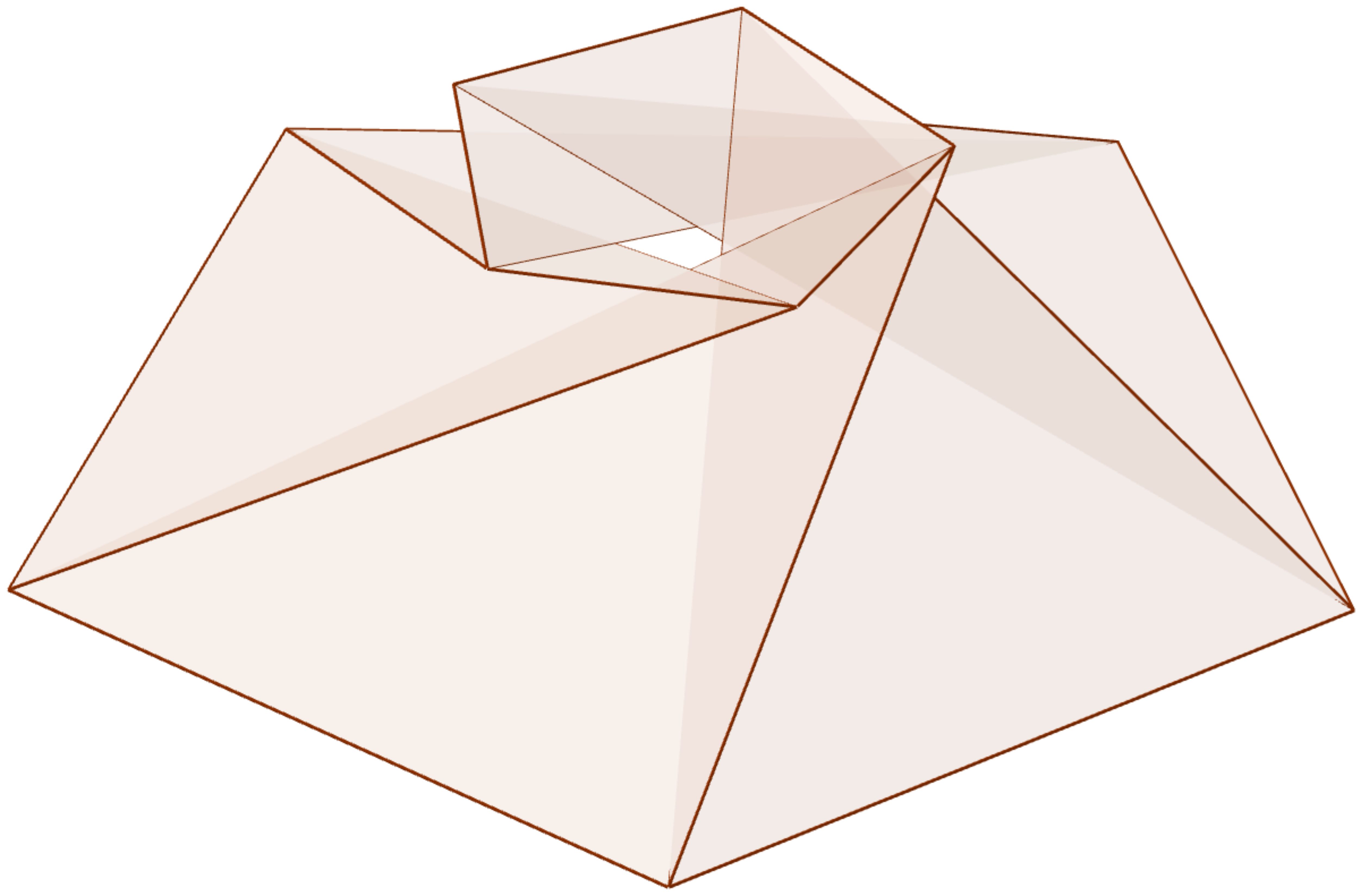}
\begin{small}
\put(5,0){b)}
\end{small}         
  \end{overpic} 
		\newline	
$\phm$ \newline	
$\phm$ \newline	
\begin{overpic}
    [width=34mm]{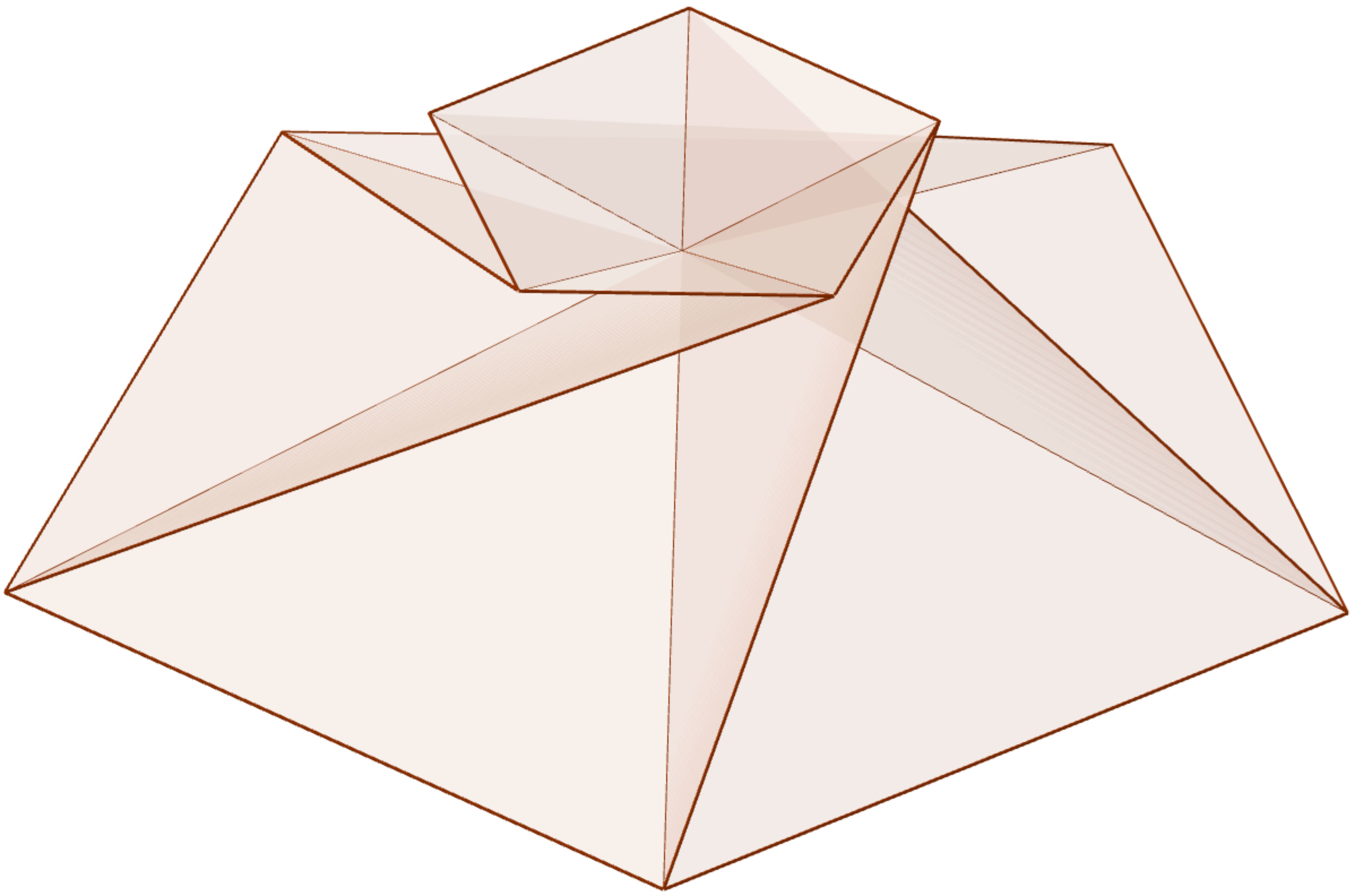}
\begin{small}
\put(5,0){d)}
\end{small}         
  \end{overpic} 	
	\newline	
$\phm$ \newline	
$\phm$ \newline	
\begin{overpic}
    [width=34mm]{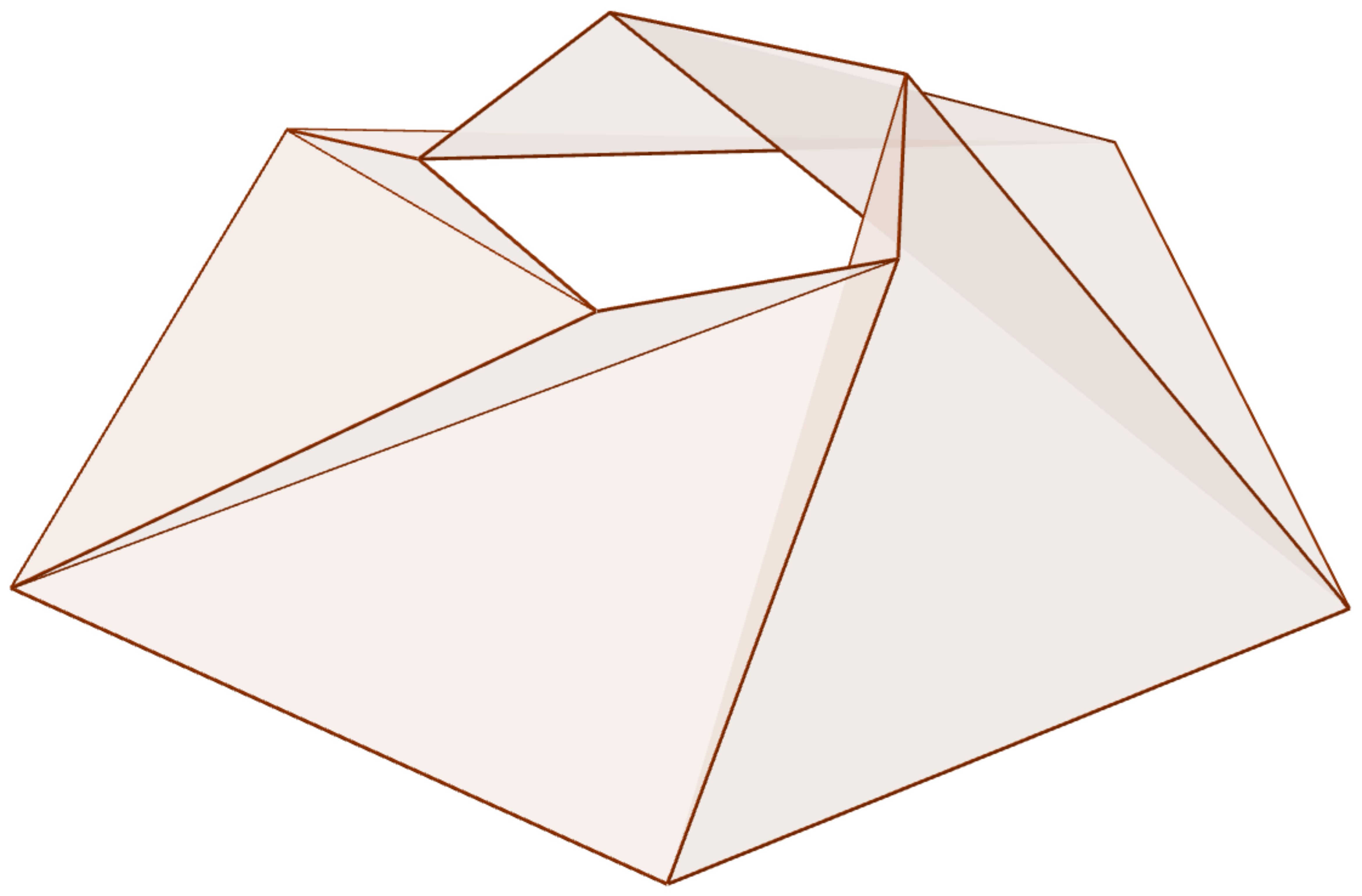}
\begin{small}
\put(5,0){f)}
\end{small}         
  \end{overpic} 	
\end{minipage}	
\hfill
\begin{minipage}{34mm} 
 \begin{overpic}
    [width=34mm]{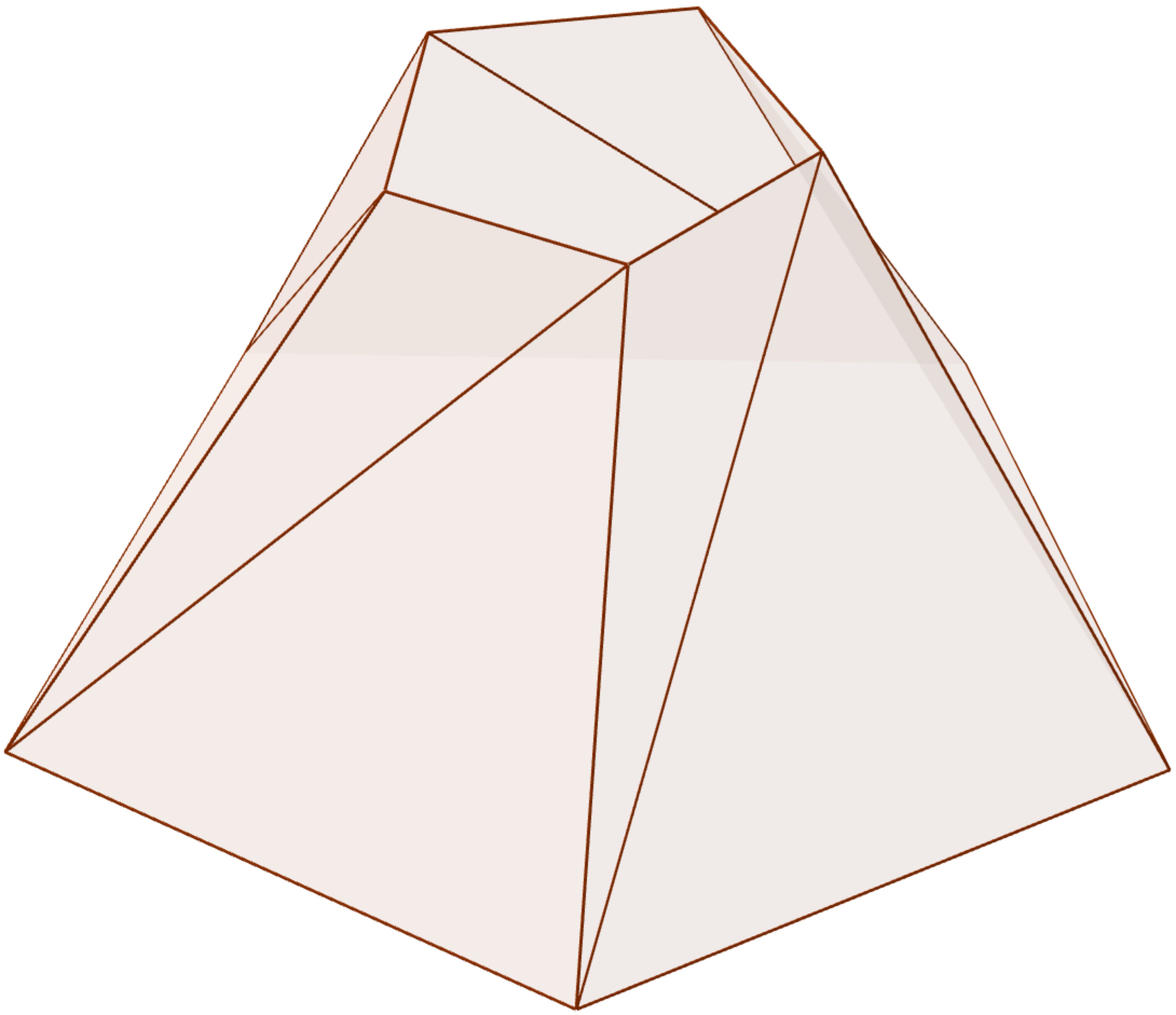}
\begin{small}
\put(5,0){c)}
\end{small}         
  \end{overpic} 
	\newline	
$\phm$ \newline	
$\phm$ \newline	
\begin{overpic}
    [width=34mm]{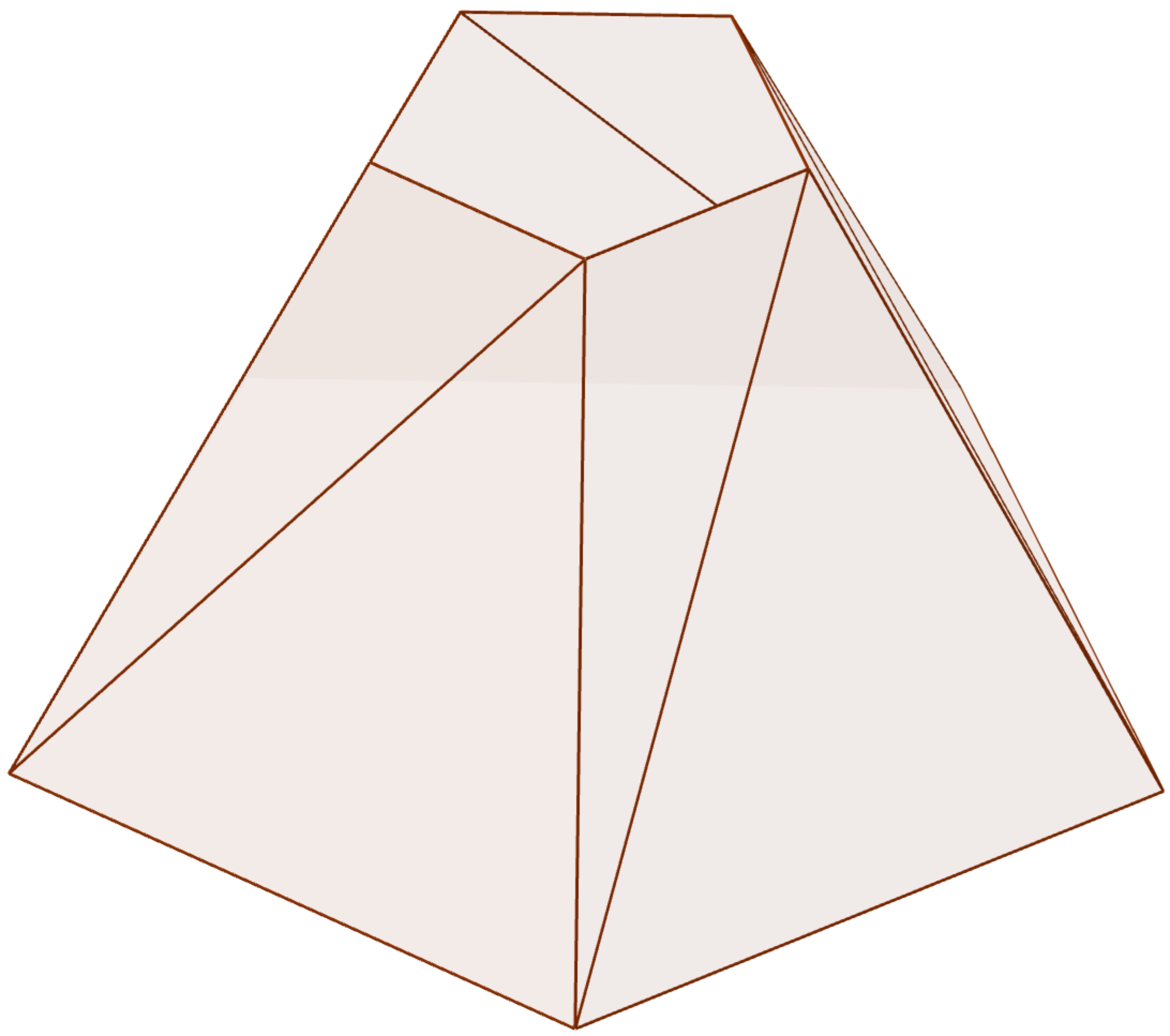}
\begin{small}
\put(5,0){e)}
\end{small}         
  \end{overpic} 	
\end{minipage}	
\caption{(a) Top and front view of a regular anti-frustum ($n=5$) visualizing the relation between the two realizations $\mathcal{R}_+$ and $\mathcal{R}_-$. Note that 
$B_{-1}^{\,\,\prime}$ and $B_{+1}^{\,\,\prime}$ are symmetric with respect to $x'$. Moreover, the interval of $\gamma\in\left]\tfrac{\pi}{2};\pi-\tfrac{\pi}{n}\right]$ corresponds 
to the cyan colored circular arc in the top view. The axonometric views of  $\mathcal{R}_-$  and  $\mathcal{R}_+$ are displayed in (b) and (c). 
The realizations $\mathcal{R}_-$  and  $\mathcal{R}_+$  for the right boundary value $\gamma=\pi-\tfrac{\pi}{n}$ are illustrated in (d) and (e).
The shaky configuration of the left boundary value $\gamma=\tfrac{\pi}{2}$ is shown in (f).
}
  \label{fig5}
\end{figure}

The computation is based on the reference frame used for 
the top/front view in Fig.\ \ref{fig5}. With respect to this frame the vertices of the realizations $\mathcal{R}_+$ and $\mathcal{R}_-$ can be coordinatized as follows:
\begin{equation}
A_{\pm 1}=(\cos{\tfrac{\pi}{n}},-\sin{\tfrac{\pi}{n}},0)^T, \quad B_{\pm 1}=(\pm r\cos{\gamma},r\sin{\gamma},h_{\pm})^T, 
\end{equation}
where we can assume without loss of generality that $h_{\pm}\geq 0$ and $0<r<1$ holds, as we set the base radius equal to $1$ (which eliminates similarities). 
  
All further vertices $A_{\pm k}$ and $B_{\pm k}$ can be obtained by applying rotations by $(k-1)\tfrac{2\pi}{n}$ to these vertices 
about the $z$-axis for $k=2,\ldots,n$. 
We rotate the triangle $A_{\pm 1}A_{\pm 2}B_{-1}$ about the line $A_{\pm 1}A_{\pm 2}$ and consider the trajectory of the point $B_{-1}$ above the $xy$-plane. 
This half-circle intersects the cylinder of rotation $\Gamma$ with radius $r$  not only in the point $B_{-1}$ but also in the point $B_{+1}$ with
\begin{equation}
h_+:=\sqrt{h_-^2-4r\cos{\gamma}\cos{\tfrac{\pi}{n}}}.
\end{equation}
This rotation also implies the interval for the angle 
$\gamma\in\left]\tfrac{\pi}{2};\pi-\tfrac{\pi}{n}\right]$:  
\begin{enumerate}[a)]
\item
In the left bound $B_{+1}$ coincides with $B_{-1}$, which is a double solution of the intersection of the circle with $\Gamma$ 
implying a shaky configuration (Fig.\ \ref{fig5}f). Moreover this already violates the assumption $\lambda_->\lambda_+$.  
\item 
Going over the right bound yields self-intersections of the anti-frustum during its snapping from $\mathcal{R}_+$ to $\mathcal{R}_-$ (Fig.\ \ref{fig5}d). 
The realization $\mathcal{R}_+$, which corresponds with the right bound, results in a frustum, where each lateral face has an additional edge 
along a quad diagonal (Fig.\ \ref{fig5}e). 
\end{enumerate}

\begin{rmk}
Alternatively, $\gamma$ can be of the interval $\gamma\in\left[\pi+\tfrac{\pi}{n}; \tfrac{3\pi}{2}\right[$ where we get 
a set of snapping anti-frusta, which is reflection symmetric with respect to the $xz$-plane to the one obtained above. 
\hfill $\diamond$
\end{rmk}

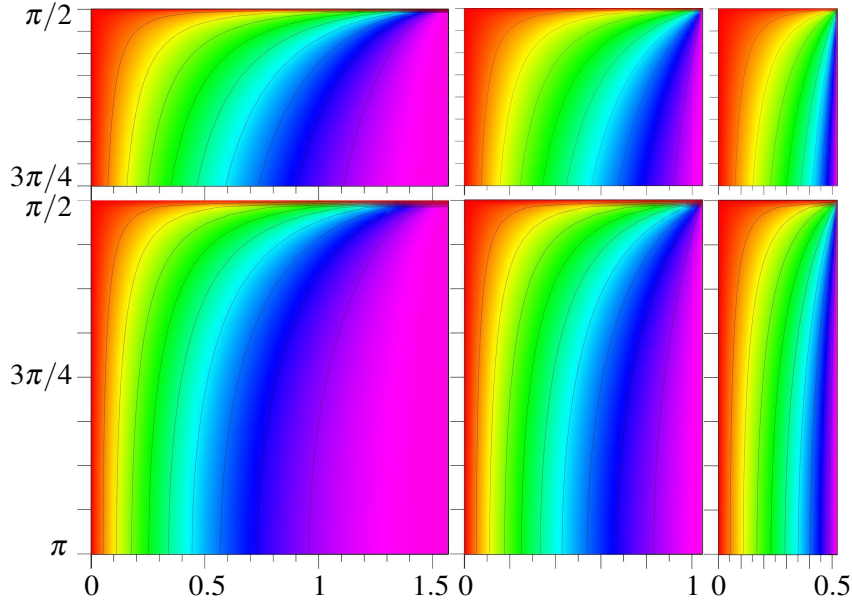
\begin{figure}[t]
\begin{center}
\begin{overpic}
    [width=100mm]{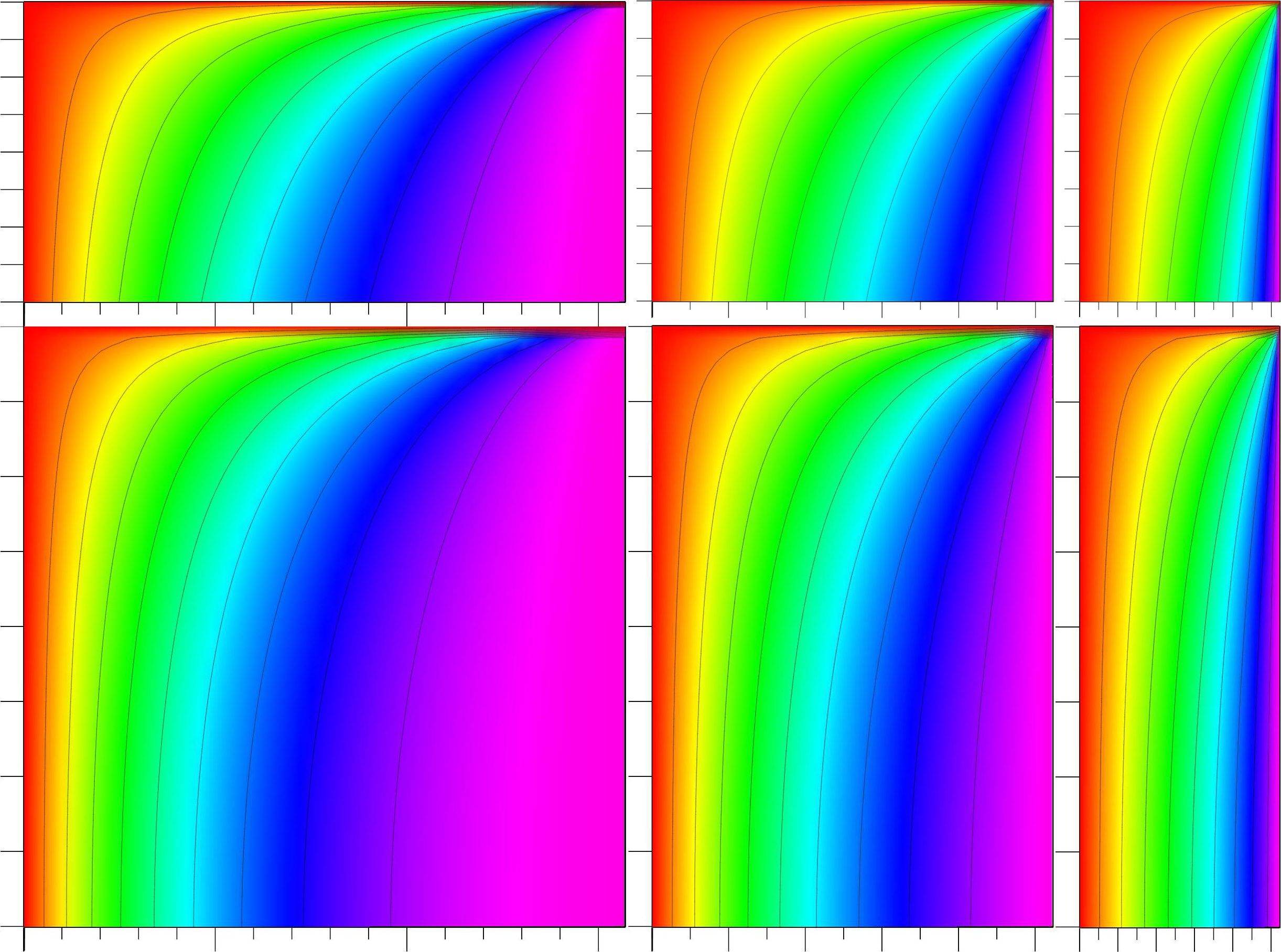}
\begin{small}
\put(1,-3.5){$0$}
\put(14.7,-3.5){$0.5$}
\put(31,-3.5){$1$}
\put(44.1,-3.5){$1.5$}
\put(50.2,-3.5){$0$}
\put(80.1,-3.5){$1$}
\put(83.5,-3.5){$0$}
\put(97,-3.5){$0.5$}
\put(-3.6,1.8){$\pi$}
\put(-8.8,24.5){$3\pi/4$}
\put(-6.9,46.8){$\pi/2$}
\put(-8.8,50.8){$3\pi/4$}
\put(-6.9,72.2){$\pi/2$}
\end{small}     
  \end{overpic} 
\end{center}	
\caption{The upper row shows the plots of $r$ for $n=4$ and $\lambda_-=\tfrac{\pi}{2}$ (left), $\lambda_-=\tfrac{\pi}{3}$ (center) and $\lambda_-=\tfrac{\pi}{6}$ (right), 
respectively. The horizontal axis corresponds to $\lambda_+\in]0;\lambda_-[$ and the vertical one to $\gamma\in\left]\tfrac{\pi}{2};\pi-\tfrac{\pi}{n}\right]$. 
The same is displayed in the lower row for the limit  $n\rightarrow\infty$. In all figures nine contour lines are given in $0.1$ steps from $0$ (magenta) to $1$ (red). 
}
  \label{fig0}
\end{figure}

The height $h_-$ and $\lambda_-$ are related by $h_-=\tfrac{1-r}{l_-}$. Moreover, one can express the radius $r$ in dependence of $l_{\pm}$ as follows: 
\begin{equation}\label{r_all}
r=
\frac{
2l_-^2l_+^2\cos{\tfrac{\pi}{n}}\cos{\gamma} + 2l_-l_+ \sqrt{\cos{\tfrac{\pi}{n}}\cos{\gamma}\left(l_-^2l_+^2\cos{\tfrac{\pi}{n}}\cos{\gamma}+l_+^2-l_-^2\right)
} +l_+^2-l_-^2
}
{
l_+^2-l_-^2
}.
\end{equation}
For the special case of flat-foldability $h_-=0$ ($\Leftrightarrow$ $\lambda_-=\tfrac{\pi}{2}$) this formula simplifies to:
\begin{equation}\label{r_flat}
r=1-2l_+\left(
l_+\cos{\tfrac{\pi}{n}}\cos{\gamma}+\sqrt{\cos{\tfrac{\pi}{n}}\cos{\gamma}\left(l_+^2\cos{\tfrac{\pi}{n}}\cos{\gamma}-1\right)}
\right).
\end{equation}
All these values of $r$ given in Eq.\ (\ref{r_all}) and Eq.\ (\ref{r_flat}), respectively, are real for $n>2$, $\gamma\in\left]\tfrac{\pi}{2};\pi-\tfrac{\pi}{n}\right]$ and 
$\lambda_{\pm}\in]0;\tfrac{\pi}{2}]$ with $\lambda_->\lambda_+$. 
Therefore by giving the input of the design parameters $\lambda_{\pm}$ there exists for every $n>2$ a one-parametric solution set (parameter $\gamma$), which is 
illustrated in Fig.\ \ref{fig0}.

%%%%%%%%%%%%%%%%%%%%%%%%%%%%%%%%%%%%%%%%%%%%%%%%%%%%%%%%%%%%%%%%%%%%%%%%%%%%%%%%%%%
%%%%%%%%%%%%%%%%%%%%%%%%%%%%%%%%%%%%%%%%%%%%%%%%%%%%%%%%%%%%%%%%%%%%%%%%%%%%%%%%%%%
%%%%%%%%%%%%%%%%%%%%%%%%%%%%%%%%%%%%%%%%%%%%%%%%%%%%%%%%%%%%%%%%%%%%%%%%%%%%%%%%%%%
%%%%%%%%%%%%%%%%%%%%%%%%%%%%%%%%%%%%%%%%%%%%%%%%%%%%%%%%%%%%%%%%%%%%%%%%%%%%%%%%%%%
%%%%%%%%%%%%%%%%%%%%%%%%%%%%%%%%%%%%%%%%%%%%%%%%%%%%%%%%%%%%%%%%%%%%%%%%%%%%%%%%%%%
%%%%%%%%%%%%%%%%%%%%%%%%%%%%%%%%%%%%%%%%%%%%%%%%%%%%%%%%%%%%%%%%%%%%%%%%%%%%%%%%%%%
%%%%%%%%%%%%%%%%%%%%%%%%%%%%%%%%%%%%%%%%%%%%%%%%%%%%%%%%%%%%%%%%%%%%%%%%%%%%%%%%%%%
%%%%%%%%%%%%%%%%%%%%%%%%%%%%%%%%%%%%%%%%%%%%%%%%%%%%%%%%%%%%%%%%%%%%%%%%%%%%%%%%%%%
%%%%%%%%%%%%%%%%%%%%%%%%%%%%%%%%%%%%%%%%%%%%%%%%%%%%%%%%%%%%%%%%%%%%%%%%%%%%%%%%%%%

\section{Spiral-motion based conical triangulation}\label{sec:spiral}

We start with a parametrization of the spiral, which reads as:
\begin{equation}\label{eq:pardisspir}
\Vkt s(\phi)=
\begin{pmatrix}
\phm re^{m\varphi}\cos\varphi \\
\phm re^{m\varphi}\sin\varphi \\\
-re^{m\varphi}\cot\lambda
\end{pmatrix}\quad\text{with} \quad r>0 \quad \text{and} \quad
m=\sin\lambda\cot\delta
\end{equation}
where $\delta$ is the angle between the spiral tangent and the corresponding generator of the cone. 
We want to consider the part of the spiral which starts below the $xy$-plane (for $\varphi=0$) and winds up (mathematically positive) to 
the origin for $\varphi \rightarrow \infty$; i.e.\ $\varphi\in\RR^+$. This is the case for $\delta \in]\tfrac{\pi}{2};\pi[$. 

We consider two different realizations $\mathcal{R}_+$ and $\mathcal{R}_-$ of our vertex set $V_i$; i.e.\
\begin{equation}
V_{\pm i}=\begin{pmatrix}
r_{\pm}p_{\pm}^i\cos(i\phi_{\pm}) \\
r_{\pm}p_{\pm}^i\sin(i\phi_{\pm}) \\
-r_{\pm}p_{\pm}^i q_{\pm}
\end{pmatrix}
\quad\text{with} \quad p_{\pm}:=e^{\phi_{\pm} \sin\lambda_{\pm}\cot\delta_{\pm}} 
\quad \text{and} \quad q_{\pm}=\cot\lambda_{\pm},
\end{equation}
where $\phi_{-}$ and $\phi_{+}$ are some fixed values out of the interval $]0;\pi[$. 
Without loss of generality we can set $r_+=1$ which again eliminates the factor of similarities. 
Clearly, the lengths of corresponding edges have to agree, which yields an infinite set $\mathcal{E}$ of 
equations $d(i,j)=0$ with
\begin{equation}
d(i,j) :=\overline{V_{+i}V_{+j}}^2-\overline{V_{-i}V_{-j}}^2
\end{equation}
for suitable pairs $(i,j)$. By looking at the structure it can be seen that its complete edge set 
can be generated by the spiral displacement of the three edges $V_{\pm 0}V_{\pm 1}$, $V_{\pm 0}V_{\pm (n-1)}$ and 
$V_{\pm 0}V_{\pm n}$, respectively. 
This shows that the relations 
\begin{align}
d(k,k+1) &= p_+^{2k}\,\overline{V_{+0}V_{+1}}^2 - p_-^{2k}\,\overline{V_{-0}V_{-1}}^2 \\
d(k,k+n-1) &= p_+^{2k}\,\overline{V_{+0}V_{+(n-1)}}^2 - p_-^{2k}\,\overline{V_{-0}V_{-(n-1)}}^2 \\
d(k,k+n) &= p_+^{2k}\,\overline{V_{+0}V_{+n}}^2 - p_-^{2k}\,\overline{V_{-0}V_{-n}}^2 
\end{align}
hold for $k\in\NN$. This already implies, that $\mathcal{E}$ can only have a solution for 
$p_-=p_+$ as $p_{\pm}>0$ has to hold (for reasons of reality). Therefore we can set $p:=p_-=p_+$, which bowls down the problem 
to the solution of the equations $d(0,1)=0$, $d(0,n-1)=0$ and $d(0,n)=0$. By using 
Chebyshev polynomials $T_i(x)$ of the first kind, which are recursively defined by:
\begin{equation}
T_{i+1}(x)=2xT_i(x)-T_{i-1}(x) \quad\text{with}\quad T_0(x)=1 \quad\text{and}\quad T_1(x)=x,
\end{equation}
we can rewrite $d(0,k)$ for $k=1,n-1,n$ under consideration of $T_i(\cos\phi_{\pm})=\cos(i\phi_{\pm})$ by
\begin{equation}\label{eq:d}
\begin{split}
d(0,k)=
&(p^{2k}q_+^2-2p^kq_+^2  - 2T_k(c_+)p^k + p^{2k} + q_+^2  + 1) - \\
&(p^{2k}q_-^2-2p^kq_-^2  - 2T_k(c_-)p^k + p^{2k} + q_-^2  + 1)r_-^2
\end{split}
\end{equation}
with $c_{\pm}=\cos\phi_{\pm}$. By the conducted substitutions $d(0,k)$ turns into an algebraic expression in the 
variables $p,q_{\pm},c_{\pm},r_-$, where $q_+$ and $q_-$ are known design inputs. In the following two subsections we distinguish the cases whether $\mathcal{R}_+$ and $\mathcal{R}_-$ 
are located on the same cone  or not.

\subsection{Two realizations on different cones}\label{sec:different}

{\bf General Case:} 
Let us assume that none of the two cones degenerates to a plane; i.e.\ $q_{\pm}\neq 0$. 
From $d(0,1)=0$ we can compute $r_-$ and 
substitute the obtained expression into $d(0,n-1)$ and $d(0,n)$, respectively, which both factor into $p$ 
and a remaining factor  $d_*(0,n-1)$ and $d_*(0,n)$, respectively. 
Finally, we eliminate $p$ from $d_*(0,n-1)$ and $d_*(0,n)$ by means of resultants, which yield 
the expression 
\begin{equation}\label{eq:fac}
q_-^4q_+^4(c_+ - 1)^6(c_- - 1)^6(c_- - c_+)^6(c_-q_+^2-c_+q_-^2+q_+^2+q_-^2 + c_- - c_+)^{2n-2}u^2,
\end{equation}
where $u=0$ is the equation of an algebraic curve\footnote{Note that the factor $u$ appears with multiplicity two in Eq.\ (\ref{eq:fac})
as for each solution point of the curve $\go h$ there exist two values for $p$, which are reciprocal to each other. 
Due to our assumptions $\delta \in]\tfrac{\pi}{2};\pi[$ and $\lambda_{\pm}\in]0;\tfrac{\pi}{2}]$ we have to pick the $p$ value 
fulfilling $1>p>0$.} 
$\go h$ of degree $n^2-9$ in $c_-$ and $c_+$ (as $q_+$ and $q_-$ are design inputs).
For reasons of reality we are only interested in the 
domain $c_{\pm}\in[-1;1]$ of $\go h$. Note that not every real curve point within this domain implies a real solution, as also $p\in]0;1[$ has to hold, 
which is obtained as common zero of $d_*(0,n)=0$ and $d_*(0,n-1)=0$ after back-substitution of $c_-$ and $c_+$. 

Note that the other factors of Eq.\ (\ref{eq:fac}) beside $u$
do not yield a solution to our problem, which is shown by the following case study: 
\begin{enumerate}
\item
$c:=c_-=c_+$: In this case the greatest common divisor (gcd) of $d_*(0,n-1)$ and $d_*(0,n)$ equals 
\begin{equation}
(p-1)^2(q_-^2-q_+^2)(c-1)(2cp-p^2-1),
\end{equation}
which can only vanish (under consideration of reality) for $q_-=q_+$, but then the realizations $\mathcal{R}_+$ and $\mathcal{R}_-$ are the same contradicting our 
assumption.
\item
$c_-\neq c_+$ and $c_-q_+^2-c_+q_-^2+q_+^2+q_-^2 + c_- - c_+=0$: This equation can be solved for $c_+$ without loss of generality. Then the gcd of $d_*(0,n-1)$ and $d_*(0,n)$ equals 
\begin{equation}
p^{n-2}(q_+^2+1)(q_-^2-q_+^2)(c_--1)^2(2q_-^2p-q_-^2p^2+2c_-p-q_-^2-p^2-1).
\end{equation}
As the vanishing of the last factor implies $r_-=0$ we only remain with $q_-=q_+$ resulting in $c_-=c_+$, a contradiction. 
\end{enumerate}
This also implies that multi-stable designs can only exist for $n>3$, which is demonstrated within the following example.

\begin{figure}[t]
\begin{overpic}
    [width=35mm]{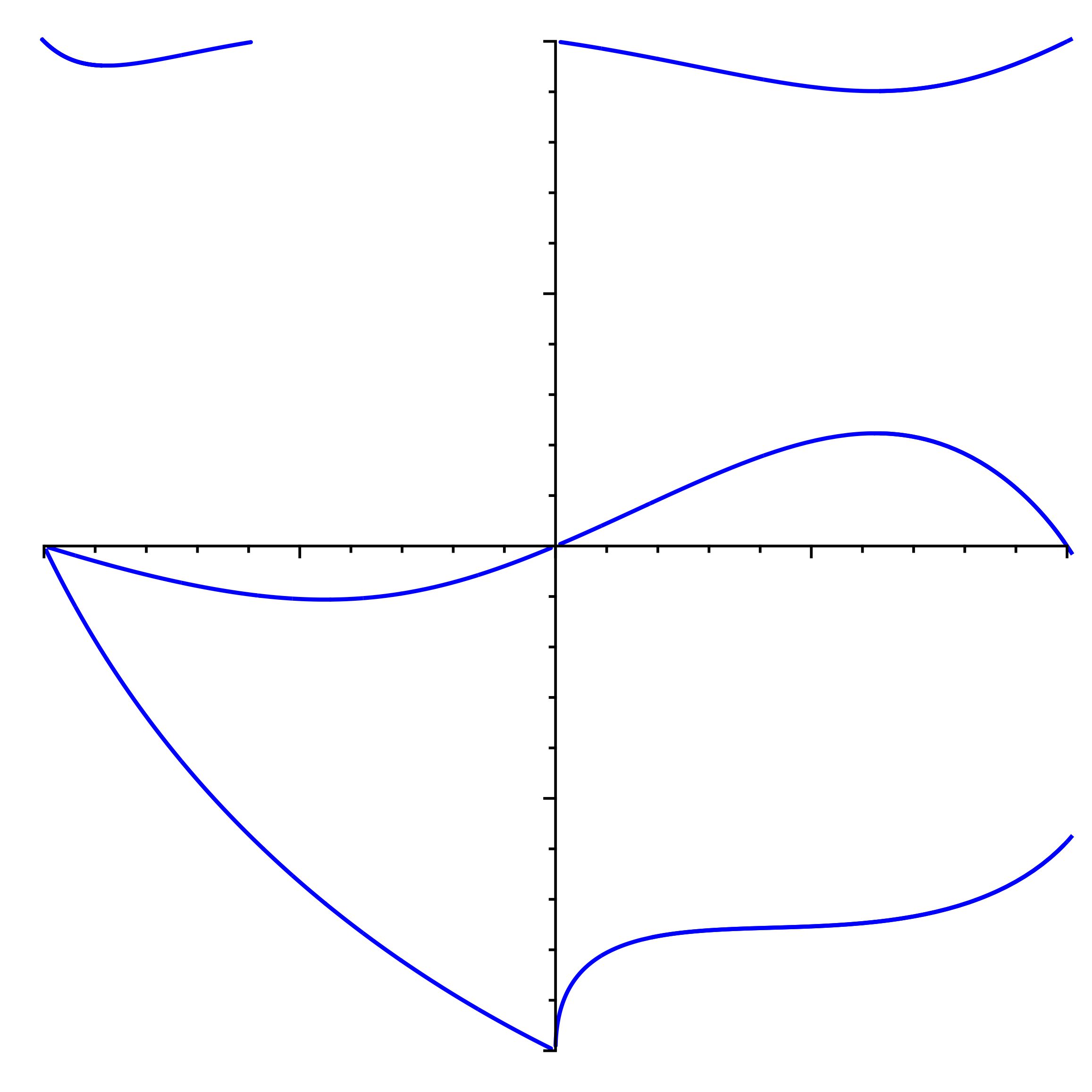}
\begin{small}
\put(0,0){a)}
\put(73,43){$c_-$}
\put(39,73){$c_+$}
\put(44.5,92){$1$}
\put(95,41){$1$}
\put(52,41){$0$}
\end{small}     
  \end{overpic} 
\hfill
 \begin{overpic}
    [width=35mm]{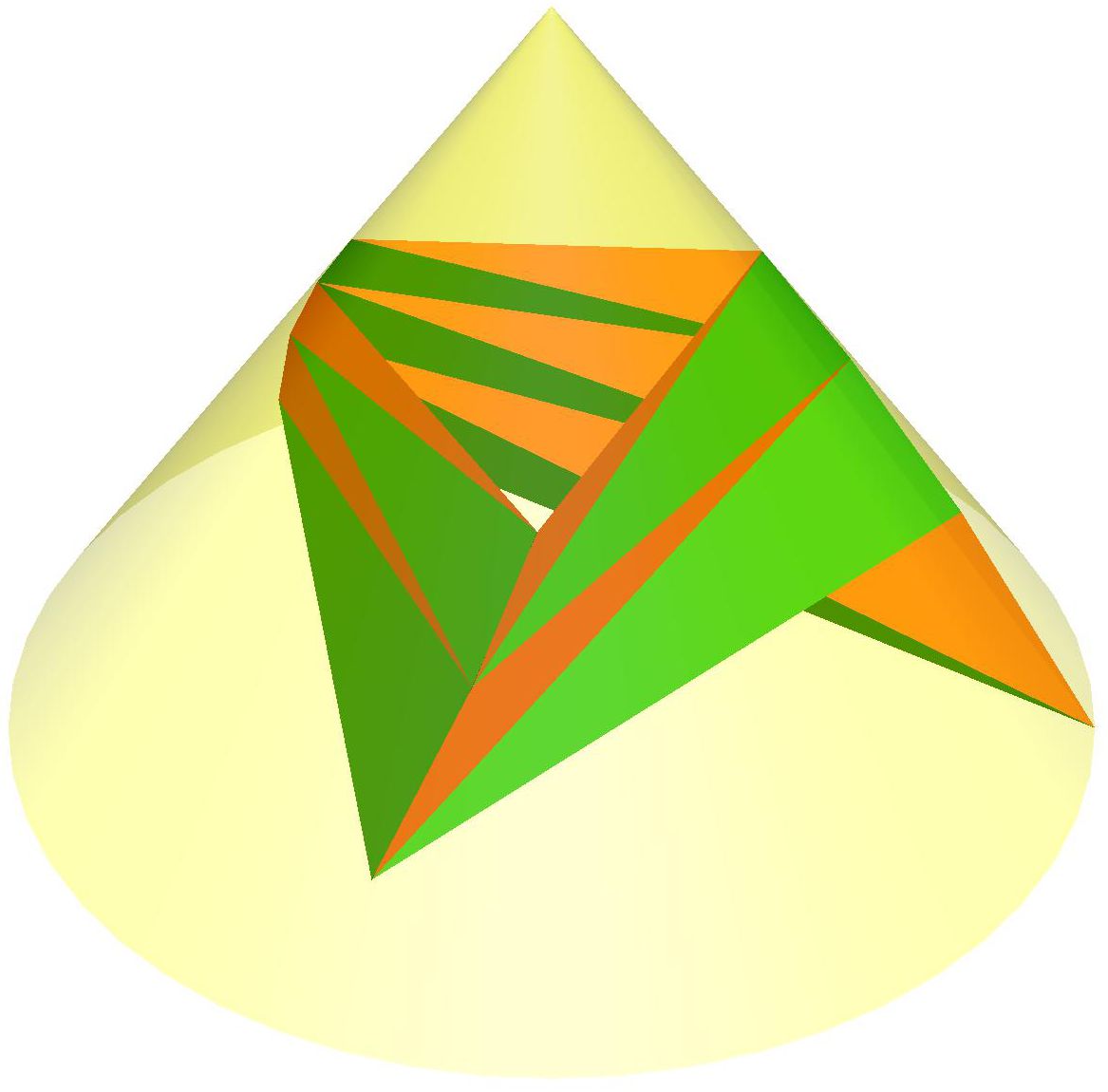}
\begin{small}
\put(0,0){b)}
\end{small}         
  \end{overpic} 
	\hfill
\begin{overpic}
    [width=35mm]{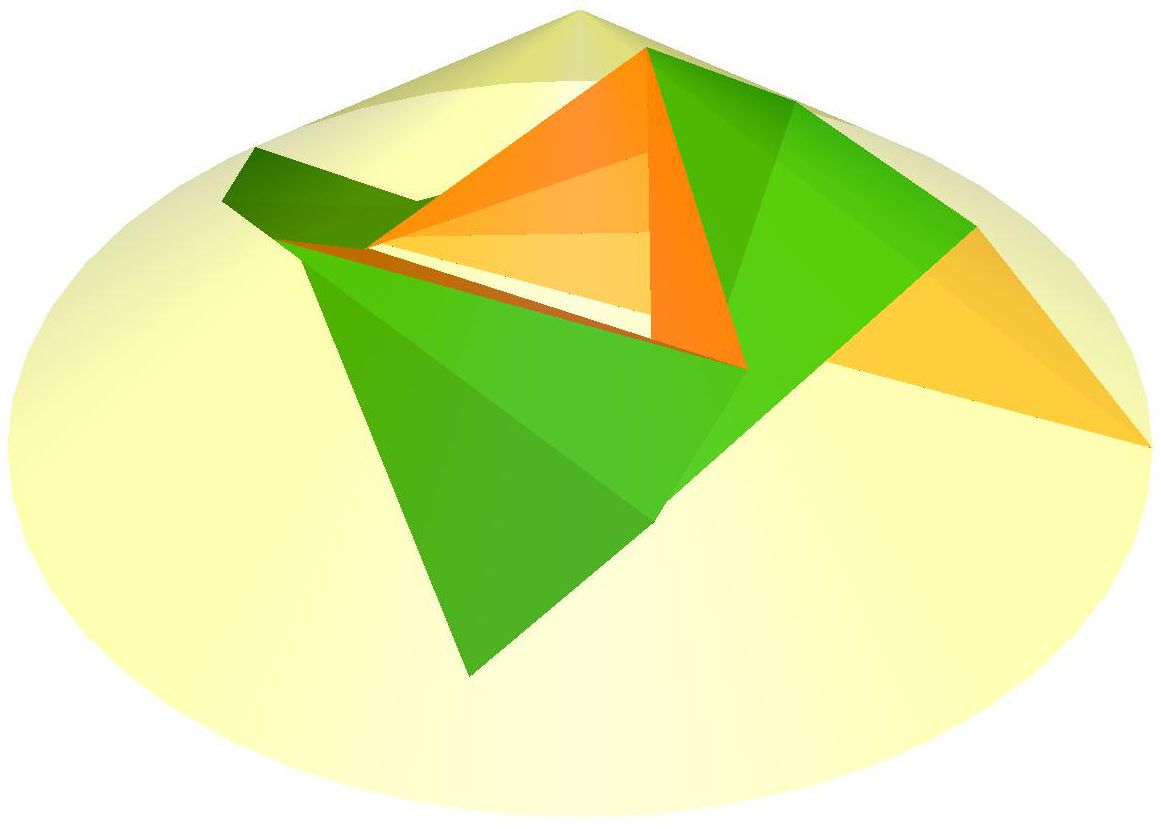}
\begin{small}
\put(0,0){c)}
\end{small}         
  \end{overpic} 	
\caption{(a) Planar septic in $c_+$ and $c_-$ within the domain $c_{\pm}\in[-1;1]$ 
(b/c) Visualization of $\mathcal{R}_{-/+}$  enclosed within the cone $\Lambda_{-/+}$, where the two different kinds of 
triangles of the triangulated cone are colored in red and green, respectively. 
}
  \label{fig6}
\end{figure}
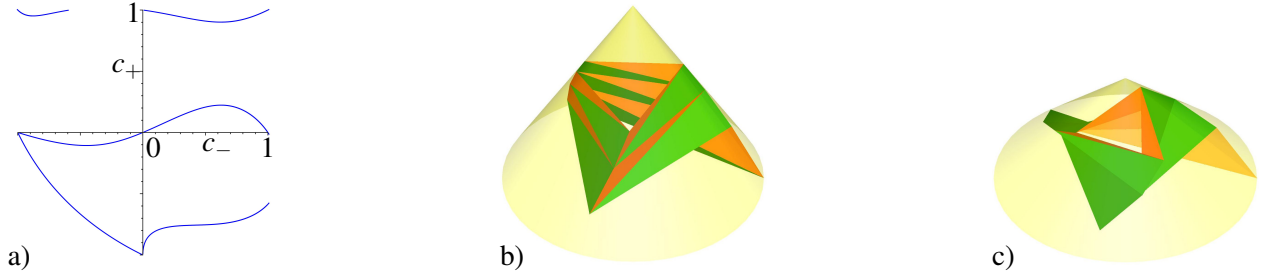

\begin{example}
For this example we set $n=4$, $\lambda_+=\tfrac{\pi}{3}$ and $\lambda_-=\tfrac{\pi}{6}$, respectively. The elimination procedure described above yields the planar 
algebraic curve of degree $7$ in $c_+$ and $c_-$, which is visualized in Fig.\ \ref{fig6}a. 
The septic point $(c_-,c_+)=(-0.55,-0.620223)$ implies $p=0.914773$. The corresponding 
realizations $\mathcal{R}_-$ and $\mathcal{R}_+$ are displayed in Fig.\ \ref{fig6}b and Fig.\ \ref{fig6}c, respectively.
\end{example}

\noindent
{\bf Special Case:} If one of the two cones degenerates to a plane we can assume that $q_+=0$ holds  without loss of generality.
Now the expressions  $d_*(0,n-1)$ and $d_*(0,n)$ factor both into 
$(2c_+p-p^2-1)$ and a further factor $d_{**}(0,n-1)$ and $d_{**}(0,n)$. 
The resultant of the later two expressions with respect to $p$ yields 
\begin{equation}
q_-^4(c_+-1)^2(c_--1)^4(c_--c_+)^2(q_-^2-c_+q_-^2+c_--c_+)^{2n-4}u^2,
\end{equation}
where $u=0$ is an algebraic curve $\go h$ of degree $(n-1)^2-4$ in $c_-$ and $c_+$. 
An analogous case study as in the general case shows that no other factor than $u$ can vanish without contradiction. 
Again multi-stable designs can only exist for $n>3$.

\subsection{Two realizations on the same cone}\label{sec:same}

{\bf General Case:} As we want to compute realizations on the same cone we can set $q:=q_-=q_+\neq 0$. 
From $d(0,1)=0$ we compute again $r_-$ and plug the obtained expression into $d(0,n-1)$ and $d(0,n)$, respectively, 
which both factor into $p(c_- - c_+)$ and a remaining factor  $d_*(0,n-1)$ and $d_*(0,n)$, respectively. 
Finally, we eliminate $p$ from $d_*(0,n-1)$ and $d_*(0,n)$ by means of resultants, which yield 
\begin{equation}
q^8(c_+ - 1)^6(c_- - 1)^6(q^2+1)^{2n-2}u^2,
\end{equation}
where $u=0$ now corresponds with an algebraic curve $\go h$ of degree $n(n-1)-6$. 
Therefore multi-stable designs can only exist for $n>3$.

\begin{figure}[t]
\begin{overpic}
    [width=35mm]{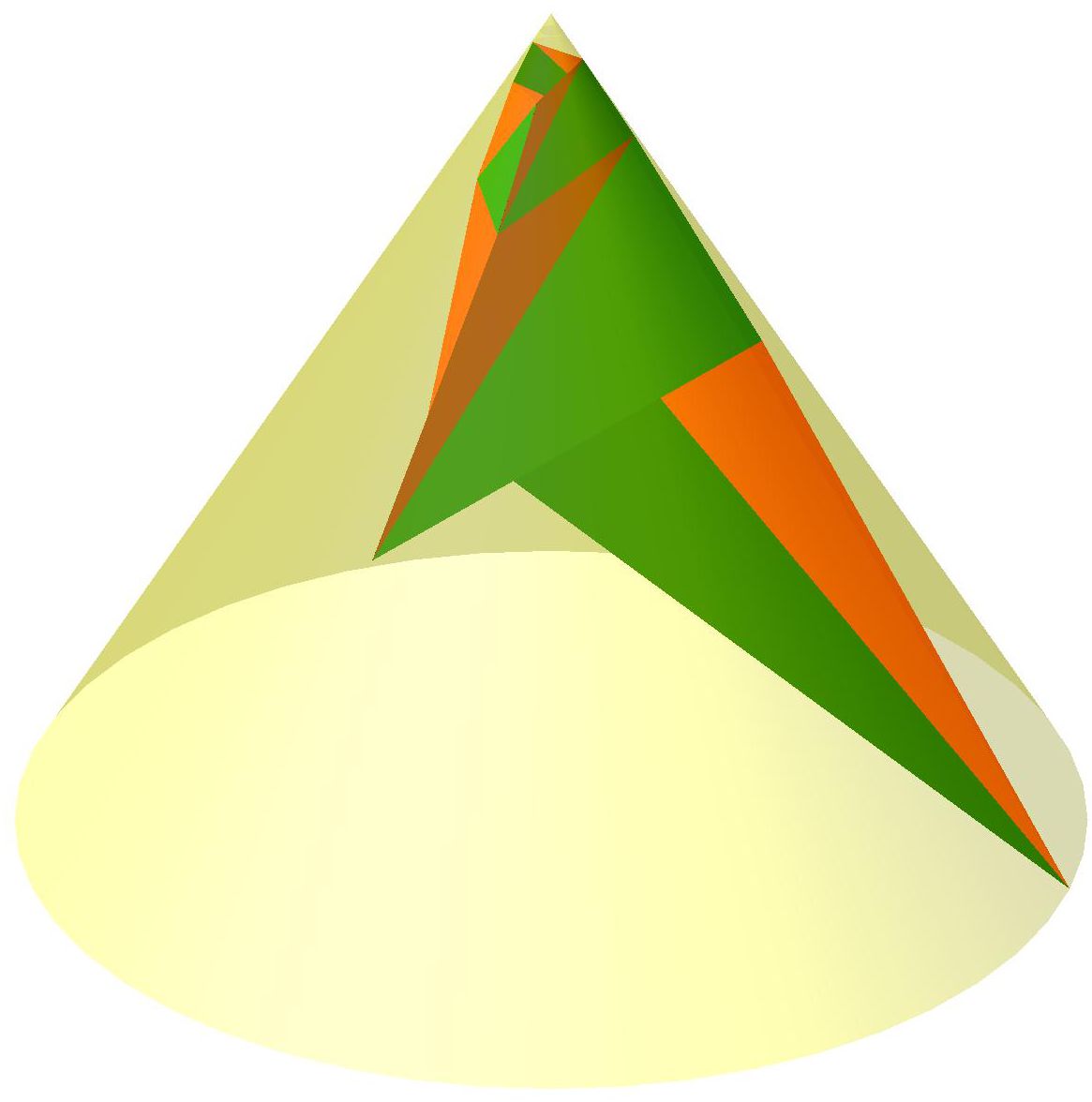}
\begin{small}
\put(0,0){a)}
\end{small}     
  \end{overpic} 
\hfill
 \begin{overpic}
    [width=35mm]{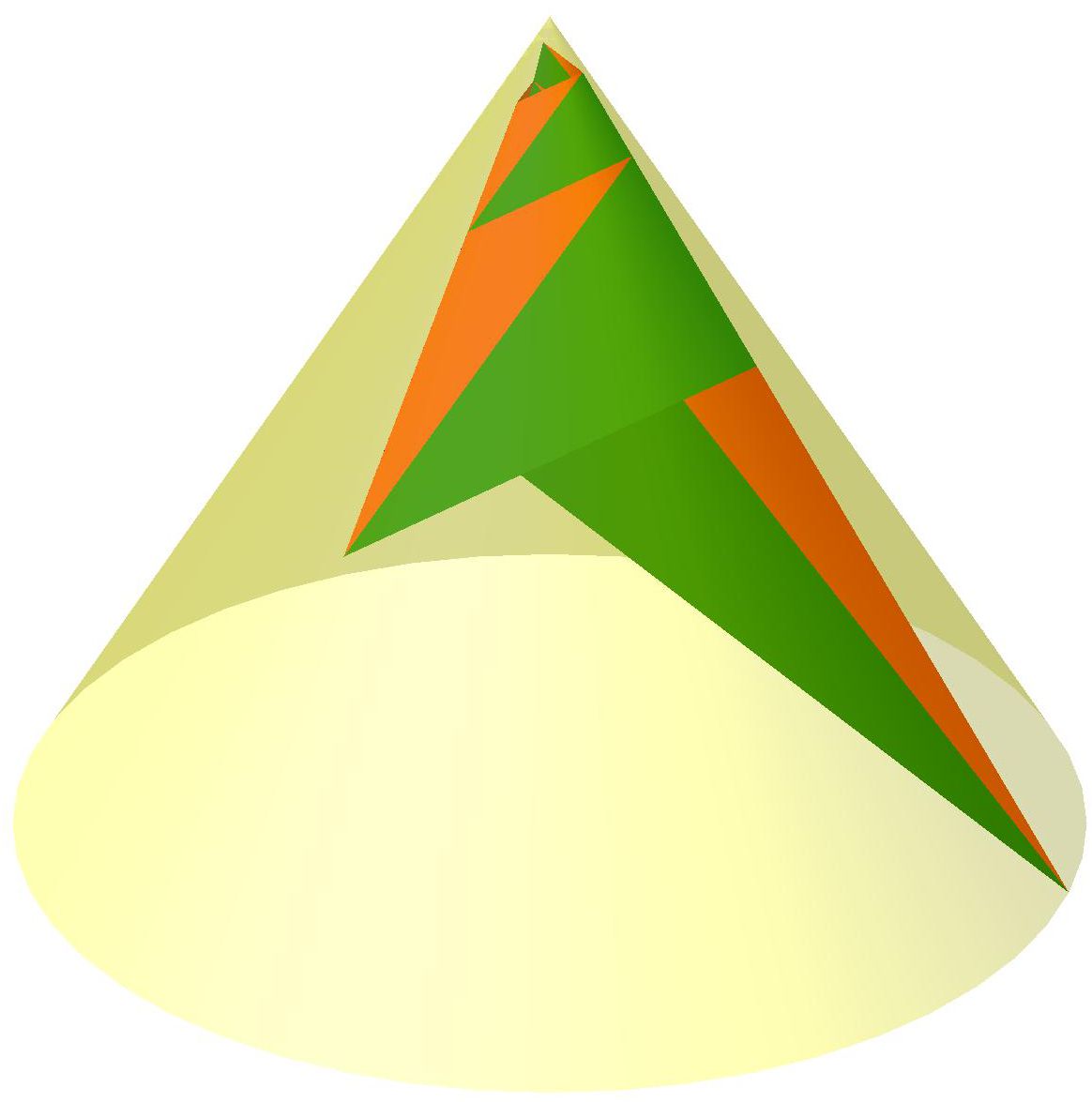}
\begin{small}
\put(0,0){b)}
\end{small}         
  \end{overpic} 
	\hfill
\begin{overpic}
    [width=35mm]{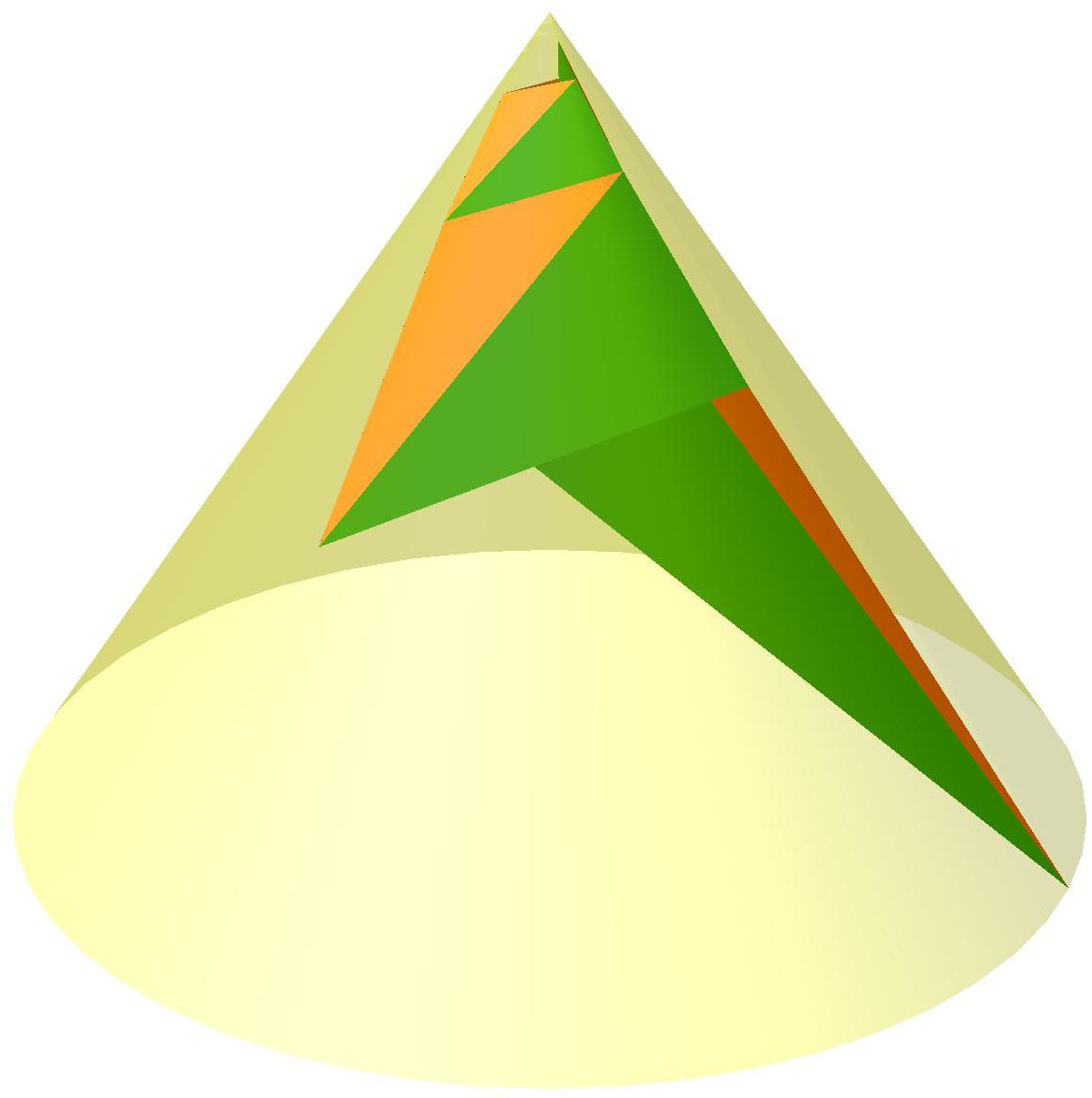}
\begin{small}
\put(0,0){c)}
\end{small}         
  \end{overpic} 	
	\\ \phm \\
\begin{overpic}
    [width=35mm]{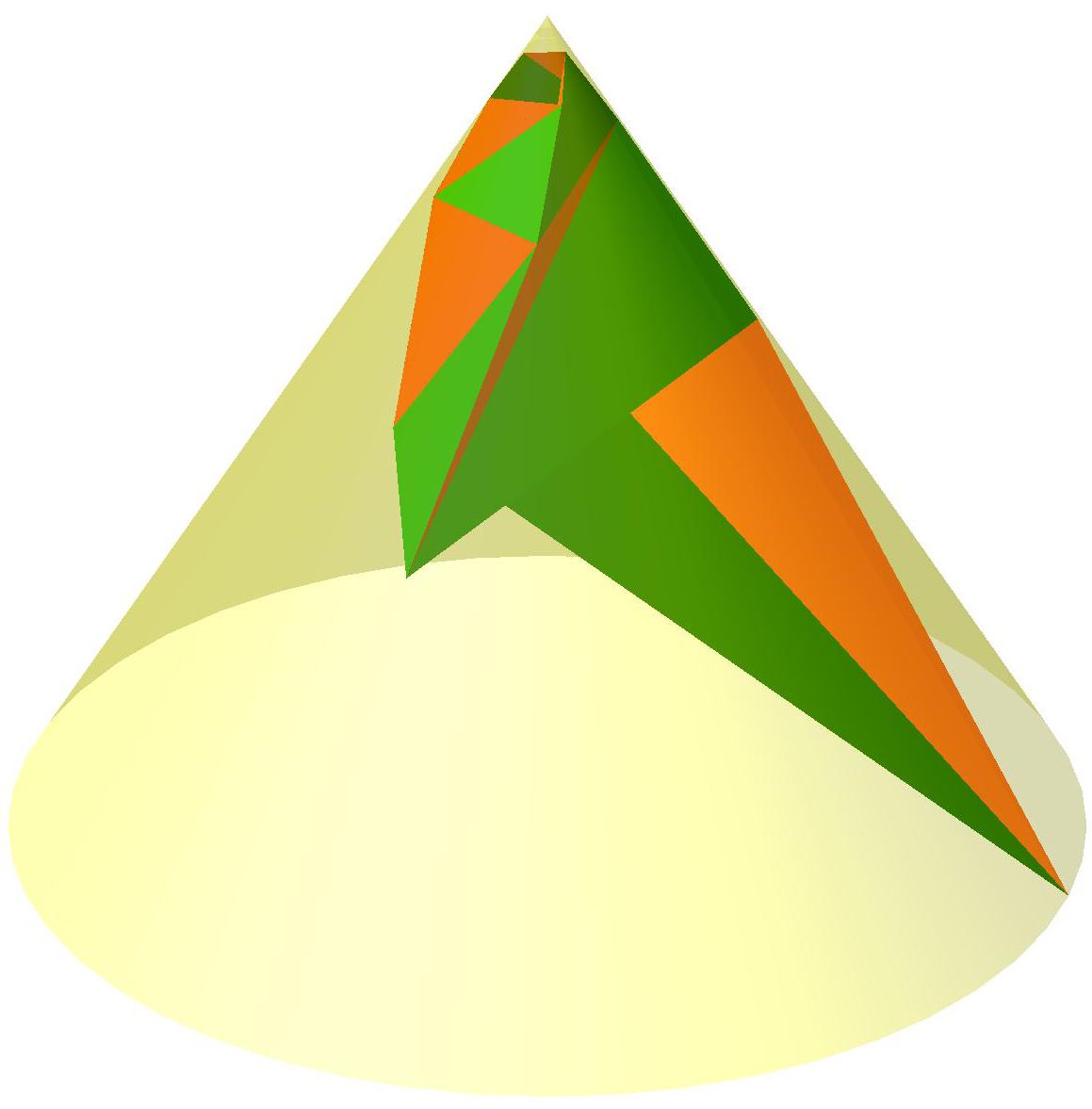}
\begin{small}
\put(0,0){d)}
\end{small}     
  \end{overpic} 
\hfill
 \begin{overpic}
    [width=35mm]{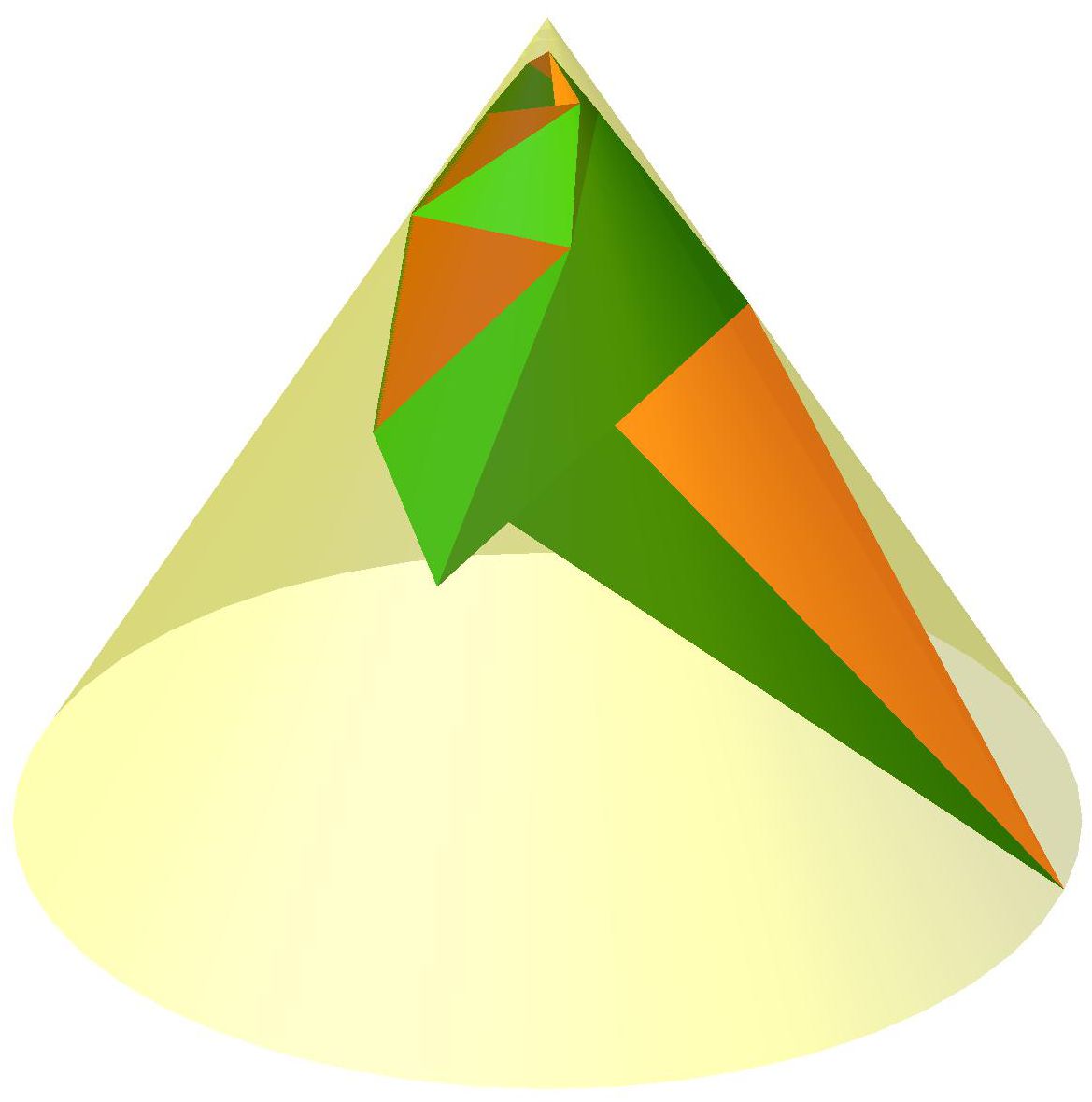}
\begin{small}
\put(0,0){e)}
\end{small}         
  \end{overpic} 
	\hfill
\begin{overpic}
    [width=35mm]{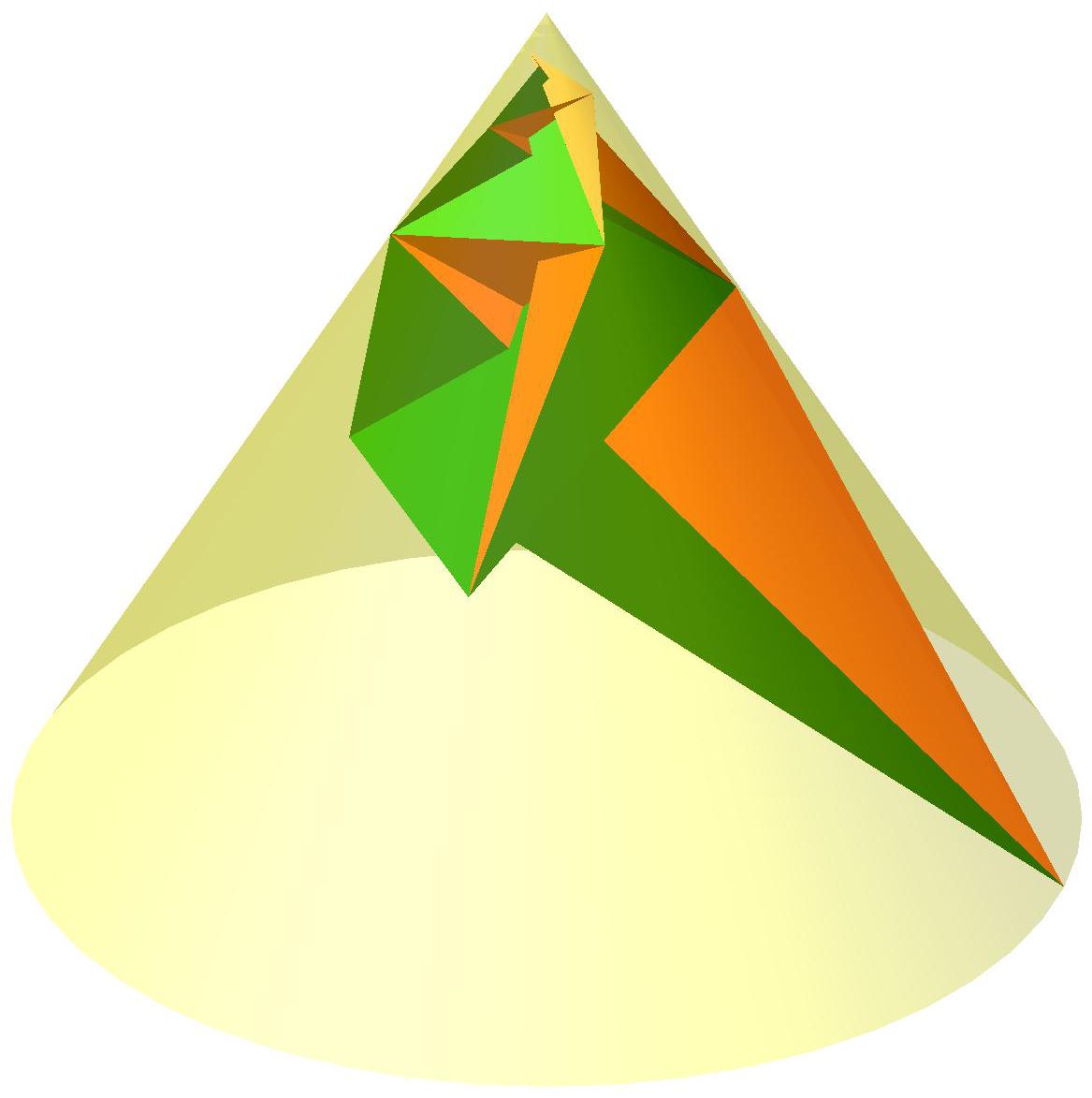}
\begin{small}
\put(0,0){f)}
\end{small}         
  \end{overpic} 	

\caption{
(a/b/c) Visualization of $\mathcal{R}_{-}^{1/2/3}$. 
The circumstance that $\mathcal{R}_-^2$ is located on a three-sided pyramid can easily be seen by the fact that the series of vertices $V_i,V_{i+3},V_{i+6}, V_{i+9}, \ldots$ ($i=0,1,2$) is located on one generator of the cone.
(d/e/f) Visualization of $\mathcal{R}_{+}^{1/2/3}$. 
}
  \label{fig7}
\end{figure}

\begin{example}\label{ex3}
In this example we want to compute realizations for $n=4$, which are located on the same cone with $\lambda_+=\lambda_-=\tfrac{\pi}{6}$. 
We consider the following three points $H^i=(c_-,c_+)$ of the resulting planar sextic $\go h$: 
\begin{equation}
\begin{split}
H^1&=(-0.55,-0.615158) \quad \text{with}\quad p=0.733163, \\ 
H^2&=(-0.50,-0.661566) \quad \text{with}\quad p=0.735983, \\ 
H^3&=(-0.45,-0.705150) \quad \text{with}\quad p=0.741100, 
\end{split}
\end{equation}
which are visualized in Fig.\ \ref{fig7}. 
The corresponding three realizations $\mathcal{R}_-^i$ for $i=1,2,3$ as well as $\mathcal{R}_+^1$ are without any self-intersections. Note that 
$\mathcal{R}_-^2$ is located on a three-sided pyramid. 
 The realization $\mathcal{R}_+^3$ has self-intersections and the realization 
$\mathcal{R}_+^2$ is the border case; i.e.\ $V_k,V_{k+1},V_{k+4},V_{k+5}$ are coplanar for $k\in\NN$.
\end{example}

\noindent
{\bf Special Case:} For the flat realization $q=0$ the expressions  $d_*(0,n-1)$ and $d_*(0,n)$ factor both into 
$(2c_+p-p^2-1)$ and a further factor $d_{**}(0,n-1)$ and $d_{**}(0,n)$. 
The resultant of the later two expressions with respect to $p$ yields $u^2$, where $u=0$ is an algebraic curve $\go h$ of 
degree $(n-2)(n-3)$ in $c_-$ and $c_+$. 
Again multi-stable designs can only exist for $n>3$.

\subsubsection{Shaky realizations}\label{subsec:shaky}

From the above approach one can easily determine the algebraic characterization of a shaky realization $\mathcal{R}_s$, by setting $c_s:=c_-=c_+$ in the 
expression of $\go h$. The zeros of this univariate polynomial in $c_s$ result in double solutions; i.e.\ the realizations  $\mathcal{R}_-$ and $\mathcal{R}_+$ coincide in $\mathcal{R}_s$. 

\begin{figure}[t]
\begin{overpic}
    [width=35mm]{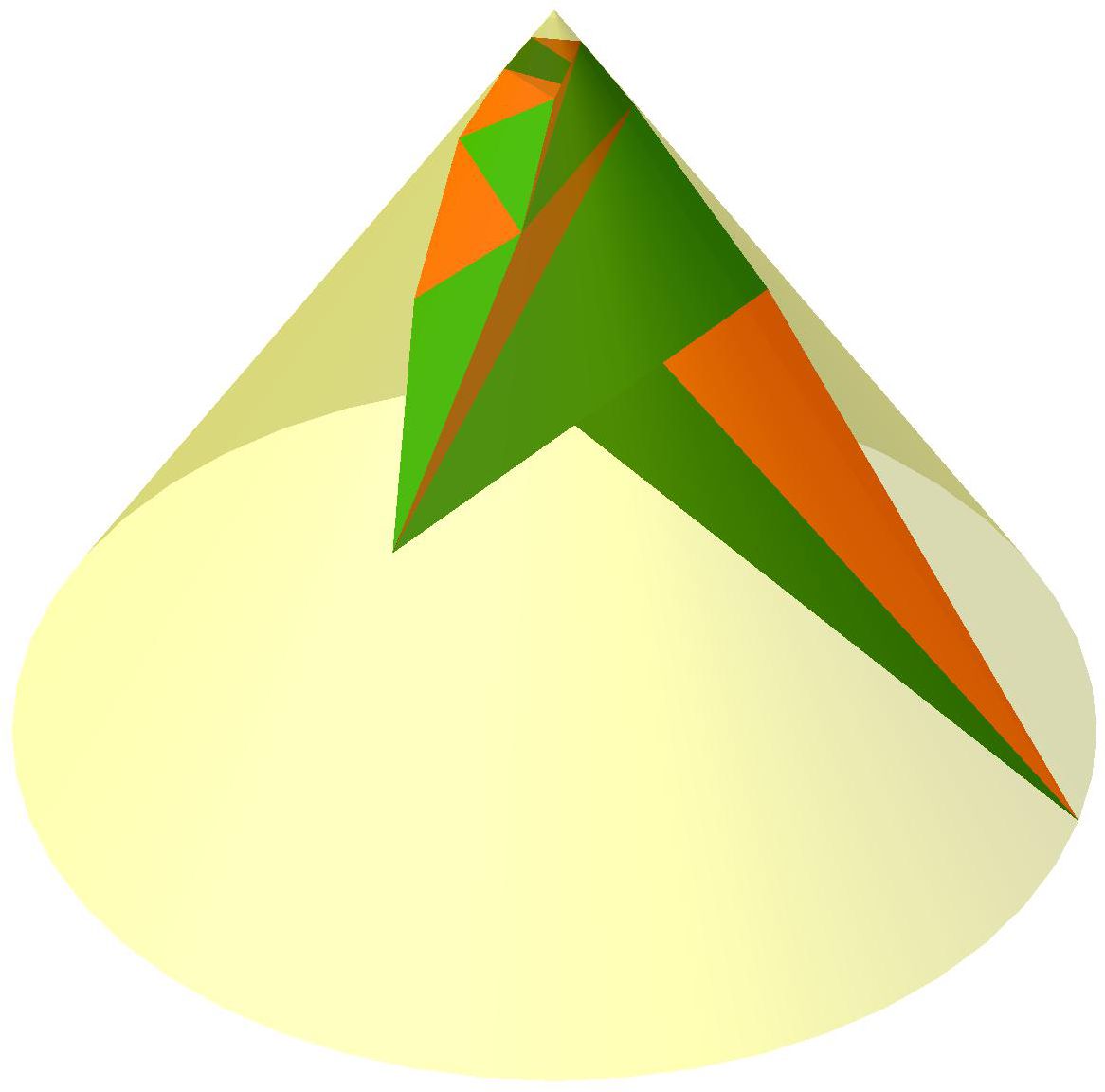}
\begin{small}
\put(0,0){a)}
\end{small}     
  \end{overpic} 
\hfill
 \begin{overpic}
    [width=35mm]{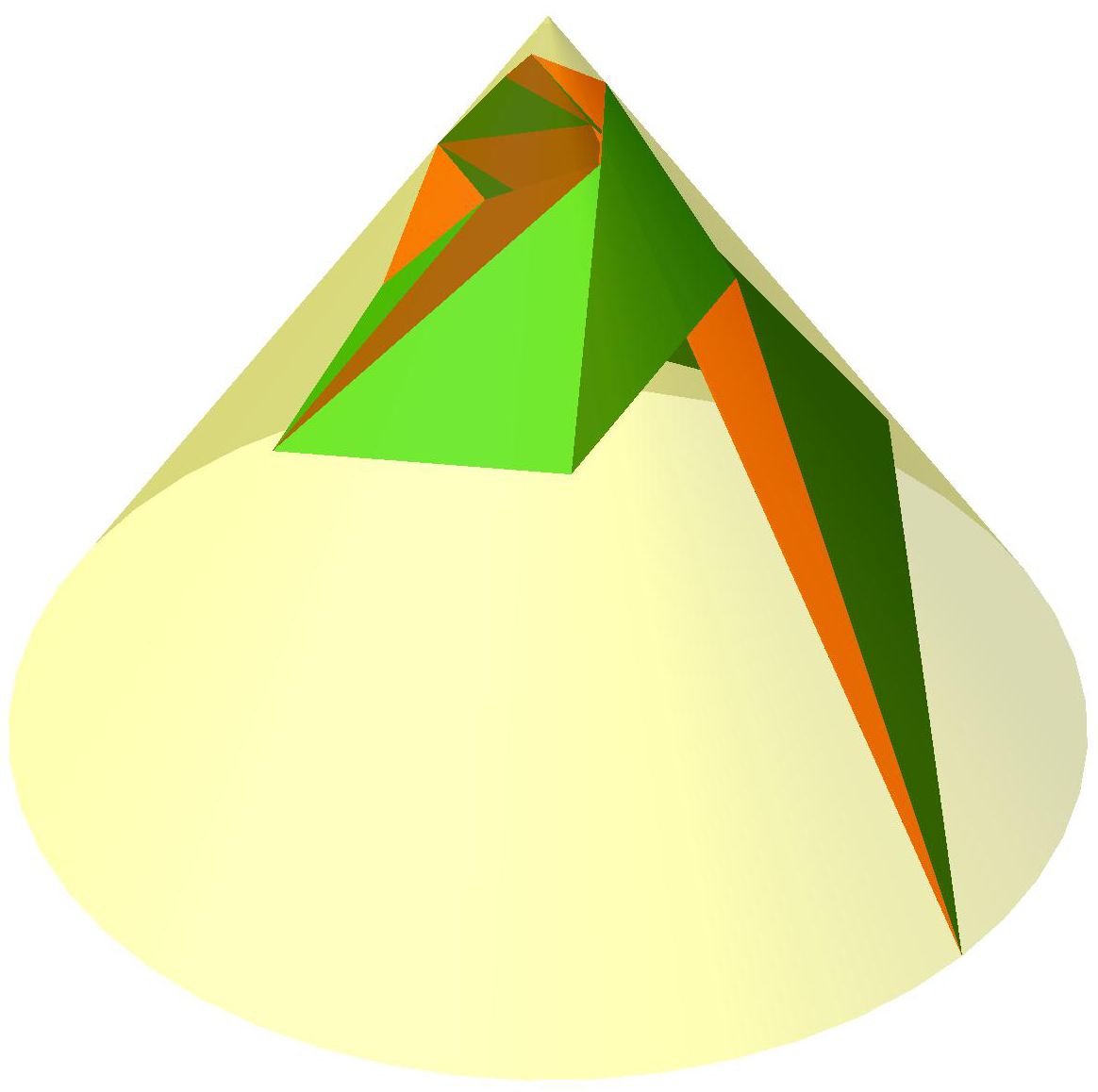}
\begin{small}
\put(0,0){b)}
\end{small}         
  \end{overpic} 
	\hfill
\begin{overpic}
    [width=35mm]{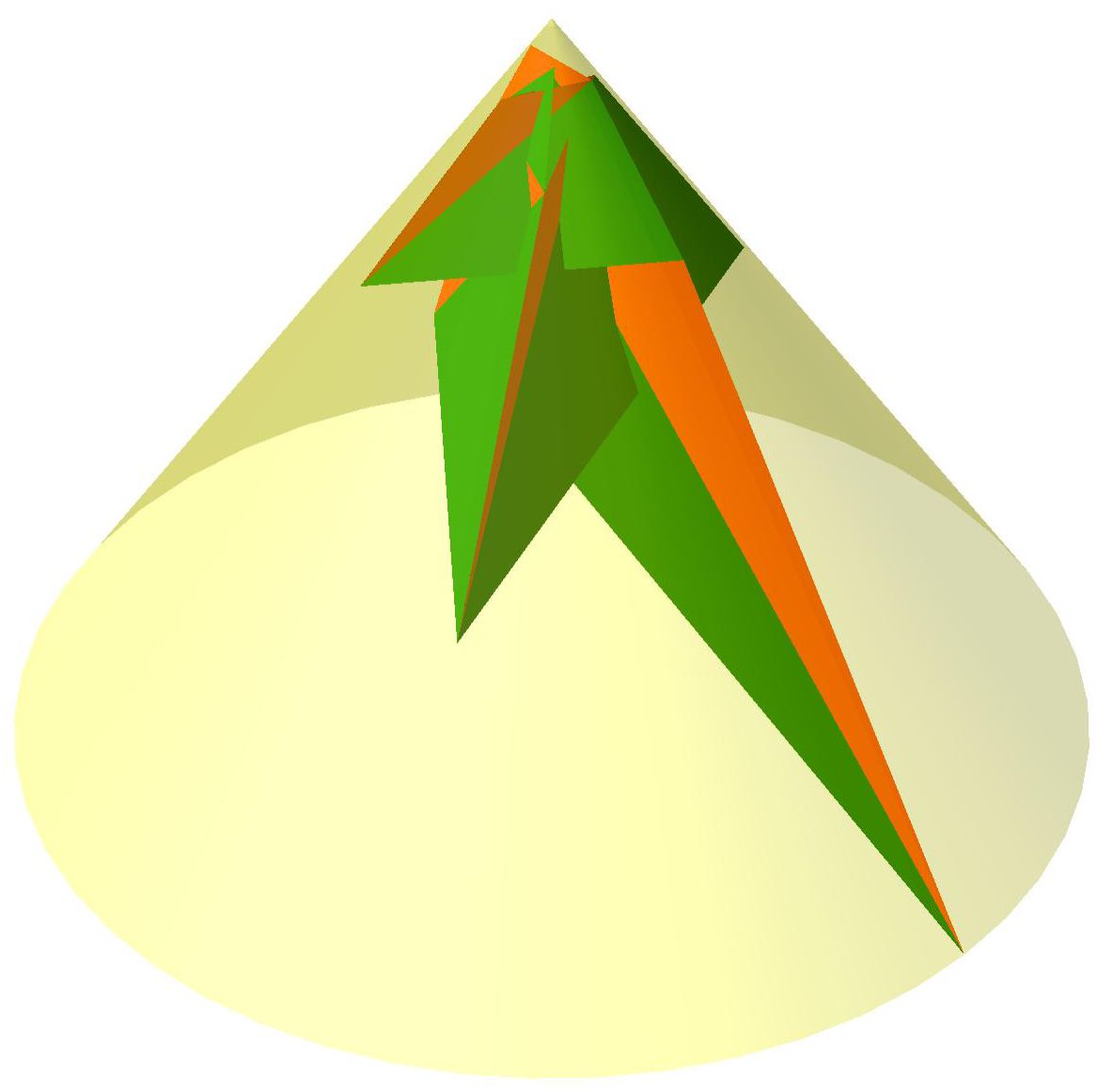}
\begin{small}
\put(0,0){c)}
\end{small}         
  \end{overpic} 	
\caption{
Shaky realization $\mathcal{R}_s$ for $n=4$ (a) and $\mathcal{R}_s^1$ and $\mathcal{R}_s^2$, respectively, for $n=6$ (b/c). Note that $\mathcal{R}_s^2$ 
has self-intersections.
}
  \label{fig8}
\end{figure}

\begin{example}\label{ex:shaky}
For the above described procedure we compute for $n=4$ and $n=6$ the shaky realizations $\mathcal{R}_s$ for $\lambda_+=\lambda_-=\tfrac{\pi}{6}$, which are 
visualized in Fig.\ \ref{fig8}. Note that for $n=6$ two shaky realizations $\mathcal{R}_s^1$ and $\mathcal{R}_s^2$ exist in contrast to the sole solution for $n=4$. The corresponding 
values for $c_s$ and $p$ are given in the following table.
\begin{center}
\begin{footnotesize}
\begin{tabular}[h]{c||c|c|c}
  $n$ & $c_s$ & $p$ &  Fig.\ \ref{fig8}   \\ \hline\hline
 $\phm$ $4$ $\phm$& $\phm -0.582928$ $\phm$& $\phm$ $0.732619$ $\phm$ & $\phm$ a $\phm$\\ \hline
 $\phm$ $6$ $\phm$& $\phm \phm0.192041$ $\phm$ & $\phm$ $0.810448$ $\phm$ & $\phm$ b $\phm$ \\ \hline
 $\phm$ $6$ $\phm$& $\phm -0.833611$ $\phm$ &$\phm$  $0.773273$ $\phm$ & $\phm$ c  $\phm$
\end{tabular}
\end{footnotesize}
\end{center}
Note that the two shaky realizations (up to scaling) for $n=6$ do not posses the same inner geometry,  which can easily be seen by computing the 
quotient of the lengths $V_0V_1$ and $V_0V_6$. 
\end{example}

By considering a flat realization ($\Leftrightarrow$ $q=0$) as a planar framework in $\RR^3$ it is trivially shaky. But by setting  $c:=c_-=c_+$  in the 
expression of $\go h$ in the special case of Subsection \ref{sec:same}, we obtain as solutions planar frameworks, which are shaky within $\RR^2$. 
A necessary condition for this case is that one of the two kinds of triangles has collinear vertices.

\begin{rmk}\label{rem:shaky2}
For the anti-frustum based conical triangulation a geometric characterization of the shaky realizations is known (cf.\ Remark \ref{rmk:shaky}).
For the spiral-motion based conical triangulation this remains an open problem, as also existing literature 
does not cover this case. 
The infinitesimal flexibility of periodic frameworks was studied in \cite{borcea} 
but only for a repetitive translational arrangement of the unit-cell. 
Moreover, with exception of the flat case  
our frameworks are located on cones of revolution in $\RR^3$. 
But until now only finite frameworks on surfaces were studied in \cite{nixon}. 
\hfill $\diamond$
\end{rmk}

\subsubsection{On self-intersection free realizations}\label{subsec:prop}

In Example \ref{ex3} the realizations $\mathcal{R}_-^2$ and $\mathcal{R}_+^2$ show up a remarkable relation, which 
is formulated within the next theorem:

\begin{thm}\label{thm:prop}
Suppose that $\mathcal{R}_-$ is a realization which is located on right pyramid over a regular $(n-1)$-gon. If there exists a second realization $\mathcal{R}_+$
located on the same cone 
then $V_k,V_{k+1},V_{k+n},V_{k+n+1}$ are coplanar for $k\in\NN$ and $n=4,\ldots,9$.
\end{thm}

\noindent Proof: 
We start with the computation of the algebraic condition $f=0$ indicating the coplanarity of $V_0,V_{1},V_{n},V_{n+1}$ with
\begin{equation}
f:=\det\begin{pmatrix} 1 & 1 & 1 & 1 \\
V_0 & V_1 & V_{n} & V_{n+1}
\end{pmatrix}.
\end{equation}
By using  Chebyshev polynomials $U_i(x)$ of the second kind, which are recursively defined by:
\begin{equation}
U_{i+1}(x)=2xU_i(x)-U_{i-1}(x) \quad\text{with}\quad U_0(x)=1 \quad\text{and}\quad U_1(x)=2x,
\end{equation}
we can rewrite $f$ under consideration of $\sin{(i\phi_{\pm})}=\sin{\phi_{\pm}}U_{i-1}(\cos{\phi_{\pm}})$ as:
\begin{equation}\label{eq:f*}
\begin{split}
&q\sin\phi_{+}\Big[\left(p^{2n+2}-p^{n+1}\right)\left(D_{n-2}(c_+)-D_{n-1}(c_+)+1\right) +\\
&\left(D_n(c_+)-D_{n-1}(c_+)-1\right)p^{2n+1} +  \left(D_{n-1}(c_+)-D_{n}(c_+)+1\right)p^{n+2}  
\Big].
\end{split}
\end{equation}
Clearly if either $q=0$ or $\sin\phi_{+}=0$ holds the complete realization $\mathcal{R}_+$ is flat and therefore the 
theorem is trivially fulfilled. Therefore we can assume $q\neq 0$ as well as $\phi_+\neq \pi$ and 
focus of the vanishing of the factor $f_*$ given in the square brackets of Eq.\ (\ref{eq:f*}). 
We proceed as in the general case of Section \ref{sec:same} and compute the expressions $d_*(0,n-1)$ and $d_*(0,n)$. 
Then we substitute $c_-$ by the analytic expression for the 
trigonometric function value $\cos{\tfrac{2\pi}{n-1}}$ according to \ref{app:trig},
which corresponds to the condition that $\mathcal{R}_-$ is located on a regular $(n-1)$-sided pyramid. 
Now we compute the three possible resultants of the expressions $f_*$, $d_*(0,n-1)$ and $d_*(0,n)$ each with respect to $p$; i.e.\ 
$Res(f_*,d_*(0,n-1))$, $Res(f_*,d_*(0,n))$ and $Res(d_*(0,n-1),d_*(0,n))$. The only factors of $Res(d_*(0,n-1),d_*(0,n))$ 
which do not appear in $Res(f_*,d_*(0,n-1))$ as well as $Res(f_*,d_*(0,n))$ are:
\begin{align}
&n=4: &\quad &(q^2+1) \label{thm:1} \\
&n=6: &\quad &(2q^2 + \sqrt{5} - 1) \label{thm:2} \\
&n=7: &\quad &(3q^2 + 2) \label{thm:3} \\
&n=8: &\quad &(w^{2/3}I\sqrt{3} + 14w^{1/3}I\sqrt{3} - 5w^{2/3} - 168q^2 - 14w^{1/3} + 112) \label{thm:4} \\
&  &\quad & (5w^{2/3}I\sqrt{3} - 14w^{1/3}I\sqrt{3} + 3w^{2/3} - 168q^2 - 42w^{1/3} + 168) \label{thm:5} \\
&n=9: &\quad & (q^2 + 2\sqrt{2} - 2)(2q^2 + \sqrt{2}) \label{thm:6} 
\end{align}
with $w:=28 + 84I\sqrt{3}$. For $n=5$ no factor remains. For $n=4,6,\ldots,9$ the above factors have no real solution for $q$, which closes the proof. 
Note that all computations were performed with {\sc Maple}. \hfill \medskip $\BewEnde$

The statement of this theorem can be extended in the following way:

\begin{thm}\label{thm:prop2}
Suppose that $\mathcal{R}_-$ is a realization which is located on right pyramid over a regular star polygon of type $\left\{\tfrac{n-1}{d} \right\}$  
with $\tfrac{n-1}{2}>d>1$. If there exists a second realization $\mathcal{R}_+$
located on the same cone then $V_k,V_{k+1},V_{k+n},V_{k+n+1}$ are coplanar for $k\in\NN$ and $n=6,\ldots,9$.
\end{thm}

\noindent Proof: 
The proof can be done in the same fashion as the one of Theorem \ref{thm:prop}. We only have to replace for type  $\left\{\tfrac{n-1}{d} \right\}$ 
the variable $c_-$ by the analytic expression for the trigonometric function value $\cos{\tfrac{2d\pi}{n-1}}$ 
according to \ref{app:trig}. 
In the following we list the factors of $Res(d_*(0,n-1),d_*(0,n))$ 
which do not appear in $Res(f_*,d_*(0,n-1))$ as well as $Res(f_*,d_*(0,n))$:
\begin{align}
&n=6, d=2: &\quad & (-2q^2 + \sqrt{5} + 1) \label{thm2:1}\\ 
&n=7, d=2: &\quad & (q^2-2) \label{thm2:2}\\
&n=8, d=2: &\quad & 
(w_1^{2/3}I\sqrt{3} + 7w_1^{1/3}I\sqrt{3} - 2w_1^{2/3} + 84q^2 - 7w_1^{1/3} - 56) \label{thm2:3} \\
& &\quad & 
(2w_1^{2/3}I\sqrt{3} - 7w_1^{1/3}I\sqrt{3} + 3w_1^{2/3} - 84q^2 - 21w_1^{1/3} + 84) \label{thm2:4} \\
&n=8, d=3: &\quad & 
(w_1^{2/3}I\sqrt{3} - 14w_1^{1/3}I\sqrt{3} + 5w_1^{2/3} + 168q^2 - 14w_1^{1/3} - 112) \label{thm2:5} \\
& &\quad & 
(5w_1^{2/3}I\sqrt{3} + 14w_1^{1/3}I\sqrt{3} - 3w_1^{2/3} + 168q^2 - 42w_1^{1/3} - 168) \label{thm2:6} \\
&n=9, d=2: &\quad &  (q^4-2q^2-1) \label{thm2:7}\\
&n=9, d=3: &\quad &  (-q^2 + 2\sqrt{2} + 2)(-2q^2 + \sqrt{2}) \label{thm2:8}  
\end{align}
with $w_1:= -\overline{w}$. Now we take a closer look at the real solutions of these factors case by case:
\begin{enumerate}[$\bullet$]
\item
Case $n=6, d=2$: We have only one possible solution $q=\tfrac{\sqrt{2 + 2\sqrt{5}}}{2}$ as $q\geq 0$ has to hold. 
Back-substitution into $d_*(0,n-1)$ and $d_*(0,n)$ shows that their gcd equals
$p^2 - \tfrac{p}{2}(\sqrt{5} - 1) +1$, which does not have real roots.
\item
Case $n=7, d=2$: We have the possible solution $q=\sqrt{2}$. Back-substitution into $d_*(0,n-1)$ and $d_*(0,n)$ shows that their gcd equals
$p^2 - p + 1$, which does not have real roots.
\item
Case $n=8, d=2$: Only the upper factor yields a possible solution, namely 
\begin{equation}
q=\tfrac{\sqrt{1176 - 21w_1^{2/3}I\sqrt{3} - 147w_1^{1/3}I\sqrt{3} + 42w_1^{2/3} + 147w_1^{1/3}}}{42}.
\end{equation}
Back-substitution into $d_*(0,n-1)$ and $d_*(0,n)$ shows that their gcd equals
\begin{equation}
pw_1^{2/3}I\sqrt{3} + 7pw_1^{1/3}I\sqrt{3} - 2pw_1^{2/3} - 7pw_1^{1/3} + 84p^2 + 28p + 84
\end{equation}
which does not have a real solution for $p$. 
\item
Case $n=8, d=3$: Each of the factors yields one possible solution, namely 
\begin{equation}
\begin{split}
q&=\tfrac{\sqrt{4704 - 42w_1^{2/3}I\sqrt{3} + 588w_1^{1/3}I\sqrt{3} - 210w_1^{2/3} + 588w_1^{1/3}}}{84}, \\
q&=\tfrac{\sqrt{7056 - 210w_1^{2/3}I\sqrt{3} - 588w_1^{1/3}I\sqrt{3} + 126w_1^{2/3} + 1764w_1^{1/3}}}{84}.
\end{split}
\end{equation}
Back-substitution into $d_*(0,n-1)$ and $d_*(0,n)$ shows that their gcd equals
\begin{equation}
\begin{split}
&(pw_1^{2/3}I\sqrt{3} - 14pw_1^{1/3}I\sqrt{3} + 5pw_1^{2/3} - 14pw_1^{1/3} + 168p^2 + 56p + 168), \\
&(pw_1^{2/3}I\sqrt{3} + 7pw_1^{1/3}I\sqrt{3} - 2pw_1^{2/3} - 7pw_1^{1/3} + 84p^2 + 28p + 84).
\end{split}
\end{equation}
Both factors do not have a real solution for $p$.
\item
Case $n=9, d=2$: We get one possible solution for $q$, namely $\sqrt{1+\sqrt{2}}$. 
Back-substitution into $d_*(0,n-1)$ and $d_*(0,n)$ shows that their gcd equals
$p^2-p\sqrt{2}  + 1$, which has no real solution for $p$.
\item
Case $n=9, d=3$: The two possible solutions are $q=\tfrac{2^{3/4}}{2}$ and $q=\sqrt{2 + 2\sqrt{2}}$.
Back-substitution into $d_*(0,n-1)$ and $d_*(0,n)$ shows that their gcd equals
$p^2 + 1$ and $p^2-p\sqrt{2}  + 1$, respectively. Both factors do not have a real solution for $p$.
\end{enumerate}
Beside the fact that we do not get real solutions for $p$, the vanishing of the related factors implies the 
vanishing of the numerator of $r$ (a contradiction). \hfill $\BewEnde$

\begin{rmk}
In the anti-prism/frustum based design of cylindrical/conical structures we assumed that the regular $n$-gons are convex  
in order to avoid self-intersections. Theoretically one can also base the anti-prism/frustum construction on 
regular star polygons. \hfill $\diamond$
\end{rmk}

It is also possible that a realization is located on a right pyramid over a regular $n$-gon or star polygon of type $\left\{\tfrac{n-1}{d} \right\}$. 
For these cases analogous statements to Theorem \ref{thm:prop} and \ref{thm:prop2} hold, which read as follows:

\begin{thm}\label{thm:prop3}
Suppose that $\mathcal{R}_-$ is a realization which is located on right pyramid over a regular $n$-gon. If there exists a second realization $\mathcal{R}_+$
located on the same cone then $V_k,V_{k+1},V_{k+n-1},V_{k+n}$ are coplanar for $k\in\NN$ and $n=4,\ldots,9$.
\end{thm}

\begin{thm}\label{thm:prop4}
Suppose that $\mathcal{R}_-$ is a realization which is located on right pyramid over a regular star polygon of type $\left\{\tfrac{n}{d} \right\}$  
with $\tfrac{n}{2}>d>1$. If there exists a second realization $\mathcal{R}_+$
located on the same cone then $V_k,V_{k+1},V_{k+n-1},V_{k+n}$ are coplanar for $k\in\NN$ and $n=5,\ldots,9$.
\end{thm}

The proofs of both theorems work similar to the ones of Theorem \ref{thm:prop} and \ref{thm:prop2}, respectively, and they are given in detail in 
\ref{app:proof3} and \medskip \ref{app:proof4}, respectively.

Now we want to explain how the Theorems \ref{thm:prop}--\ref{thm:prop4} are related to the self-intersection problem. For this purpose we distinguish following two types of 
self-intersections: If triangular faces with a common vertex intersect each other, then we call this self-intersection {\it local} otherwise {\it global}. 
We consider the vertex star of $V_k$ of our polyhedral surface. By the edges $V_{k-1}V_k$ and $V_kV_{k+1}$ the vertex star is split into two sets of 
three triangles; an upper set and a lower one. As the vertices are located on a spiral it is impossible that triangles of the upper set intersect 
triangles of the lower set and vice versa. Without loss of generality we can focus on the three triangles of the upper set of the vertex star. 
Let us assume that we start with a self-intersection free vertex star and deform it continuously such that the three triangles of the upper set
intersect each other. As $\phi$ ranges within the interval $]0;\pi[$ the deformation has to pass a configuration, where two adjacent triangles of the upper set are coplanar 
enclosing a dihedral angle of zero degrees, which is referred as the {\it boundary of local self-intersection}.
This can only happen either along the edge $V_{k}V_{k+n-1}$ or along the edge $V_kV_{k+n}$  for $k\in\NN$.
A necessary condition for this is the coplanarity condition formulated in Theorems \ref{thm:prop}--\ref{thm:prop2} and Theorems \ref{thm:prop3}--\ref{thm:prop4}, respectively. 
This coplanarity is not sufficient as the possibility remains that the dihedral angle equals $\pi$, but in this cases the realization has to be located on a right pyramid again. 
The following two theorems show that this is not possible:

\begin{thm}\label{thm:prop5}
Suppose that $\mathcal{R}_-$ is a realization which is located on a right pyramid over a regular $(n-1)$-gon. If there exists a second realization $\mathcal{R}_+$
located on the same cone then it is on the boundary of local self-intersection for $n=4,\ldots ,9$.
\end{thm}

\noindent Proof: 
As $\mathcal{R}_-$  is located on a right pyramid over a  regular $(n-1)$-gon the 
realization $\mathcal{R}_+$ of Theorem \ref{thm:prop} cannot be located on 
a right pyramid over a regular star polygon of type $\left\{\tfrac{n-1}{d} \right\}$ as in this case not 
$V_k,V_{k+1},V_{k+n},V_{k+n+1}$ have to be coplanar but $V_k,V_{k+1},V_{k+n-1},V_{k+n}$. 
Therefore only the possibilities remain that  $\mathcal{R}_+$ is located on a right pyramid over a regular: 

\begin{enumerate}[$\bullet$]
\item
$n$-gon: 
We compute the expressions $d_*(0,n-1)$ and $d_*(0,n)$ as in the general case of Section \ref{sec:same}. 
Then we substitute $c_-$ and  $c_+$ by the analytic expression for $\cos{\tfrac{2\pi}{n-1}}$ and $\cos{\tfrac{2\pi}{n}}$, respectively, according to \ref{app:trig}. 
We proceed by calculating the resultant of $d_*(0,n-1)$ and $d_*(0,n)$ with respect to $p$. 
It can be seen by direct computations that the resulting expressions in $q$ have only zero as real solution for $n=4,\ldots ,9$. 
Then it can easily be checked that for $q=0$ the gcd of $d_*(0,n-1)$ and $d_*(0,n)$ has no real solution for $p$, which finishes this part of the proof. 
\item
star polygon of type $\left\{\tfrac{n}{d} \right\}$: 
We compute the expressions $d_*(0,n-1)$ and $d_*(0,n)$ as in the general case of Section \ref{sec:same}. 
Then we substitute $c_-$ and  $c_+$ by the analytic expression for $\cos{\tfrac{2\pi}{n-1}}$ and $\cos{\tfrac{2d\pi}{n}}$, respectively, according to \ref{app:trig}. 
We have to check this for the pairs of $n$ and $d$ listed in the proof of Theorem \ref{thm:prop4} 
given in \ref{app:proof4}. 
We compute the resultant of $d_*(0,n-1)$ and $d_*(0,n)$ with respect to $p$. In all cases the resulting expression has $q=0$ as a real solution, but 
there exist further solutions which are listed below:
\begin{align}
&n=5, d=2: &\quad & q=\tfrac{1}{2}\sqrt{2 + 2\sqrt{5}}\\ 
&n=6, d=2: &\quad & q=\sqrt{2}\\
&n=7, d=2: &\quad &  q=\tfrac{1}{42}\sqrt{147(1 - I\sqrt{3})w_1^{1/3} - 21w_1^{2/3}I\sqrt{3} + 42w_1^{2/3} + 1176} \\
&n=7, d=3: &\quad & q= \tfrac{1}{84}\sqrt{588(1 + I\sqrt{3})w_1^{1/3} - 42w_1^{2/3}I\sqrt{3} - 210w_1^{2/3} + 4704} \\
& &\quad &   q = \tfrac{1}{84}\sqrt{(1764 - 588I\sqrt{3})w_1^{1/3} - 210w_1^{2/3}I\sqrt{3} + 126w_1^{2/3} + 7056} \\
&n=8, d=2: &\quad &  q=\sqrt{1 + \sqrt{2}}\\
&n=8, d=3: &\quad &  q = \tfrac{1}{2}2^{3/4} \quad \text{and} \quad q=\sqrt{2 + 2\sqrt{2}}  \\
&n=9, d=2: &\quad &  q =  \tfrac{1}{4}\sqrt{16 + 2v_1^{2/3}I\sqrt{3} - 2v_1^{2/3} + 8v_1^{1/3}} \\
&n=9, d=3: &\quad & q = \tfrac{\sqrt{2v_3^{1/3}(v_3^{2/3} + 4v_3^{1/3} + 12)}}{2v_3^{1/3}} \quad \text{with} \quad v_3=12(3 + I\sqrt{3}) \\
& &\quad &   q = \tfrac{\sqrt{v_3^{1/3}(12I\sqrt{3} - v_3^{2/3}I\sqrt{3} - v_3^{2/3} + 8v_3^{1/3} - 12)}}{2v_3^{1/3}} 
\end{align}
\begin{align}
&n=9, d=4: &\quad & q = \tfrac{1}{12}\sqrt{6v_2^{2/3}I\sqrt{3} + 6v_2^{2/3} + 24v_2^{1/3} -48} \\
& &\quad &   q =  \tfrac{1}{4}\sqrt{16 + 2v_2^{2/3}I\sqrt{3} - 4v_2^{1/3}I\sqrt{3} - 2v_2^{2/3} + 4v_2^{1/3}} \\
& &\quad &   q =   \tfrac{1}{2}\sqrt{v_2^{2/3}I\sqrt{3} + v_2^{1/3}I\sqrt{3} + 2v_2^{2/3} + 8 + 5v_2^{1/3}} \qquad\qquad\qquad\quad
\end{align}
It can easily be checked that for all solutions the gcd of $d_*(0,n-1)$ and $d_*(0,n)$ has no real solution for $p$, which finishes the proof. \hfill $\BewEnde$
\end{enumerate}

\begin{thm}\label{thm:prop6}
Suppose that $\mathcal{R}_-$ is a realization which is located on a right pyramid over a regular $n$-gon. If there exists a second realization $\mathcal{R}_+$
located on the same cone then it is on the boundary of local self-intersection for $n=4,\ldots ,9$.
\end{thm}

The proof of Theorem \ref{thm:prop6} works similar to the one of Theorem \ref{thm:prop5} and its details are given in \ref{app:proof6}.

\begin{rmk} \label{rem:moren}
One can also extend Theorems \ref{thm:prop5} and \ref{thm:prop6} to conical structures located on right pyramids over a regular
star polygons (cf.\  Theorems \ref{thm:prop2} and \ref{thm:prop4}), which would not make much sense as these structures 
cannot be realized as panel-hinge models without global self-intersection. Nevertheless, these structures can be of interest for 
some applications if they are materialized as bar-joint frameworks.

Clearly, one can also try to prove  the Theorems \ref{thm:prop}--\ref{thm:prop6} for further values of $n>9$, which is a cumbersome 
task due to the increasing degree of involved polynomials but the  procedure is straight forward. 
Due to the obtained results, it is very likely that these six theorems hold for all $n>3$ but an elegant proof 
(preferably a pure geometric one) of this conjecture remains open.  
\hfill $\diamond$
\end{rmk}

The algebraic curve $\go h$  in the $(c_-,c_+)$-plane (cf.\ opening of Section \ref{sec:same}) visualizes the set of conical frameworks (some of the points 
do not correspond to realizations as the corresponding $p$-values are not within the interval $]0;1[$), 
which is symmetric with respect to the line $c_-=c_+$ characterizing shaky frameworks. Therefore a snap 
of the framework on the cone corresponds to a path along the curve $\go h$ between the point $(a,b)$ and $(b,a)$. 
As a consequence such a snap of the framework on the cone has to pass a shaky realization $\mathcal{R}_s$. 

\begin{rmk}\label{rem:snap}
This is in accordance with known results on snapping frameworks (cf.\ \cite{nawr1,nawr2}).
The shaky pose $\mathcal{R}_s$  can be used to compute the snappability of these structures (see Section \ref{sec:roughspiral}).
\hfill $\diamond$
\end{rmk}

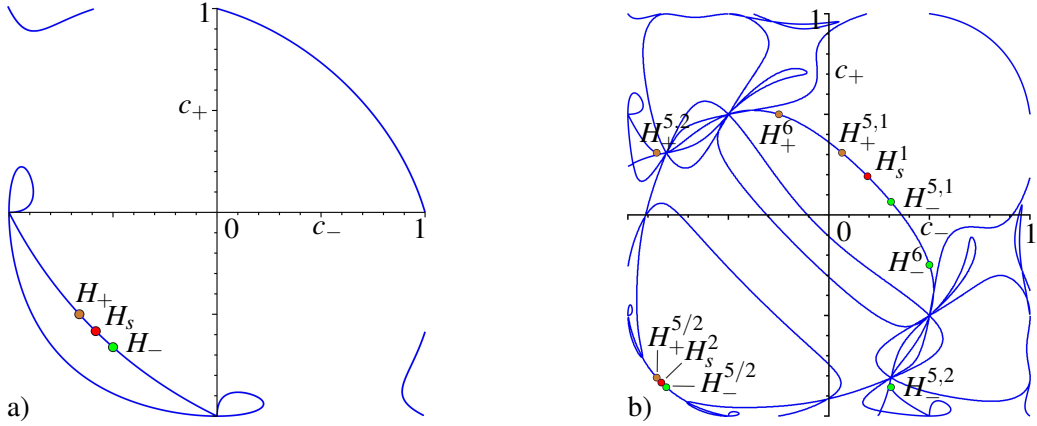
\begin{figure}[t]
\hfill
\begin{overpic}
    [width=55mm]{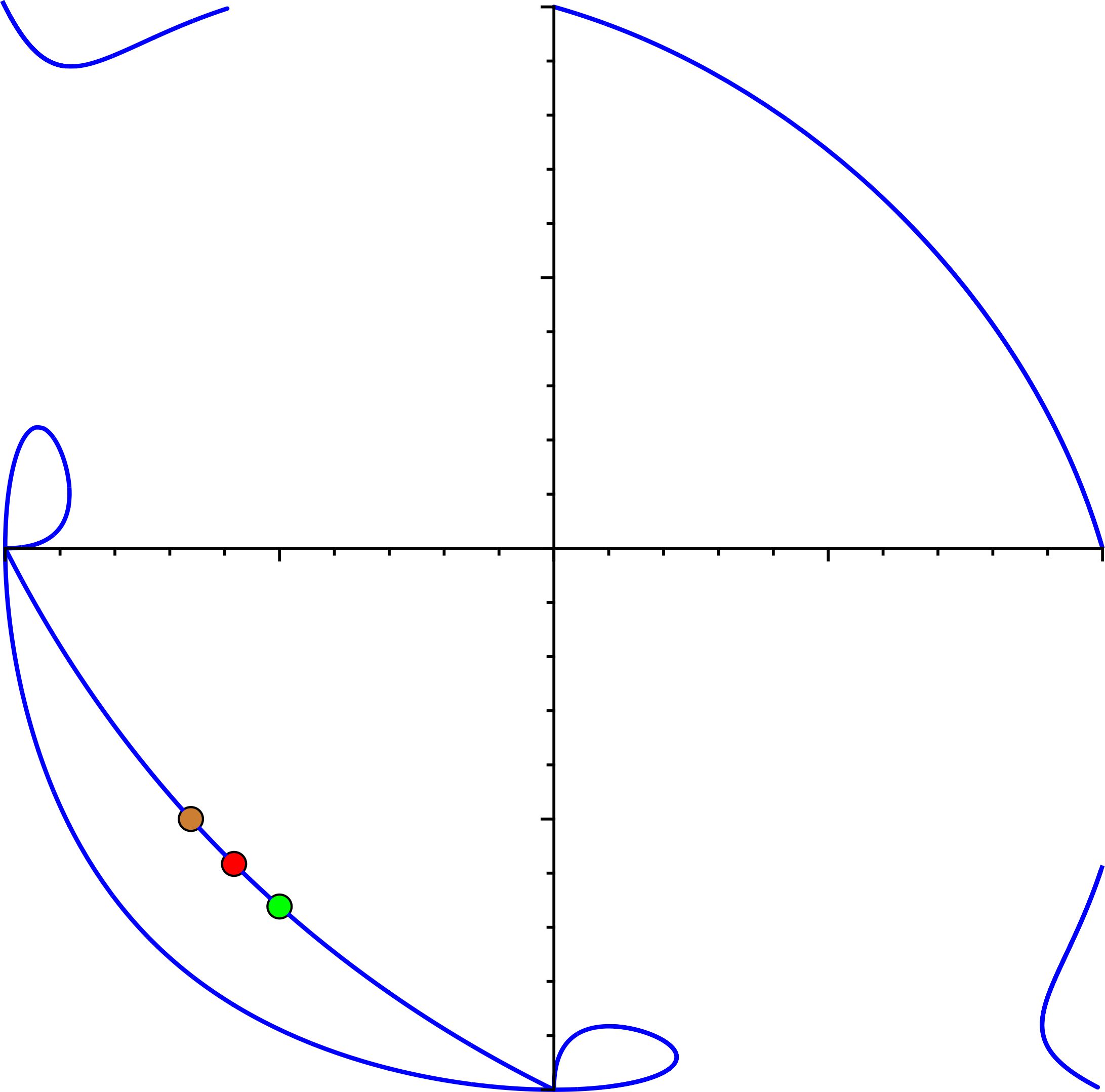}
\begin{small}
\put(0,0){a)}
\put(73,44){$c_-$}
\put(41,73){$c_+$}
\put(45.5,95){$1$}
\put(97,43){$1$}
\put(52,43){$0$}
\put(28.5,15.5){$H_-$}
\put(23.5,22){$H_s$}
\put(16.5,27.5){$H_+$}
\end{small}     
  \end{overpic} 
\hfill\hfill
\begin{overpic}
    [width=53mm]{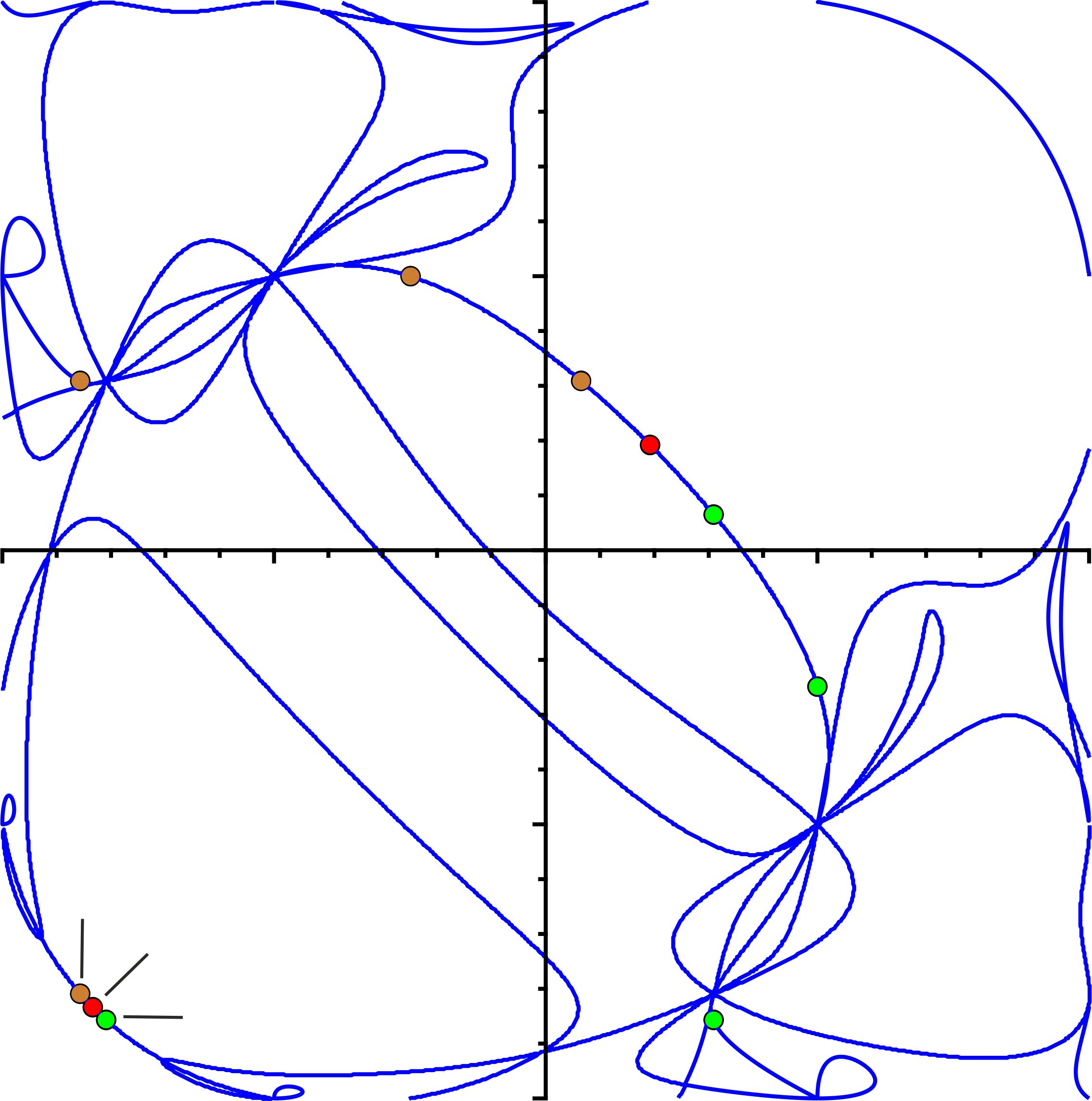}
\begin{small}
\put(0,0){b)}
\put(72,45){$c_-$}
\put(51,84){$c_+$}
\put(44.5,96){$1$}
\put(97.5,43){$1$}
\put(51.5,43.5){$0$}
\put(18,6){$H_-^{5/2}$}
\put(14,14){$H_s^2$}
\put(5.,17.5){$H_+^{5/2}$}
\put(61,61){$H_s^1$}
\put(68,52){$H_-^{5,1}$}
\put(52,68){$H_+^{5,1}$}
\put(64.2,36){$H_-^{6}$}
\put(32.4,67.5){$H_+^{6}$}
\put(68,5){$H_-^{5,2}$}
\put(3,68){$H_+^{5,2}$}
\end{small}     
  \end{overpic} 
	\hfill $\phm$
\caption{(a/b) Algebraic curve $\go h$ within the domain $c_{\pm}\in[-1;1]$ of Example \ref{ex:simple_snap}/\ref{ex:advanced_snap}. 
}
  \label{fig9a}
\end{figure}

Therefore the bounds of local self-intersection free realizations are given by $\mathcal{R}_-$ located on right pyramid over a regular 
$(n-1)$-gon or $n$-gon and the shaky realization $\mathcal{R}_s$. As a consequence, only these intervals can contain feasible candidates for 
a global self-intersection free snapping between two realizations. To make the things more clear, we consider the following two examples: 

\begin{example}\label{ex:simple_snap}
We start with the most simple example of $n=4$ for $\lambda_{\pm}=\tfrac{\pi}{6}$. In this case we already visualized the realizations $\mathcal{R}_-$
and $\mathcal{R}_+$ of Theorem \ref{thm:prop} in Fig.\ \ref{fig7}b,e. Moreover the shaky realization $\mathcal{R}_s$ is given in Fig.\ \ref{fig8}a. 
The corresponding curve $\go h$ where the three corresponding points $H_-$, $H_+$ and $H_s$ are highlighted is displayed in Fig.\ \ref{fig9a}a. 
Note that in the case $n=4$ no star pyramids (cf.\ Theorems \ref{thm:prop2} and \ref{thm:prop4})  exist and that for 
$c_-=0$ of Theorem \ref{thm:prop3} we obtain $c_+=\pm 1$ implying $p=1$, a contradiction. 
As a consequence the interval of self-intersection free snapping realizations corresponds to points of the 
curve segment of $\go h$ bounded by $H_-$ and $H_s$. 
\end{example}

\begin{example}\label{ex:advanced_snap}
Now we want to consider the more sophisticated example given by $n=6$ and $\lambda_{\pm}=\tfrac{\pi}{6}$. The two shaky realizations $\mathcal{R}_s^1$ and 
$\mathcal{R}_s^2$ are already displayed in Fig.\ \ref{fig8}b,c. There are two realizations $\mathcal{R}_-^{5,i}$ $(i=1,2)$ which are located on right 
pyramids over a regular $5$-gon displayed in Fig.\ \ref{fig9b}a,c. Their corresponding realizations $\mathcal{R}_+^{5,i}$ according to Theorem \ref{thm:prop} 
are illustrated in  Fig.\ \ref{fig9b}b,d. There is one realization $\mathcal{R}_-^{6}$ which is located on right pyramids over a regular $6$-gon 
shown in Fig.\ \ref{fig9b}g. The related realization $\mathcal{R}_-^{6}$ according to Theorem \ref{thm:prop3} is given in Fig.\ \ref{fig9b}h. 
There does not exist a pair of snapping realizations, where one is located on a right pyramid over a regular star polygon of type $\left\{\tfrac{6}{2} \right\}$, 
but such a pair exists for type $\left\{\tfrac{5}{2} \right\}$, which is denoted by $\mathcal{R}_-^{5/2}$ and $\mathcal{R}_+^{5/2}$ (cf.\ Fig.\ \ref{fig9b}e,f). 

\begin{figure}[t]
 \begin{overpic}
    [width=35mm]{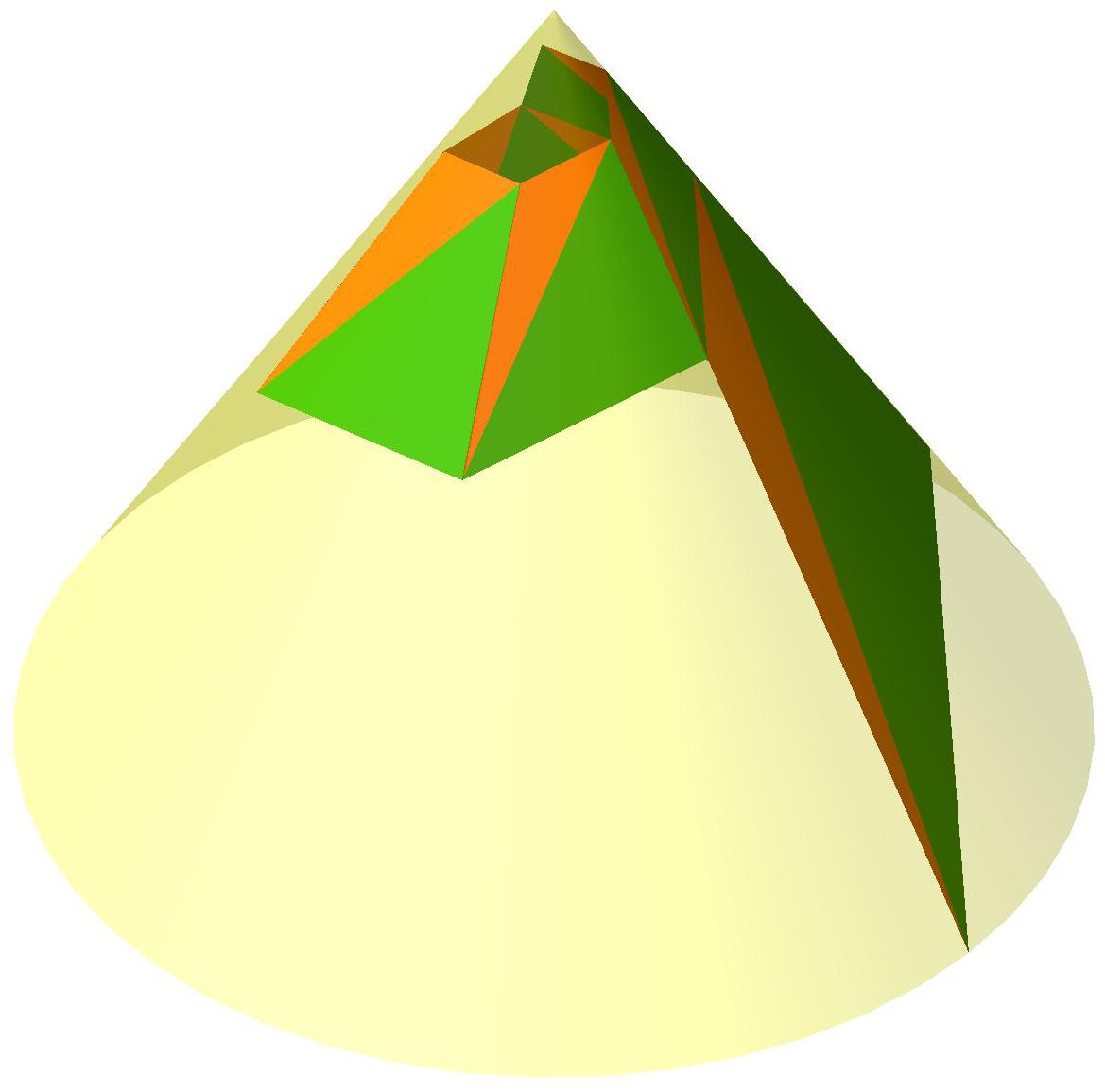}
\begin{small}
\put(0,0){a)}
\end{small}         
  \end{overpic} 
	\hfill
\begin{overpic}
    [width=35mm]{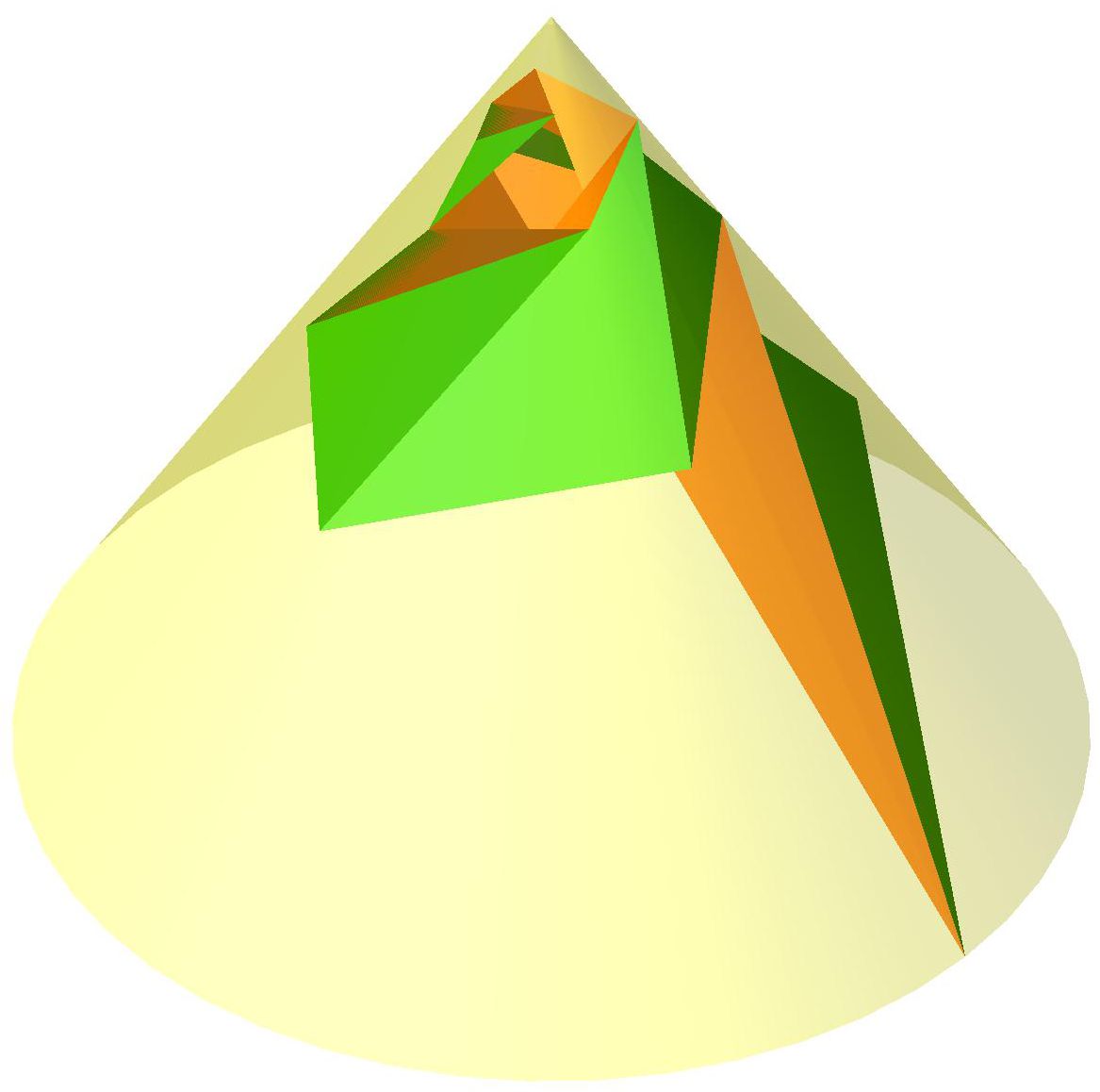}
\begin{small}
\put(0,0){b)}
\end{small}         
  \end{overpic}
			\hfill
		\begin{overpic}
    [width=35mm]{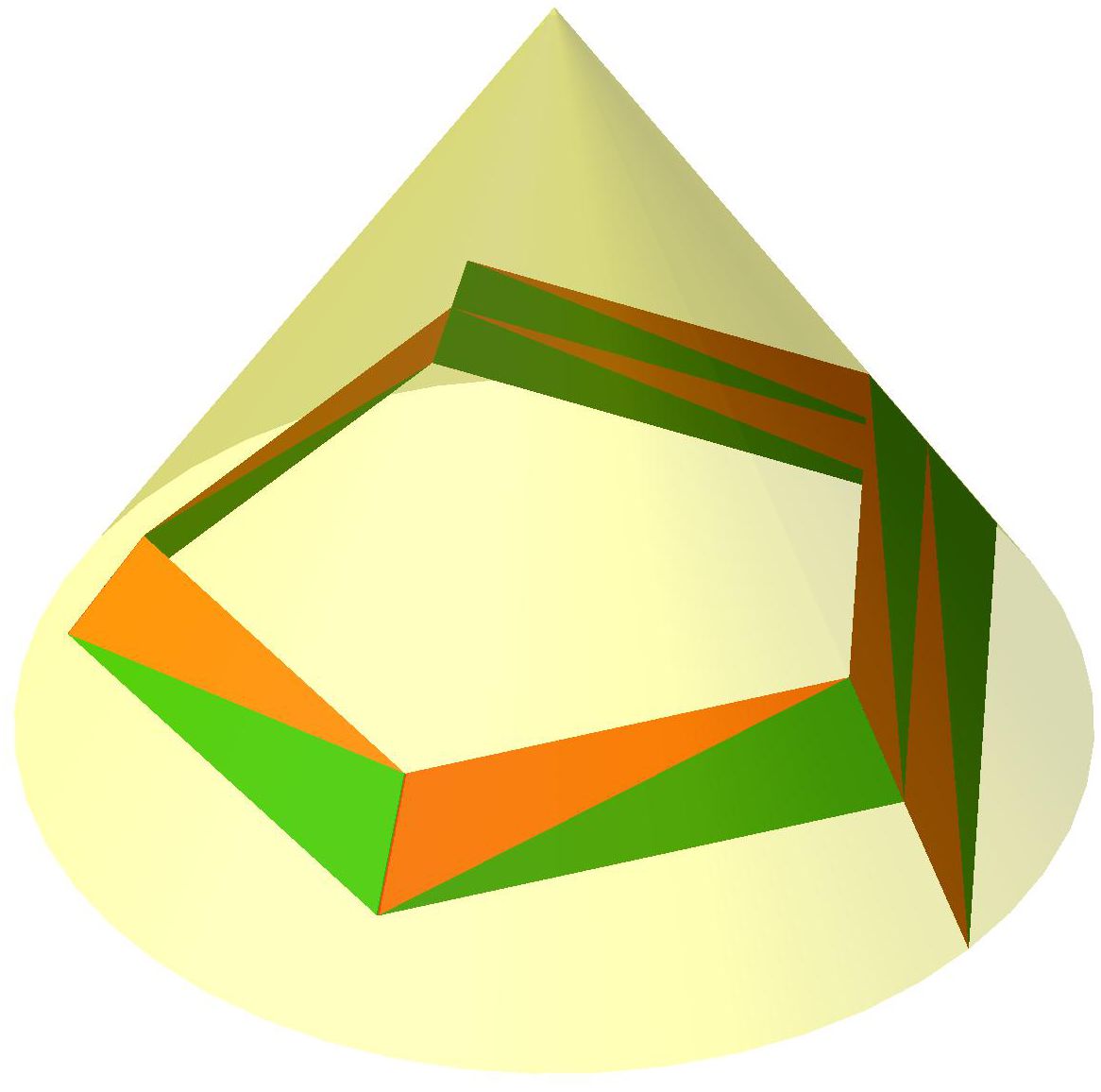}
\begin{small}
\put(0,0){c)}
\end{small}         
  \end{overpic} 
		\hfill
\begin{overpic}
    [width=35mm]{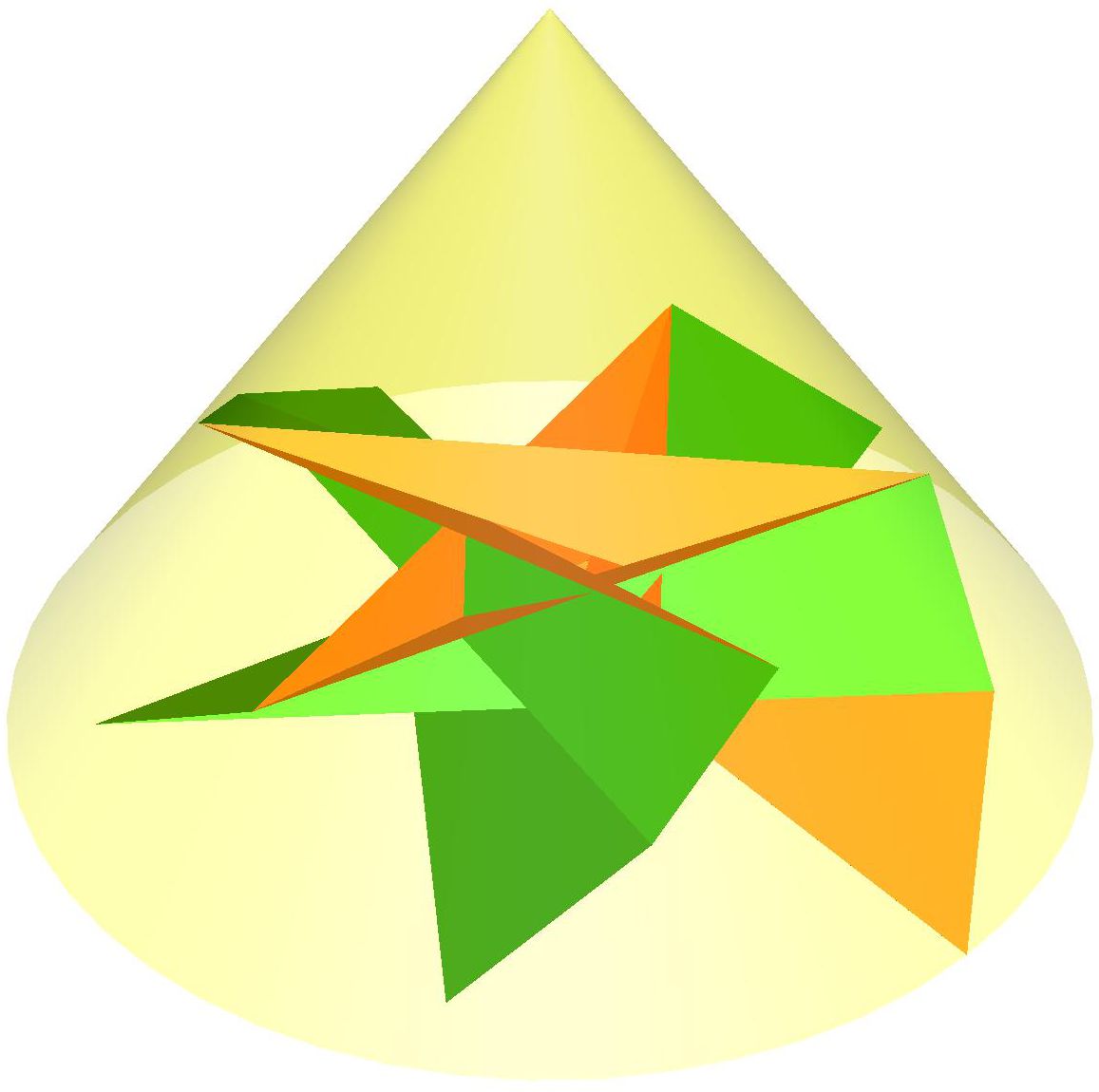}
\begin{small}
\put(0,0){d)}
\end{small}         
  \end{overpic} 
		\\ \phm \\	
\begin{overpic}
    [width=35mm]{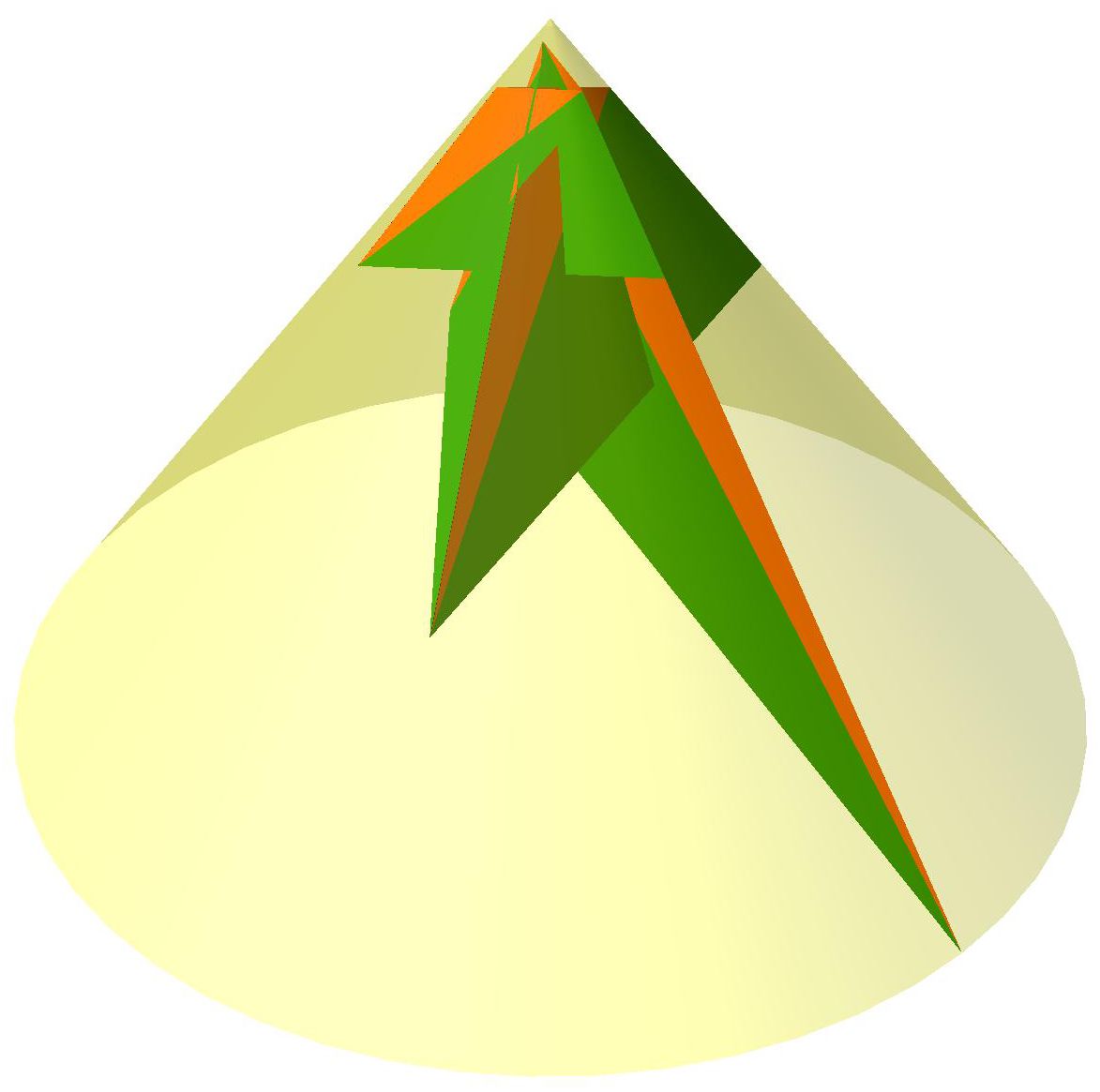}
\begin{small}
\put(0,0){e)}
\end{small}     
  \end{overpic}
\hfill
\begin{overpic}
    [width=35mm]{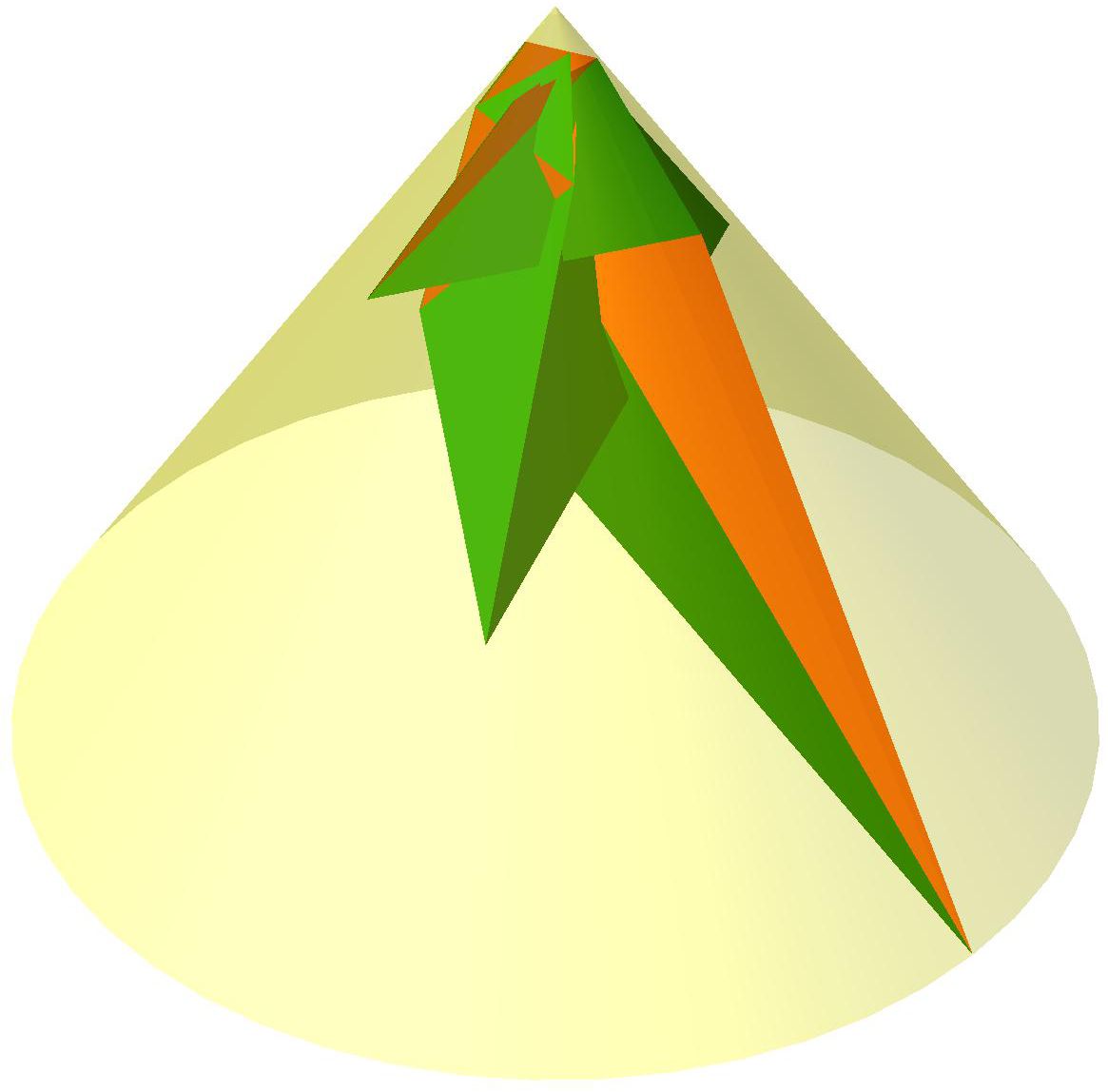}
\begin{small}
\put(0,0){f)}
\end{small}     
  \end{overpic} 
\hfill
	 \begin{overpic}
    [width=35mm]{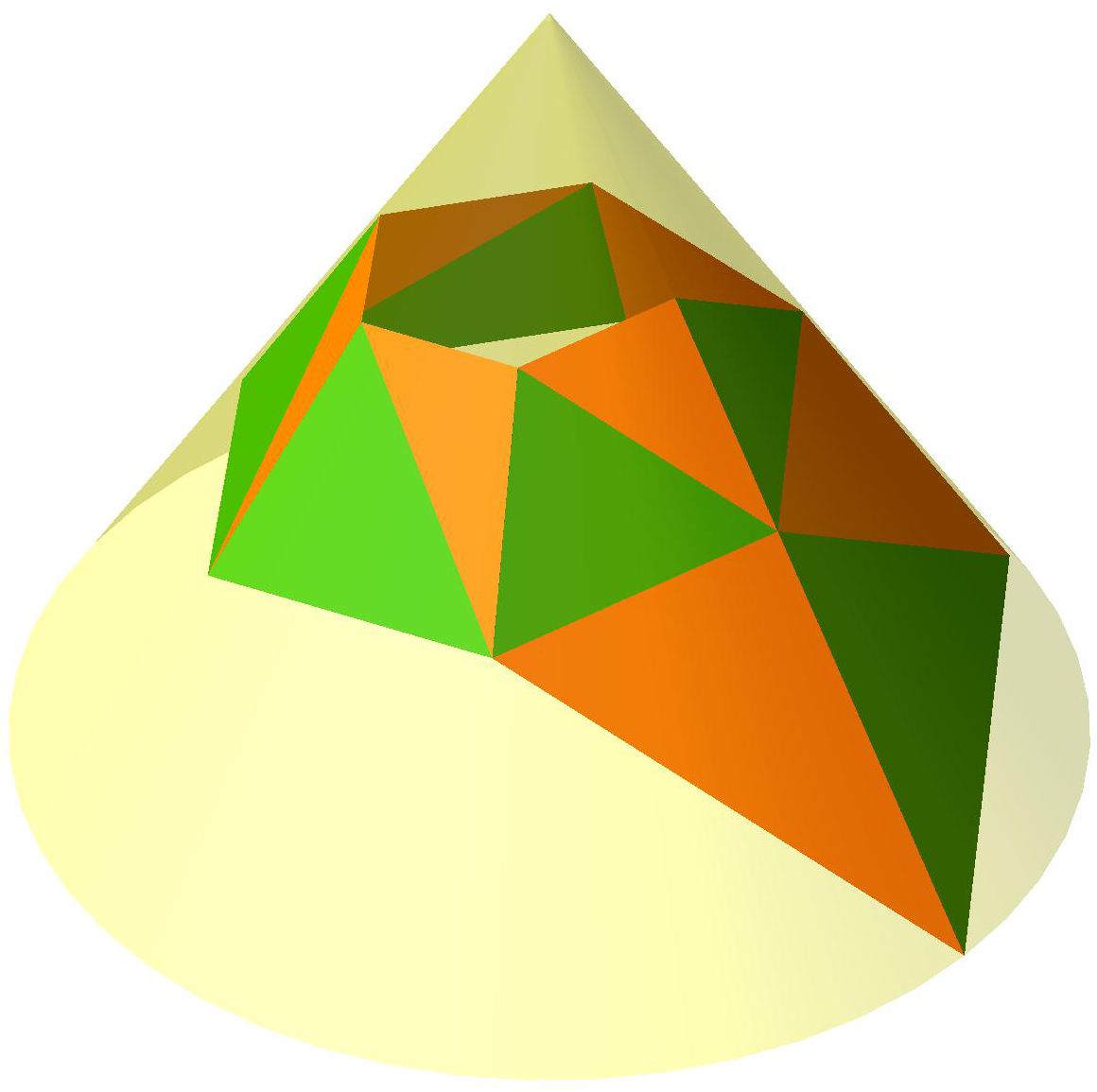}
\begin{small}
\put(0,0){g)}
\end{small}         
  \end{overpic} 
\hfill
 \begin{overpic}
    [width=35mm]{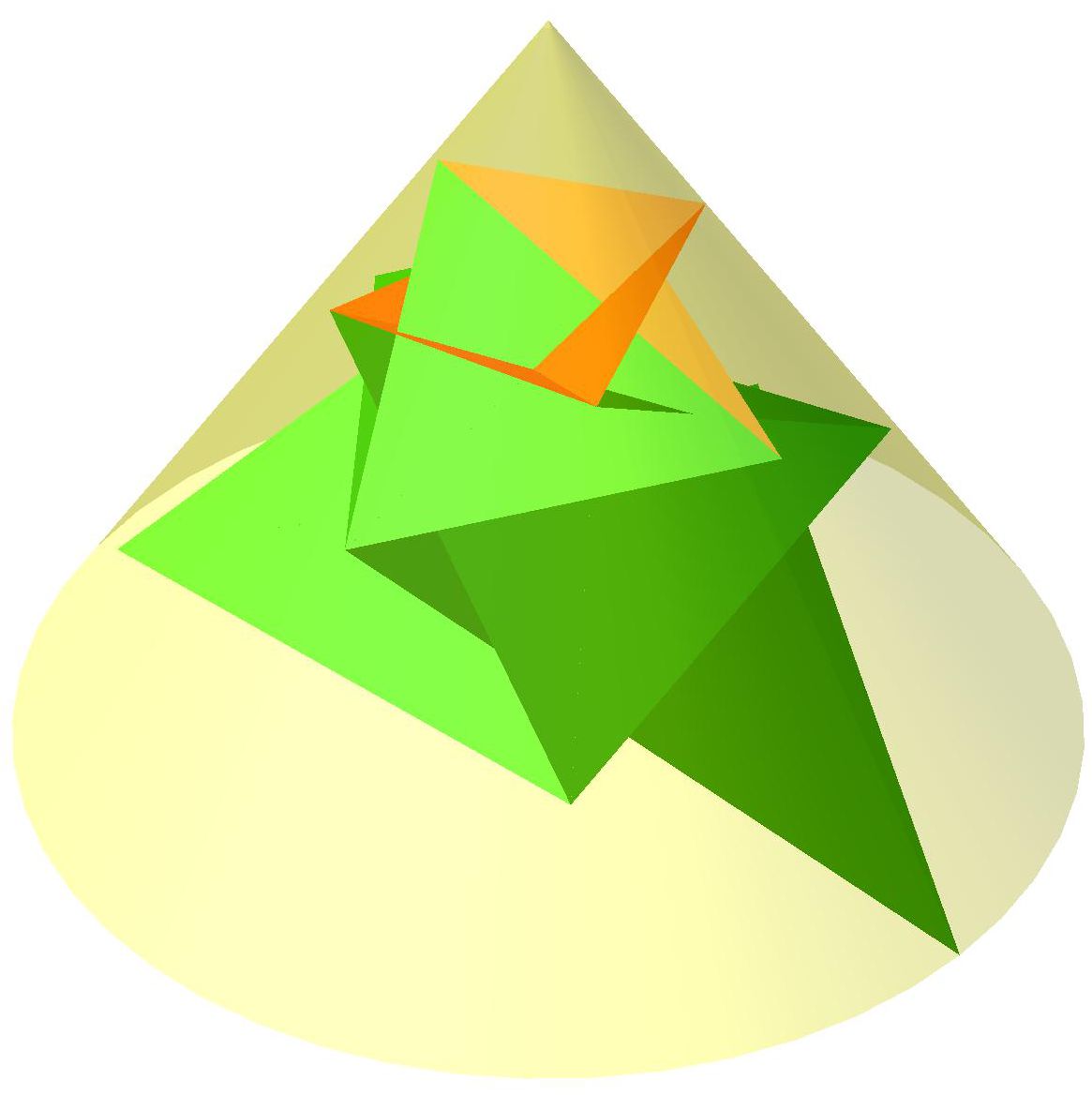}
\begin{small}
\put(0,0){h)}
\end{small}         
  \end{overpic}
\caption{
In (a/c) the two realizations $\mathcal{R}_-^{5,i}$ located on a $5$-pyramid are displayed and in (b/d) their corresponding realizations $\mathcal{R}_+^{5,i}$ 
are given for $i=1,2$. The realization $\mathcal{R}_-^{5/2}$ on the $\left\{\tfrac{5}{2} \right\}$ star polygon pyramid is displayed in (e) and its corresponding realization 
$\mathcal{R}_+^{5/2}$ in (f). The realization $\mathcal{R}_-^{6}$ located on a $6$-pyramid is illustrated in (g) and its related realizations $\mathcal{R}_+^{6}$ 
is shown in (h). 
}
  \label{fig9b}
\end{figure}     

The curve $\go h$ is displayed in Fig.\ \ref{fig9a}b, where the corresponding points 
\begin{equation}
\begin{split}
H_-^{5,1}&=(\tfrac{\sqrt{5}-1}{4} , 0.065071) \quad \text{with}\quad p=0.819582, \\ 
H_-^{5,2}&=(\tfrac{\sqrt{5}-1}{4} ,-0.858230) \quad \text{with}\quad p=0.966477, \\ 
H_-^{6}&=(\tfrac{1}{2} ,-0.248762) \quad \text{with}\quad p=0.905087, \\ 
H_-^{5/2}&=(-\tfrac{1+\sqrt{5}}{4} ,-0.856766) \quad \text{with}\quad p=0.774327, 
\end{split}
\end{equation}
of the realizations are highlighted. 
For the coordinates of the points $H_s^1$ and  $H_s^2$ we refer to the table given in Example \ref{ex:shaky}. 
As  $\mathcal{R}_s^{2}$ has self-intersections (cf.\ Fig.\ \ref{fig8}c), a self-intersection free snap has to pass 
the self-intersection free shaky realization $\mathcal{R}_s^{1}$ (cf.\ Fig.\ \ref{fig8}b). 
By the above study it can be seen that the interval of the global self-intersection free snapping realizations corresponds to points of the 
curve segment of $\go h$ bounded by $H_-^{5,1}$ and $H_s^1$. 

Note that for the determination of self-intersection free snapping realizations, there is no need of considering realizations located on 
right pyramids over star polygons. In this example it was done for demonstration of Theorem \ref{thm:prop2}. 
\end{example}

\subsection{Three realizations}\label{sec:tristable}

The existence of a one-parametric solution curve $\go h$ for the design of bi-stable conical structures allows 
theoretically the design of tri-stable\footnote{Note that this was not possible in Section \ref{sec:frusta}
due to the bi-stable behavior of a regular anti-frustum.} conical structures. In the following we only want to sketch a way how this can be done, 
as in practice we were not able to find examples where all three realizations are free of self-intersections.

We compute again $r_-$ from $d(0,1)=0$ and substitute the obtained expression into $d(0,n-1)$ and $d(0,n)$, respectively, which both factor into $p$ 
and $d_*(0,n-1)$ and $d_*(0,n)$, respectively. 
Now we calculate the resultant of the later two factors with respect to $c_-$, which results in the 
expression $g_-$ depending on $p$ and $c_+$ beside the design parameters $q_+$ and $q_-$. 

If we index the third realization by the symbol ${\circ}$, then we can just substitute $q_-$ by $q_\circ$ in  $g_-$ to get the 
expression $g_\circ$. Finally one can compute the resultant of  $g_-$ and $g_\circ$ with respect to $c_+$ in order to get a 
univariate polynomial in $p$.

\begin{figure}[t]
\begin{overpic}
    [width=35mm]{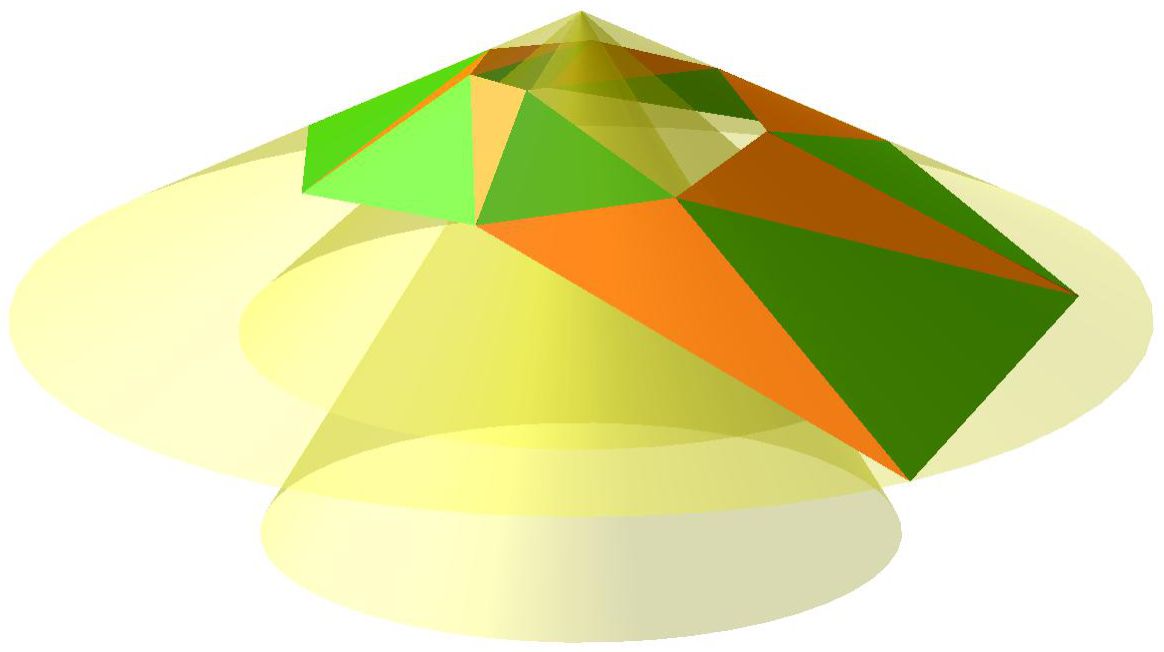}
\begin{small}
\put(0,0){a)}
\end{small}     
  \end{overpic} 
\hfill
 \begin{overpic}
    [width=35mm]{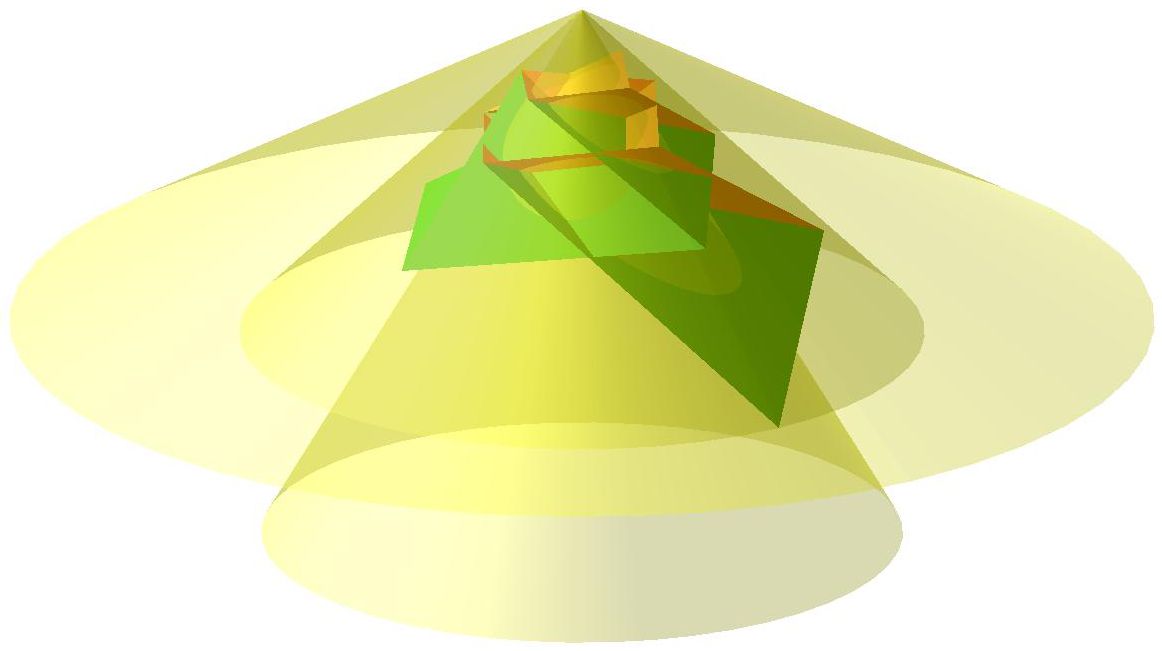}
\begin{small}
\put(0,0){b)}
\end{small}         
  \end{overpic} 
	\hfill
\begin{overpic}
    [width=35mm]{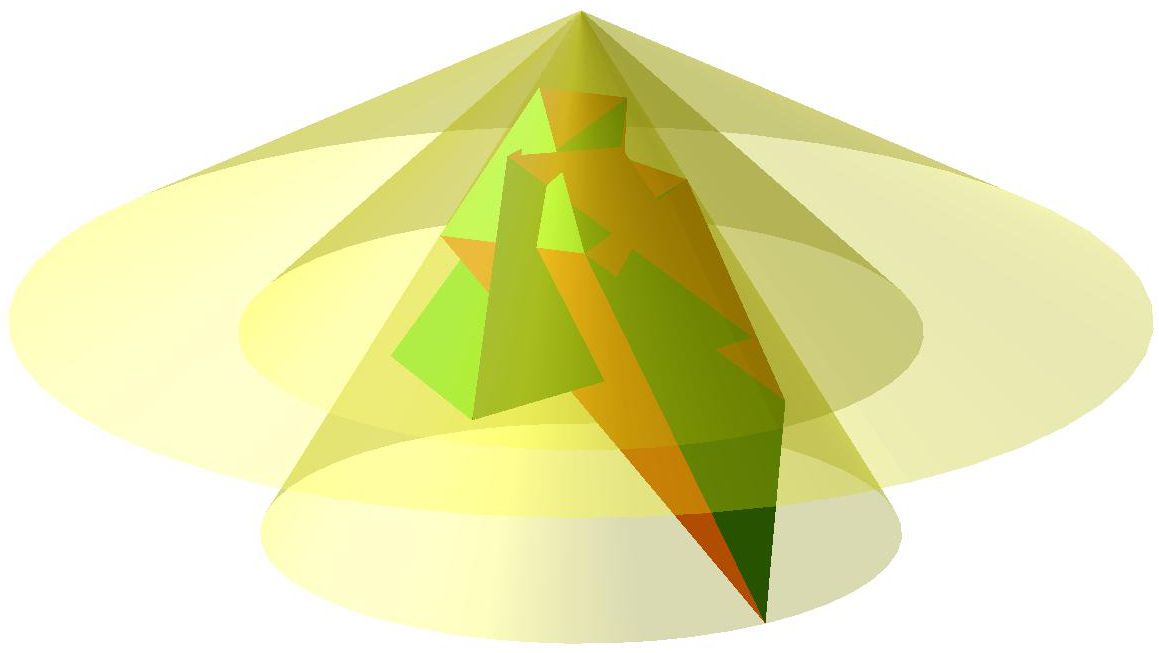}
\begin{small}
\put(0,0){c)}
\end{small}         
  \end{overpic} 	
\caption{The three realizations $\mathcal{R}_+$ (a), $\mathcal{R}_{\circ}$ (b) and $\mathcal{R}_-$ (c), respectively. 
}
  \label{fig10}
\end{figure}

\begin{example}
The given three half apex angles are $\lambda_+=\tfrac{\pi}{3}$,  $\lambda_{\circ}=\tfrac{\pi}{4}$ and $\lambda_-=\tfrac{\pi}{6}$. 
For $n=7$ one solution obtained by the procedure described above reads as:
\begin{equation}
p=0.872272, \phm c_+=0.642073,\phm c_{\circ}=-0.015748 \phm c_-=-0.140929,
\end{equation}
which is also illustrated for the closed strip of $2n$ triangles in Fig.\ \ref{fig10}. Note that only $\mathcal{R}_+$ 
is free of self-intersections.
\end{example}

\begin{rmk}\label{rem:2oncone}
If one wants to have three realizations on the same cone ($\Leftrightarrow$ $q_-=q_+$), the discriminant\footnote{It equals 
 up to a factor (coefficient of the highest power of $c_+$ in $g_-$) the resultant of $g_-$ and $\partial g_-/ \partial c_+$ with respect to $c_+$.
} of $g_-$ with respect to $c_+$ has to vanish. This is a univariate polynomial in $p$. 
Until now we were not able to find an example with three realizations on the same cone. 
This brings us to conjecture that only two realizations (over $\RR$) exits on the same cone. 
\hfill $\diamond$ 
\end{rmk}

%%%%%%%%%%%%%%%%%%%%%%%%%%%%%%%%%%%%%%%%%%%%%%%%%%%%%%%%%%%%%%%%%%%%%%%%%%%%%%%%%%%%%%%%%%%%%%%%%%%%%%%%%%%%%
%%%%%%%%%%%%%%%%%%%%%%%%%%%%%%%%%%%%%%%%%%%%%%%%%%%%%%%%%%%%%%%%%%%%%%%%%%%%%%%%%%%%%%%%%%%%%%%%%%%%%%%%%%%%%
%%%%%%%%%%%%%%%%%%%%%%%%%%%%%%%%%%%%%%%%%%%%%%%%%%%%%%%%%%%%%%%%%%%%%%%%%%%%%%%%%%%%%%%%%%%%%%%%%%%%%%%%%%%%%

\subsection{Orthogonal Cross Sections} \label{sec:cross}

For a self-intersection free realization of a polyhedral cylinder based on a helical arrangement of triangles (cf.\ Fig.\ \ref{fig2}c) 
it was proven by Wittenburg \cite{wittenburg} (for $n=3,4,5,6$) that the area of the cross section (orthogonal to the cylinder axis) does not 
depend on the cut height\footnote{\label{foot:flow}This property is maybe of interest for the usage of such structures as tubes (stents) as the 
flow velocity remains constant under the assumption of a constant volumetric flow.}. 
Therefore, the volume of the cylinder can easily be computed as the area of the 
cross section times its height. 
In the following we show that an analogous property also holds for a polyhedral cone based on a spiral arrangement of triangles. 

To do so, we first compute the cross sectional area of a realization $\mathcal{R}$, which is free of self-intersections. 
Due to the kinematic construction of the triangulated structures we can restrict ourselves to cross sections, which 
intersect the line-segment $V_nV_{n+1}$. Then the intersection polygon consists of the following ordered sequence of points:
\begin{equation}
\begin{split}
E_1&:=V_nV_{n+1}\cap\varepsilon, \phm E_2:=V_1V_{n+1}\cap\varepsilon, \phm E_3:=V_{n+1}V_2\cap\varepsilon, \\ 
E_4&:=V_2V_{n+2}\cap\varepsilon,\phm \ldots \phm E_{2n-1}:=V_{2n-1}V_n\cap\varepsilon,\phm E_{2n}:= V_2V_{2n}\cap\varepsilon,
\end{split}
\end{equation}
where $\varepsilon$ denotes the cross sectional plane. Based on these points one can compute the area $A$ by the following well-known formula:
\begin{equation}
A:=\tfrac{1}{2} |\det(E_1,E_2)+\det(E_2,E_3)+\det(E_3,E_4)+\ldots+\det(E_{2n-1},E_{2n})+\det(E_{2n},E_{1})|.
\end{equation}
By denoting the distance of $\varepsilon$ to the cone apex by $h$ (cut height) we can formulate our theorem as follows:

\begin{thm}\label{thm:7}
For any realization $\mathcal{R}$ the expression $A/h^2$ is constant; i.e.\ it does not depend on the cut height $h$, for $n=3,\ldots , 6$.
\end{thm}

\noindent Proof: 
We parametrize $E_1$ by $V_n+\alpha(V_{n+1}-V_n)$ which also determines the height in dependence of $\alpha$ 
as $h=-rp^nq(\alpha p-\alpha+1)$ holds. Straight forward computation yields the following expressions for $A/h^2$ 
for $n=3,\ldots , 6$: 
\begin{align}
&n=3: &\quad & \tfrac{K}{(p^2 + p + 1)q^2} \quad \text{with} \quad K:=s(c - 1)(2cp - p^2 - 1)\\
&n=4: &\quad & \tfrac{2K\big[(c+1)p^2 + (2c + 1)p + c + 1\big]}{(p^2 + p + 1)(p^2 + 1)q^2} \\
&n=5: &\quad & \tfrac{2K\big[(2c^2 + 2c + 1)p^4 + (4c^2 + 5c + 1)p^3 + (8c^2 + 6c + 1)p^2 + (4c^2 + 5c + 1)p + 2c^2 + 2c + 1\big]}{(p^2 + 1)(p^4 + p^3 + p^2 + p + 1)q^2} \\
&n=6: &\quad & \text{\begin{scriptsize}$K\big[(8c^3 + 8c^2 + 2c + 2)p^6 + (16c^3 + 20c^2 + 8c + 1)p^5 + (32c^3 + 40c^2 + 8c)p^4 + $\end{scriptsize}} \notag \\
& &\quad & \text{\begin{scriptsize}$(48c^3 + 44c^2 + 4c - 1)p^3 + (32c^3 + 40c^2 + 8c)p^2 + (16c^3 + 20c^2 + 8c + 1)p + $\end{scriptsize}}  \\ 
& &\quad & \text{\begin{scriptsize}$8c^3 + 8c^2 + 2c + 2\big]/\big[(p^4 + p^3 + p^2 + p + 1)(p^2 + p + 1)(p^2 - p + 1)q^2\big]$\end{scriptsize}}. \notag 
\end{align}
As these expressions are independent of the parameter $\alpha$ the theorem  is proven. \hfill $\BewEnde$

\begin{rmk}\label{rem:moren2}
Clearly, one can also try to prove Theorem \ref{thm:7} for further values of $n>6$ as the  procedure is straight forward. 
In analogy to Remark \ref{rem:moren} we conjecture that this theorem holds for all $n>3$ but a geometric proof for this is also missing. \hfill $\diamond$
\end{rmk}

Note that Theorem \ref{thm:7} holds for any realization and for all $\alpha\in\RR$. 
But only for self-intersection free realizations and $\alpha\in[0;1]$ a geometric interpretation for 
$A$ as cross sectional area is available. 
Thus for self-intersection free realizations also the volume can easily be computed as the area $A$ of the base times 
the height $h$ divided by three\footnote{It is somehow astonishing that the volume of the infinite structure can 
be computed without any limit process.}. 

\begin{figure}[t]
\quad\,
\begin{overpic}
    [height=33mm]{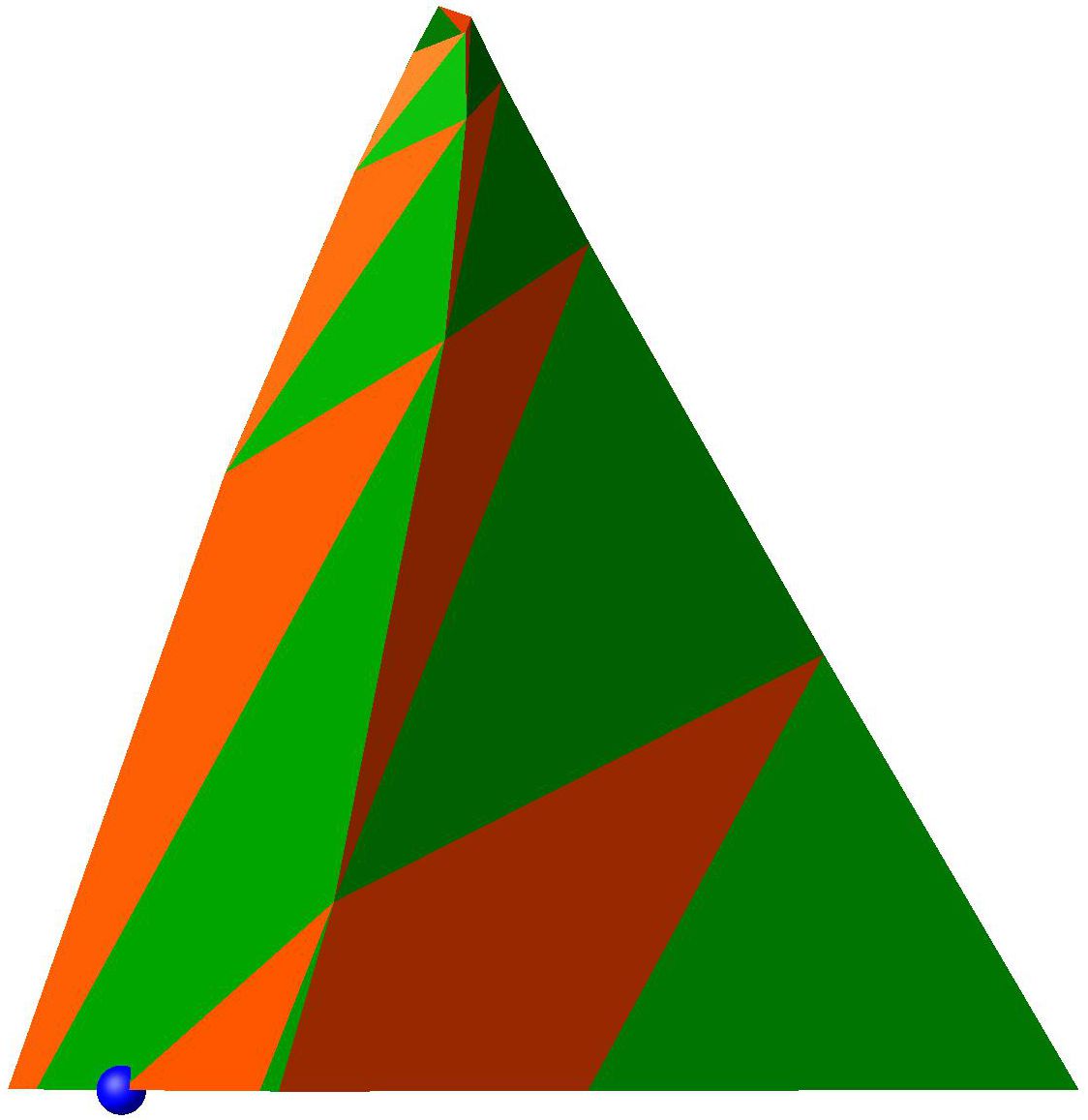}
\begin{small}
\put(-11,0){a)}
\put(11,-7){$E_1$}
\end{small}     
  \end{overpic} 
\hfill
 \begin{overpic}
    [height=33mm]{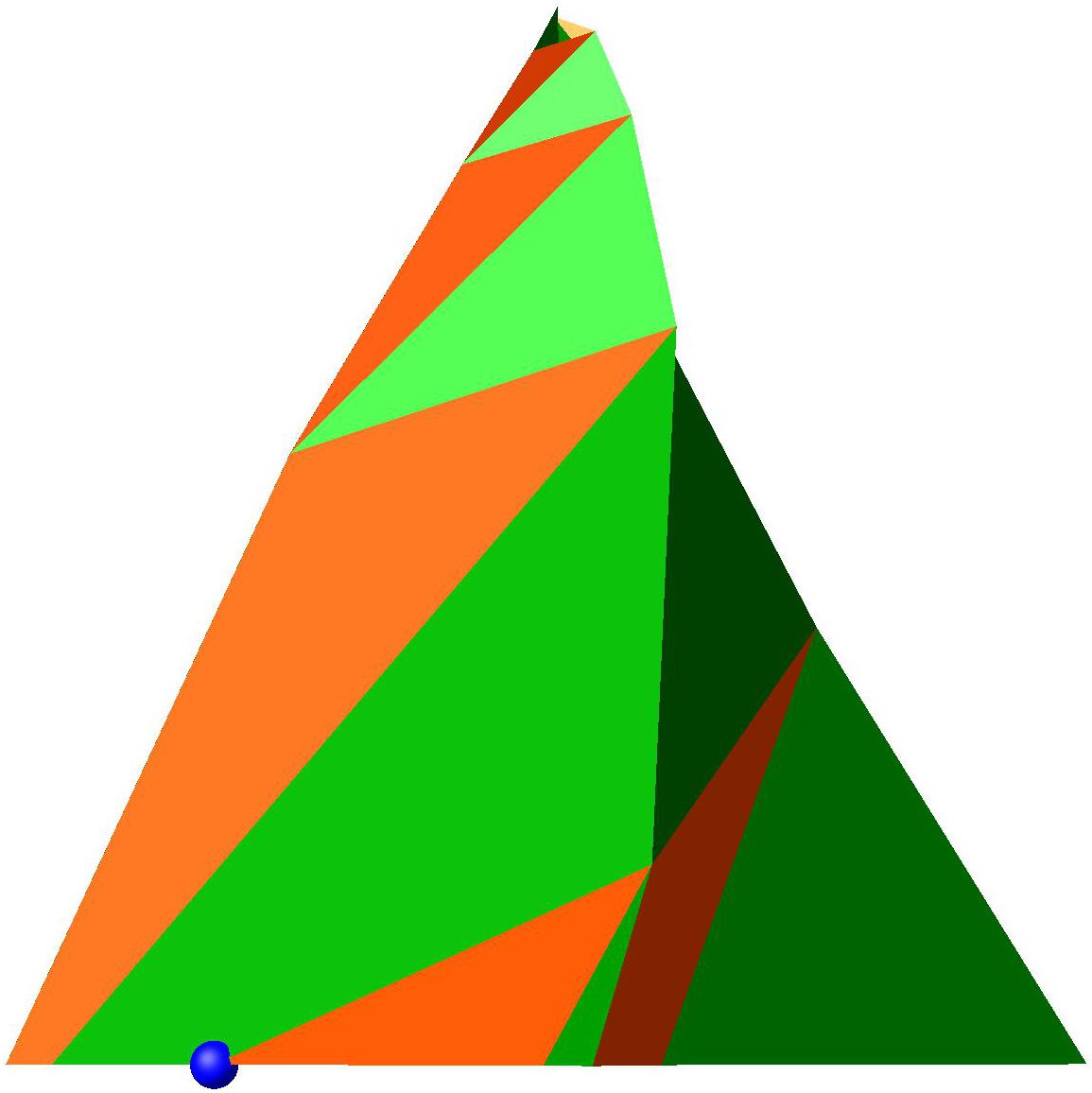}
\begin{small}
\put(-11,0){b)}
\put(9,-7){$E_1$}
\end{small}         
  \end{overpic} 
	\hfill
\begin{overpic}
    [height=33mm]{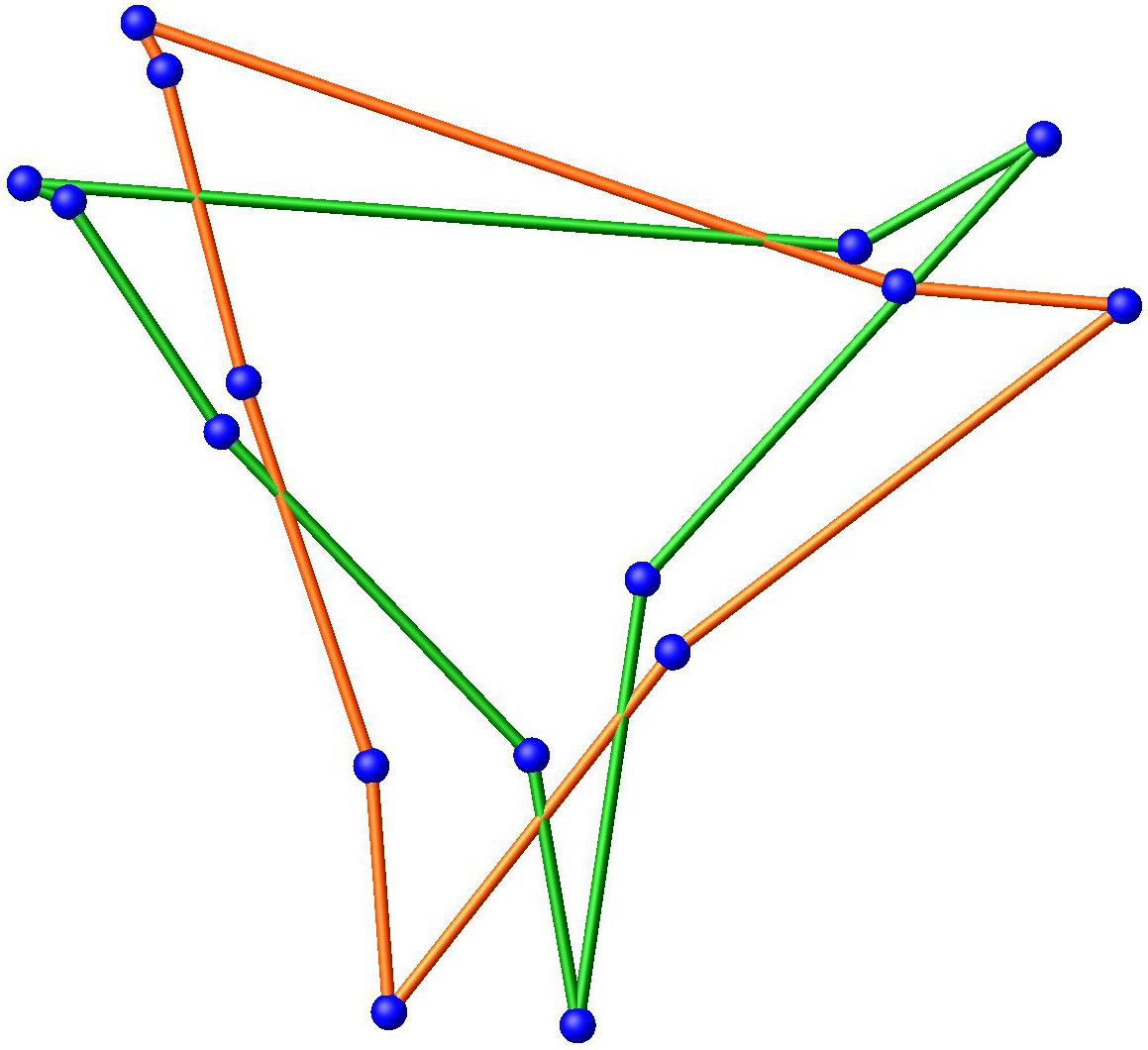}
\begin{small}
\put(4,0){c)}
\put(9.5,47.5){$E_1$}
\put(23.5,57){$E_1$}
\put(30,85){$\mathcal{R}_-^1\cap\varepsilon$}
\put(55,8){$\mathcal{R}_+^1\cap\varepsilon$}
\end{small}         
  \end{overpic} 	
\caption{Front view of the realizations $\mathcal{R}_-^1$ (a) and $\mathcal{R}_+^1$ (b) cut at the height $h=-0.45$, where the point $E_1$ of 
the base polygon is highlighted in blue. (c) Top view of these two base polygons, whose vertices are labeled counter-clockwise starting at $E_1$.
}
  \label{fig11}
\end{figure}
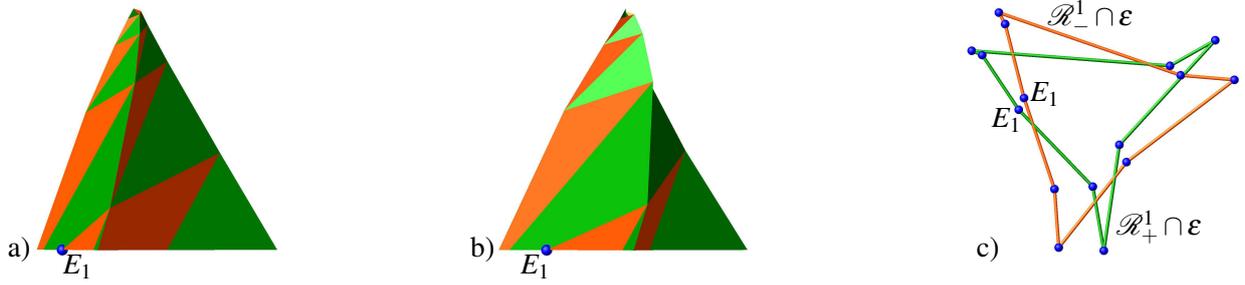

\begin{example}\label{ex:cross_sec}
We consider the two self-intersection free realizations $\mathcal{R}_-^1$ and $\mathcal{R}_+^1$ of Example \ref{ex3}, which are 
visualized in Fig.\ \ref{fig7}a,d. We cut both realizations at the height $h=-0.45$ and consider the conical structures above (cf.\ Fig.\ \ref{fig11}a,b). 
Both base polygons are visualized in Fig.\ \ref{fig11}c and their corresponding areas equals $A_-=0.072578$ and $A_+=0.050793$, respectively. 
This example also shows that the cutting  of realizations located on the same cone with one and the same plane yields 
cross sections, which differ in their areas\footnote{Following Footnote \ref{foot:flow} these snapping structures are maybe of interest for application as 
bi-stable conical diffusers or nozzles.}. Finally, we computed the half apex  angles $\mu_-=18.663188^{\circ}$ and $\mu_+=15.778341^{\circ}$, respectively, 
of the cones of revolution having the same cross sectional area as $\mathcal{R}_-^1$ and $\mathcal{R}_+^1$ at each cut height. 
\end{example}

\begin{rmk} 
The cross section property of Theorem \ref{thm:7} does not hold for the anti-frustum based conical triangulation given in Section \ref{sec:frusta}. 
This can for example easily be seen by looking at Fig.\ \ref{fig5}d, where the base polygon and platform polygon have some non-zero area but for 
one cross section in between it is zero. \hfill $\diamond$
\end{rmk}

%%%%%%%%%%%%%%%%%%%%%%%%%%%%%%%%%%%%%%%%%%%%%%%%%%%%%%%%%%%%%%%%%%%%%%%%%%%%%%%%%%%%%%%%%%%%%%%%%%%%%%%%%%%%%%%%%%%%%%%%%%%%%%%%%%%%%%%%%%%%%%%%%%%%%%%%%%

\section{Snappability computation for triangulated cones}\label{sec:snap}

With exception of \cite{sharma}, which is reviewed latter on, no mathematical studies on the deformation of triangulated cones are known to the
author.
Therefore we proceed with a short literature overview on the 
approaches used for the related case of cylindrical structures, which split up into analytical studies and numerical simulations. 

\begin{enumerate}
\item
The analytical studies are either based on a panel-hinge model using virtual folds and creases \cite{pagano} or on the 
interpretation of the structure as a bar-joint framework \cite{cai1,cai2,guestI}, 
where the bars can be categorized into three sets with respect to their length. In the latter studies it is assumed that only the bars of one of these 
sets can deform\footnote{\label{yes}All in the same way in order to  keep up the symmetry of the structure.} during the 
transmission between the different realizations. 
\item
The numerical simulations are either based on finite element methods (using the software LS-DYNA \cite{wu,zhao} or ANSYS \cite{moshtaghzadeh} for triangle shell elements or 
ABAQUS \cite{cai1,cai2} for bar elements) or on a force method. The latter approach is again based on a bar-joint model, where either all three sets 
of bars can change their lengths \cite{liu2017,guestII} or just two of them  \cite{kidambi,wu}. 
\end{enumerate}
In the already above mentioned publication \cite{sharma} an anti-frustum is modeled as a bar-joint framework and  its deformation is studied analytically under 
the assumption that only the length of one set of bars is allowed to change (cf.\ Footnote \ref{yes}). 
Moreover, a single anti-frustum and a conical structure consisting of three anti-frusta is also simulated using a nonlinear finite element software. 

Within the paper we use the analytic approach of the so-called snappability, 
which measures the snapping capability of the structure and is defined as follows \cite{nawr1,nawr2}:

\begin{definition}
If an isostatic framework snaps out of a realization $\mathcal{R}$
by applying the minimum strain energy (using the concept of Green-Lagrange strain) needed to it, then the
corresponding deformation has to pass a shaky realization $\mathcal{R}_s$ at
the maximum state of deformation. The snappability for the realization $\mathcal{R}$ is defined as 
the strain energy density of $\mathcal{R}_s$ divided by the Young modulus. 
\end{definition}
 
For the triangulated cone the mentioned energy can be computed for two different interpretations;
namely as bar-joint structure or as panel-hinge structure. 
In both cases $\mathcal{R}_s$ is determined as a critical point of the corresponding energy function $U_{total}$, 
which is polynomial and does not differ in the number of unknowns and their occurring degrees. 
Therefore the computational complexity is for both approaches the same, but we restrict here to the 
bar-joint model as it can be formulated within a few lines in order to render the paper 
self-contained\footnote{The interested reader is 
referred to \cite{nawr1,nawr2} for the formulation of the panel-hinge model.}.

Given is a bar, which is deformed in length. Before the deformation its length is denoted by $L$ and 
afterwards by $L'$. Then the elastic strain energy $U(L,L')$ of the deformed bar can be computed as 
\begin{equation}
U(L,L'):=\frac{E\,\Area_{\diameter}}{8L^3}\Big(L'^2-L^2\Big)^2,
\end{equation}
where $E$ is the Young modulus and $\Area_{\diameter}$ the cross-sectional area of the bar. In this context it should be noted that we assume that all 
bars of the conical structure have the same cross-sectional area.

Also for conical structures the set of edges can be subdivided into three subsets 
where edges of the same subset can be transformed into each other by a spiral displacement. 
Due to the periodicity of the structure, it is enough to study the deformation of 
three representative bars to conclude the snappability of the complete conical structure by 
considering the limit of a geometric series. 
In the following we demonstrate this concept 
on basis of some exemplary conical triangulations based on anti-frusta and 
spiral-motions, respectively.

%%%%%%%%%%%%%%%%%%%%%%%%%%%%%%%%%%%%%%%%%%%%%%%%%%%%%%%%

\subsection{Snappability of anti-frustum based conical triangulations}\label{sec:snap_frust}

In this subsection we study anti-frustum based conical triangulations with a flat-foldability ($\Leftrightarrow$ $\lambda_-=\tfrac{\pi}{2}$). 
The advantage of a flat realization is that it is save with respect to high loads/pressures. 
Beside this self-protecting property, deployable structures are of great interest for many practical applications.

We consider anti-frusta as building blocks of our conical structure, where the 
sequence of radii of the regular $n$-gons equals $1, r, r^2, \ldots$. 
We discuss two approaches which differ in their computational complexity due to the number of 
involved unknowns. They are both based on the following parametrization of the 
above mentioned shaky configuration $\mathcal{R}_s$:
\begin{equation}\label{eq:ansatz}
A_{1s}=(\rho \cos{\tfrac{\pi}{n}},-\rho\sin{\tfrac{\pi}{n}},0)^T, \quad A_{2s}=(\rho \cos{\tfrac{\pi}{n}},\rho\sin{\tfrac{\pi}{n}},0)^T, \quad B_{1s}=(0,\rho r,\rho h_s)^T,
\end{equation}
with $r$ of Eq.\ (\ref{r_flat}) and the unknowns $\rho>0$ and $h_s>0$, respectively. 
This ansatz is valid as from Section \ref{sec:frusta} 
we already know that for a shaky configuration the $x$-coordinate of $B_{1s}$ has to be zero. 
Based on Eq.\ (\ref{eq:ansatz}) we can determine $U_{total}$ as the limit of a geometric series where $r$ is the common 
ration between adjacent terms; i.e.\
\begin{equation}
U_{total}=\frac{n\left[U(\overline{A_{\pm1}A_{\pm2}},\overline{A_{1s}A_{2s}})+U(\overline{A_{\pm1}B_{\pm1}},\overline{A_{1s}B_{1s}})+
U(\overline{A_{\pm2}B_{\pm1}},\overline{A_{2s}B_{1s}})\right]}{1-r}.
\end{equation}
As the snappability results from the total elastic energy density, we also have to compute the volume $\Vol_{total}$ of the bar structure, which reads as:
\begin{equation}
\Vol_{total}=\frac{\Area_{\diameter}n\left[\overline{A_{\pm1}A_{\pm2}}+ \overline{A_{\pm1}B_{\pm1}} + \overline{A_{\pm2}B_{\pm1}}\right]}{1-r}.
\end{equation}
Finally the snappability index $\varsigma$ of the structure can be computed as ${U_{total}}/(E\Vol_{total})$ 
according to \cite{nawr1,nawr2}.

\begin{figure}[t]
\hfill
\begin{overpic}
    [height=50mm]{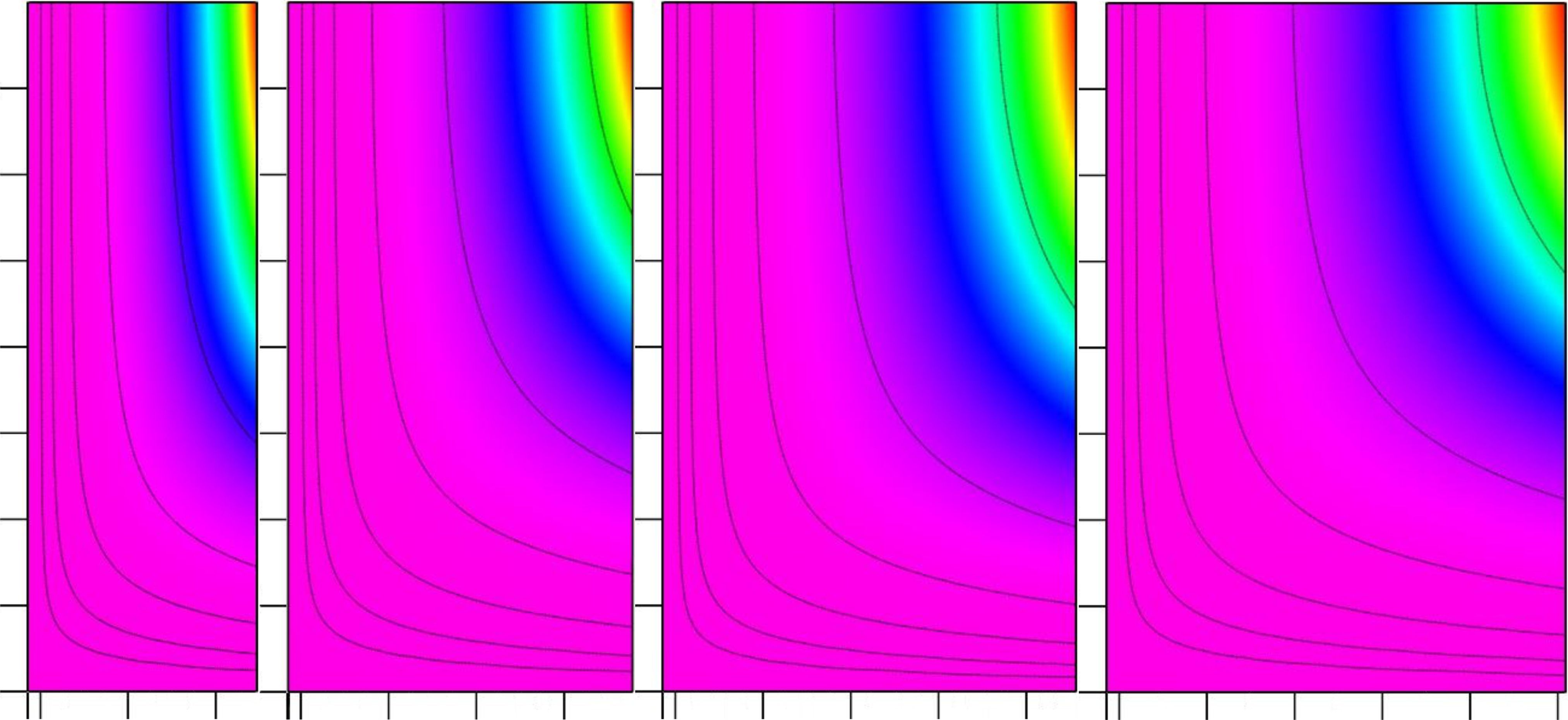}
\begin{small}
\put(-6.5,1){$\pi/2$}
\put(-6.5,23){$\pi/4$}
\put(-1,44){$0$}
\put(0.5,-3.5){$\tfrac{\pi}{2}$}
\put(13,-3.){$2$}
\put(17,-3.5){$\tfrac{\pi}{2}$}
\put(29.5,-3.){$2$}
\put(41,-3.5){$\tfrac{\pi}{2}$}
\put(53.5,-3.){$2$}
\put(63,-3.){$2.4$}
\put(69.5,-3.5){$\tfrac{\pi}{2}$}
\put(82,-3.){$2$}
\put(91.5,-3.){$2.4$}
\end{small}     
  \end{overpic} 
\hfill $\phm$
	 \\
$\phm$
\caption{The graphs of $\varsigma$ for $n=3,4,5,6$ (from left to right), where the contour lines are at the heights $10^{k}$ for $k=-8,\ldots,-3$ (from magenta to red).  
The horizontal axis corresponds to $\gamma\in\left]\tfrac{\pi}{2};\pi-\tfrac{\pi}{n}\right]$ and the vertical one to $\lambda_+\in\left]0;\tfrac{\pi}{2}\right[$.
}
  \label{fig12}
\end{figure}
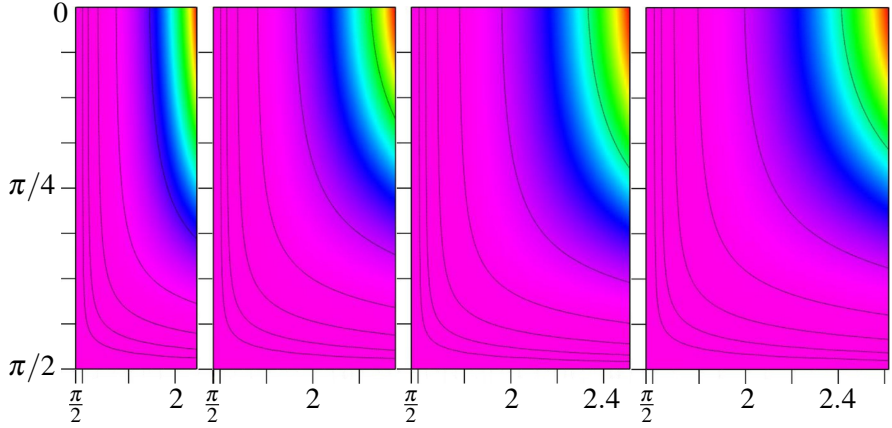

\subsubsection{Rough computation}\label{sec:redcomp}

\begin{figure}[b]
\qquad
\begin{overpic}
    [height=33mm]{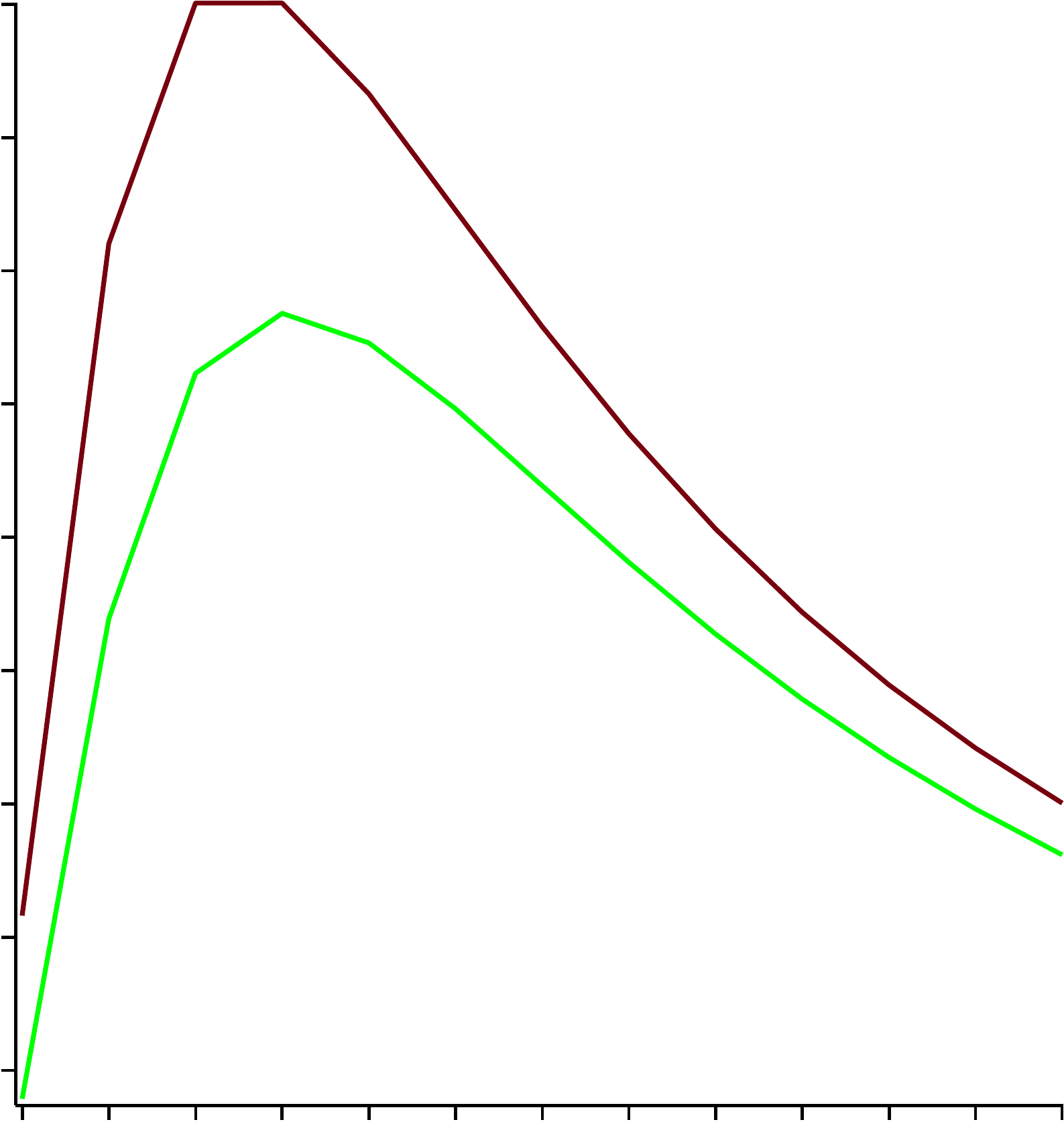}
\begin{small}
\put(-24,-10){a)}
\put(0,-10){$3$}
\put(15,-10){$5$}
\put(33,-10){$n$}
\put(51,-10){$10$}
\put(90,-10){$15$}
\put(-15,2){$\tfrac{4}{10^{4}}$}
\put(-15,37){$\tfrac{1}{10^{3}}$}
\put(-7,65){$\varsigma$}
\put(-15,97){$\tfrac{2}{10^{3}}$}
\end{small}     
  \end{overpic} 
\hfill
 \begin{overpic}
    [height=33mm]{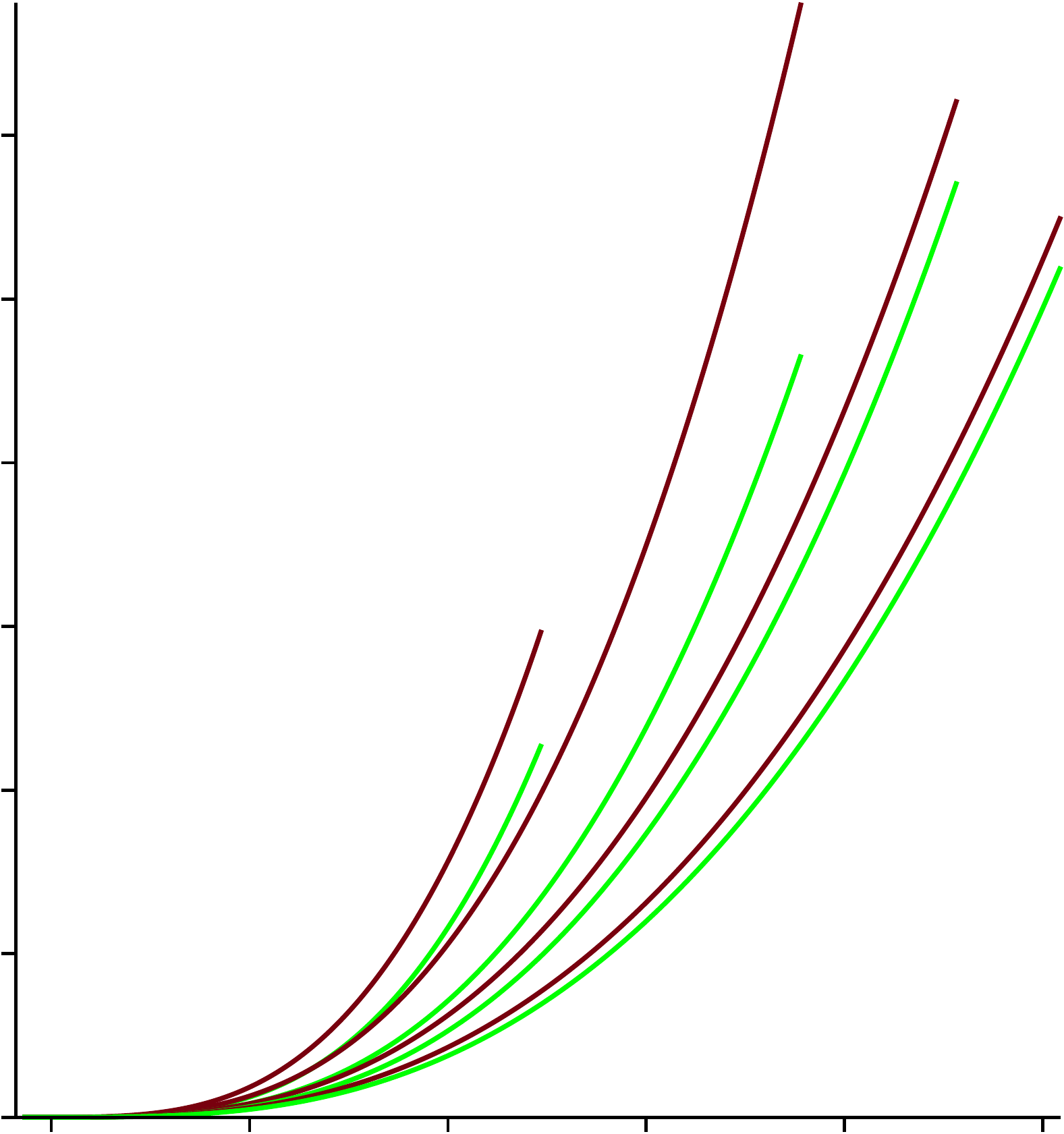}
\begin{small}
\put(-24,-10){b)}
\put(1,-10){$\tfrac{\pi}{2}$}
\put(19,-10){$\gamma$}
\put(37,-10){$2$}
\put(68,-10){$2.4$}
\put(-7,-2){$0$}
\put(-15,13){$\tfrac{1}{10^{4}}$}
\put(-7,65){$\varsigma$}
\put(-15,85){$\tfrac{6}{10^{4}}$}
\end{small}         
  \end{overpic} 
	\hfill
\begin{overpic}
    [height=33mm]{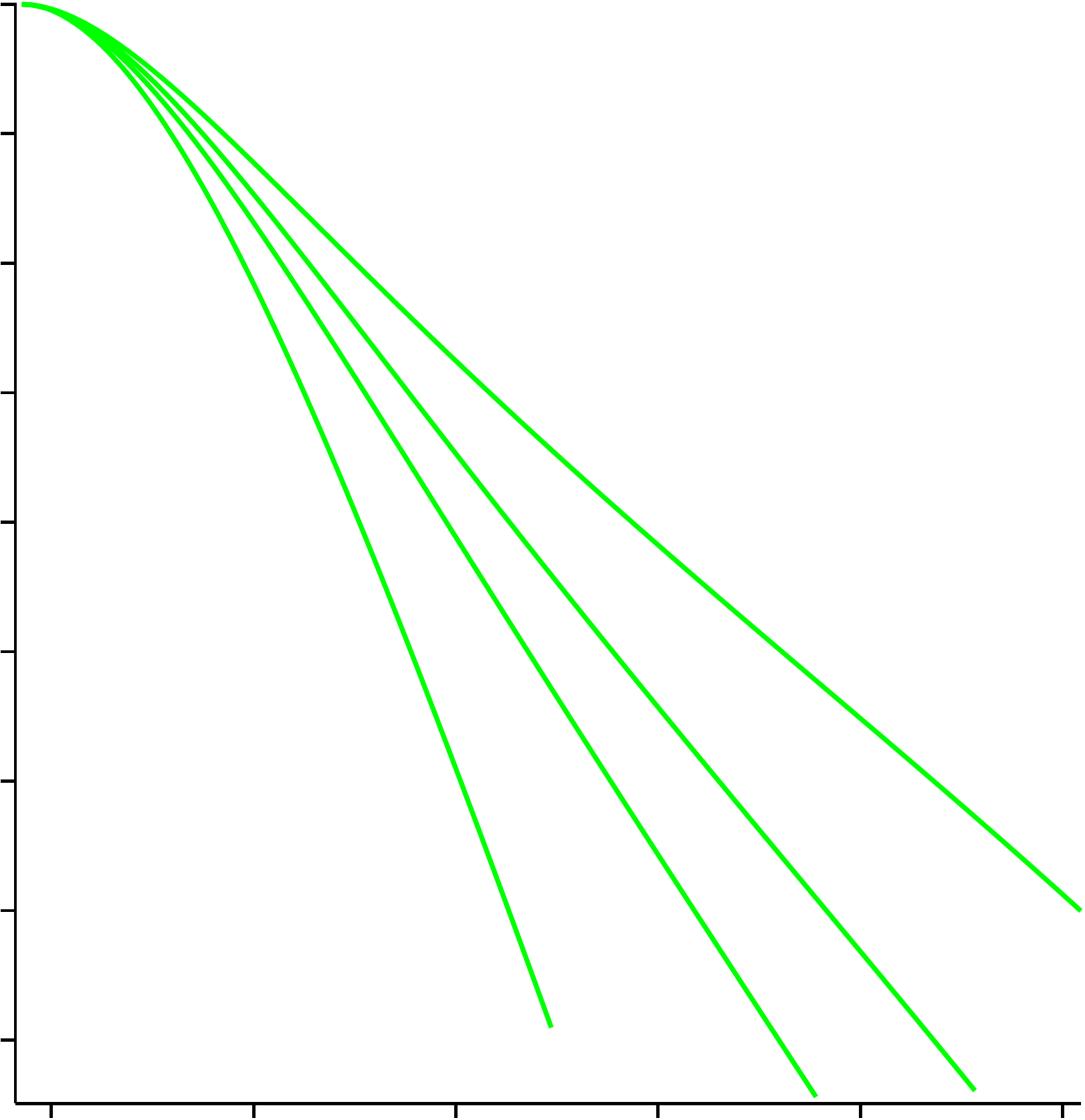}
\begin{small}
\put(-26,-10){c)}
\put(1,-10){$\tfrac{\pi}{2}$}
\put(20,-10){$\gamma$}
\put(39,-10){$2$}
\put(71,-10){$2.4$}
\put(-6,95){$1$}
\put(-8,65){$\rho$}
\put(-22.5,39){$0.99$}
\end{small}         
  \end{overpic} 	
	\hfill
	 \\
$\phm$
\caption{(a) Highest values of $\varsigma$ for $n=3,\ldots,15$. The red/green curve is obtained by the method of 
Section \ref{sec:redcomp} and Section \ref{sec:greedcomp}, respectively. Note that in both cases the peak is obtained for 
$n=6$.
(b) The graph of $\varsigma$ for $n=3,4,5,6$  and $\lambda_+=\tfrac{\pi}{4}$ with respect to the 
method of Section \ref{sec:redcomp} (red curves) and Section \ref{sec:greedcomp} (green curves), respectively, 
where $\gamma$ ranges within the interval  $\left]\tfrac{\pi}{2};\pi-\tfrac{\pi}{n}\right]$.
The corresponding values of $\rho$ are illustrated in (c).
}
\label{fig13}
\end{figure}
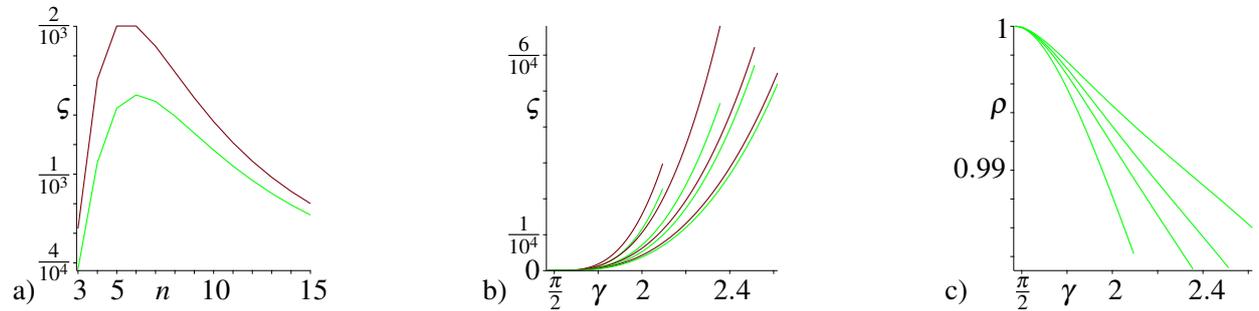

A first estimate (upper border) on the snappability of the cone structure can be obtained by assuming
that the bars of the regular $n$-gons are undeformable; i.e.\ $\rho=1$ in Eq.\ (\ref{eq:ansatz}). 
Despite this assumption, the presented approach is less restrictive than all the analytic studies 
of the cylindrical case reviewed above \cite{cai1,cai2,guestI}. 
Due to $\rho=1$ the total elastic energy function $U_{total}$ only depends on $h_s$. The partial derivative 
$\tfrac{\partial U_{total}}{\partial h_s}$ has a unique solution for $h_s> 0$, which can be computed without 
specifying $n$, $\lambda_+$ and $\gamma$. Therefore we obtain a formula of $\varsigma$ in dependence of these 
unknowns. For $n=3,4,5,6$ the graph of $\varsigma$ for $\lambda_+\in\left]0;\tfrac{\pi}{2}\right[$ and 
$\gamma\in\left]\tfrac{\pi}{2};\pi-\tfrac{\pi}{n}\right]$ is illustrated in Fig.\ \ref{fig12}. 
For each $n$ we get the highest $\varsigma$-value for $\gamma=\pi-\tfrac{\pi}{n}$ when the cone converges towards 
a cylinder; i.e.\ $\lambda_+\rightarrow 0$, which matches with our intuitive expectations.
The run of these highest $\varsigma$-values for $n=3,\ldots, 15$ is illustrated in Fig.\ \ref{fig13}a.

\subsubsection{Improved computation}\label{sec:greedcomp}

Within this more accurate approach we relax the assumption on the bars of the regular $n$-gons as follows: 
They are all scaled by the same factor $\rho>0$ during the snapping procedure. 
This relation ensures that the uniformity of the cone structure does not get violated 
during the snap.

For the determination of $\rho$ and $h_s$ we compute the resultant of 
$\tfrac{\partial U_{total}}{\partial h_s}$ and $\tfrac{\partial U_{total}}{\partial \rho}$ with respect to $h_s$ 
which has a unique solution for $\rho$. Its back-substitution into 
$\tfrac{\partial U_{total}}{\partial h_s}$ results in an expression, where $h_s$ appears only quadratically. 
Therefore $h_s>0$  is also determined uniquely. Due to the length of the obtained expressions for $\rho$ and $h_s$ a 
symbolic computation of $\varsigma$ in its full generality (i.e.\ without specifying $n$, $\lambda_+$ and $\gamma$) 
is not viable. For $n=3,4,5,6$ and $\lambda_+=\tfrac{\pi}{4}$ the resulting $\varsigma$-values are computed 
for $\gamma\in\left]\tfrac{\pi}{2};\pi-\tfrac{\pi}{n}\right]$ and displayed in Fig.\ \ref{fig13}b. 
The corresponding values of $\rho$ are illustrated in Fig.\ \ref{fig13}c.

Moreover we computed also the run of the highest $\varsigma$-values for $n=3,\ldots, 15$, which are obtained for 
$\gamma=\pi-\tfrac{\pi}{n}$ and $\lambda_+\rightarrow 0$ (cf.\ Fig.\ \ref{fig13}a).

%%%%%%%%%%%%%%%%%%%%%%%%%%%%%%%%%%%%%%%%%%%%%%%%%%%%%%%%%%%%%%%%%%%%%%%%%%%%%%%

\subsection{Snappability of spiral-motion based conical triangulation}

In this subsection we consider spiral-motion based conical triangulations where the two realizations $\mathcal{R}_+$ and $\mathcal{R}_-$  are 
located on the same cone $\Lambda_{\pm}$. These structures are of interest as they can change the
cross-sectional area while their conical shape is preserved (cf.\ Example \ref{ex:cross_sec}).

As in the study \cite{guestI} by Guest and Pellegrino on the analogue 
construction for the cylindrical case, we assume that the folding process is {\it uniform}; i.e.\ 
the vertices remain on a spiral curve during the deformation, in order to make an analytical study feasible. 
Therefore the vertices $V_{is}$ of the shaky configuration $\mathcal{R}_s$ associated with the snap between 
$\mathcal{R}_+$ and $\mathcal{R}_-$ can be parametrized as follows (cf.\ Eq.\ (\ref{eq:pardisspir})):  
\begin{equation}
V_{is}=\begin{pmatrix}
r_s p_s^i\cos(i\phi_s) \\
r_s p_s^i\sin(i\phi_s) \\
-r_sp_s^i q_s
\end{pmatrix}
\quad\text{with} \quad p_s:=e^{\phi_s \sin\lambda_s\cot\delta_s} 
\quad \text{and} \quad q_s=\cot\lambda_s
\end{equation}
for $i\in\NN$. Now $U_{total}$ can be computed as 
\begin{equation}
\sum_{k=0}^{\infty}\left[U(\overline{V_{\pm k}V_{\pm (k+1)}}, \overline{V_{ks}V_{(k+1)s}})+
U(\overline{V_{\pm k}V_{\pm (k+n-1)}}, \overline{V_{ks}V_{(k+n-1)s}})+
U(\overline{V_{\pm k}V_{\pm (k+n)}}, \overline{V_{ks}V_{(k+n)s}})\right],
\end{equation}
where each  sub-total can be split up in a sum of geometric series as follows: 
\begin{equation}
\begin{split}
\sum_{k=0}^{\infty}U(\overline{V_{\pm k}V_{\pm (k+t)}}, \overline{V_{ks}V_{(k+t)s}})
&=\frac{E\Area_{\diameter}}{8}\sum_{k=0}^{\infty}\left[
p_{\pm}^k\overline{V_{\pm 0}V_{\pm t}} -
2\tfrac{p_s^{2k}}{p_{\pm}^k}\tfrac{\overline{V_{0s}V_{ts}}^2}{\overline{V_{\pm 0}V_{\pm t}}}  +
\tfrac{p_s^{4k}}{p_{\pm}^{3k}}\tfrac{\overline{V_{0s}V_{ts}}^4}{\overline{V_{\pm 0}V_{\pm t}}^3} 
\right] \\
&= \frac{E\Area_{\diameter}}{8}\left[
\tfrac{\overline{V_{\pm 0}V_{\pm t}}}{1-p_{\pm}}-
2\tfrac{p_{\pm}\overline{V_{0s}V_{ts}}^2}{(p_{\pm}-p_s^2)\overline{V_{\pm 0}V_{\pm t}}} + 
\tfrac{p_{\pm}^3\overline{V_{0s}V_{ts}}^4}{(p_{\pm}^3-p_s^4)\overline{V_{\pm 0}V_{\pm t}}^3}
\right]
\end{split}
\end{equation}
for $t=1,n-1,n$ under the assumption that the individual series converge\footnote{Note that the relation $1>p_s,p_{\pm}>0$ holds true. 
If at least one of the geometric series diverges then this would mean that infinite energy is needed for the snap of the infinite structure. 
In this case the snappability would be infinite.}
; i.e.\ $\tfrac{p_s^{2}}{p_{\pm}}$<1 and $\tfrac{p_s^{4}}{p_{\pm}^{3}}<1$. 

Under consideration of $\Vol_{total}$, whose computation is straightforward; i.e.\ 
\begin{equation}
\Vol_{total}=\Area_{\diameter}
\sum_{k=0}^{\infty}\left[\overline{V_{\pm k}V_{\pm (k+1)}}+
\overline{V_{\pm k}V_{\pm (k+n-1)}}+
\overline{V_{\pm k}V_{\pm (k+n)}}\right]=
\tfrac{\Area_{\diameter}\left[\overline{V_{\pm 0}V_{\pm 1}}+
\overline{V_{\pm 0}V_{\pm (n-1)}}+
\overline{V_{\pm 0}V_{\pm n}}\right]}{1-p_{\pm}}
\end{equation}
the snappability index of this infinite structure results from $\varsigma={U_{total}}/(E\Vol_{total})$.
In the following we distinguish again two approaches.

%%%%%%%%%%%%%%%%%%%%%%%%%%%%%%%%%%%%%%%%%%%%%%%%%%%%%%%%%%%%%%%%%%%%%%%%%%%%%%%

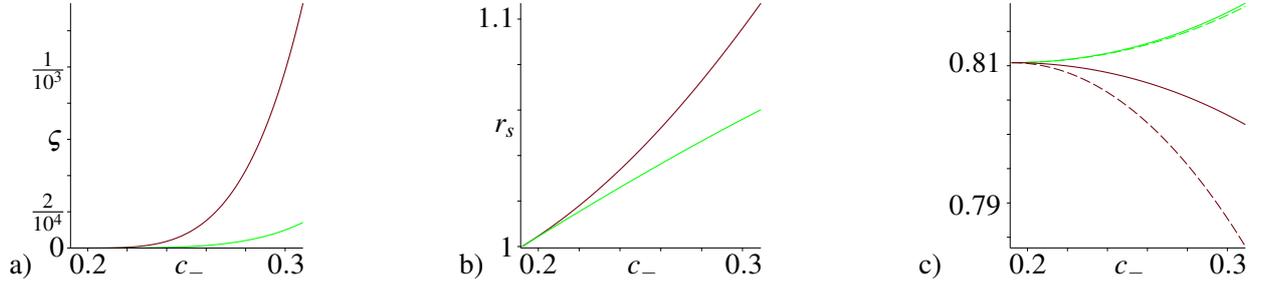
\begin{figure}[t]
\qquad
\begin{overpic}
    [height=33mm]{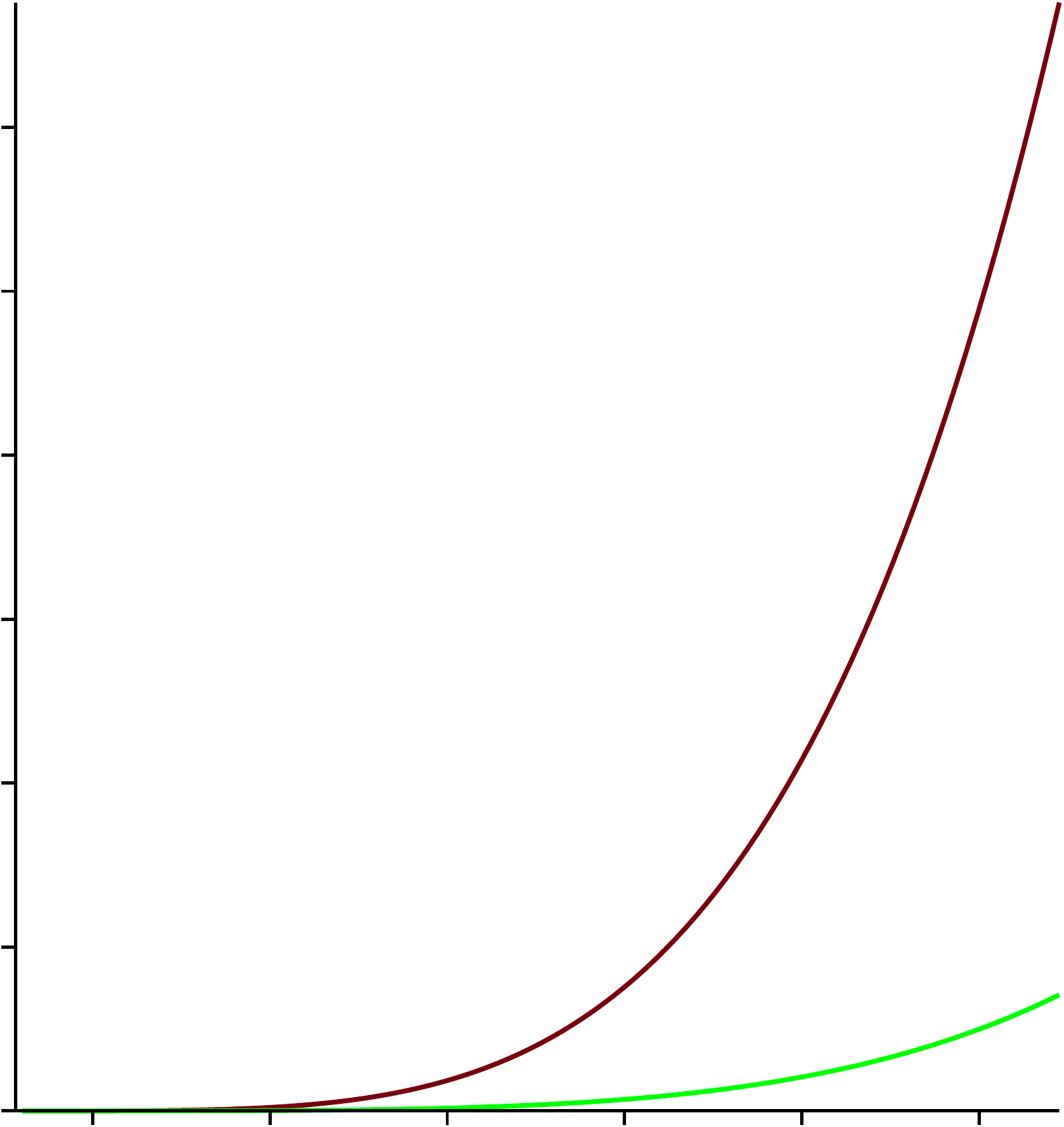}
\begin{small}
\put(-23,-9){a)}
\put(1,-9){$0.2$}
\put(43,-9){$c_-$}
\put(80,-9){$0.3$}
\put(-15,13){$\tfrac{2}{10^{4}}$}
\put(-7,-2){$0$}
\put(-8,43){$\varsigma$}
\put(-15,70){$\tfrac{1}{10^{3}}$}
\end{small}     
  \end{overpic} 
\hfill
 \begin{overpic}
    [height=33mm]{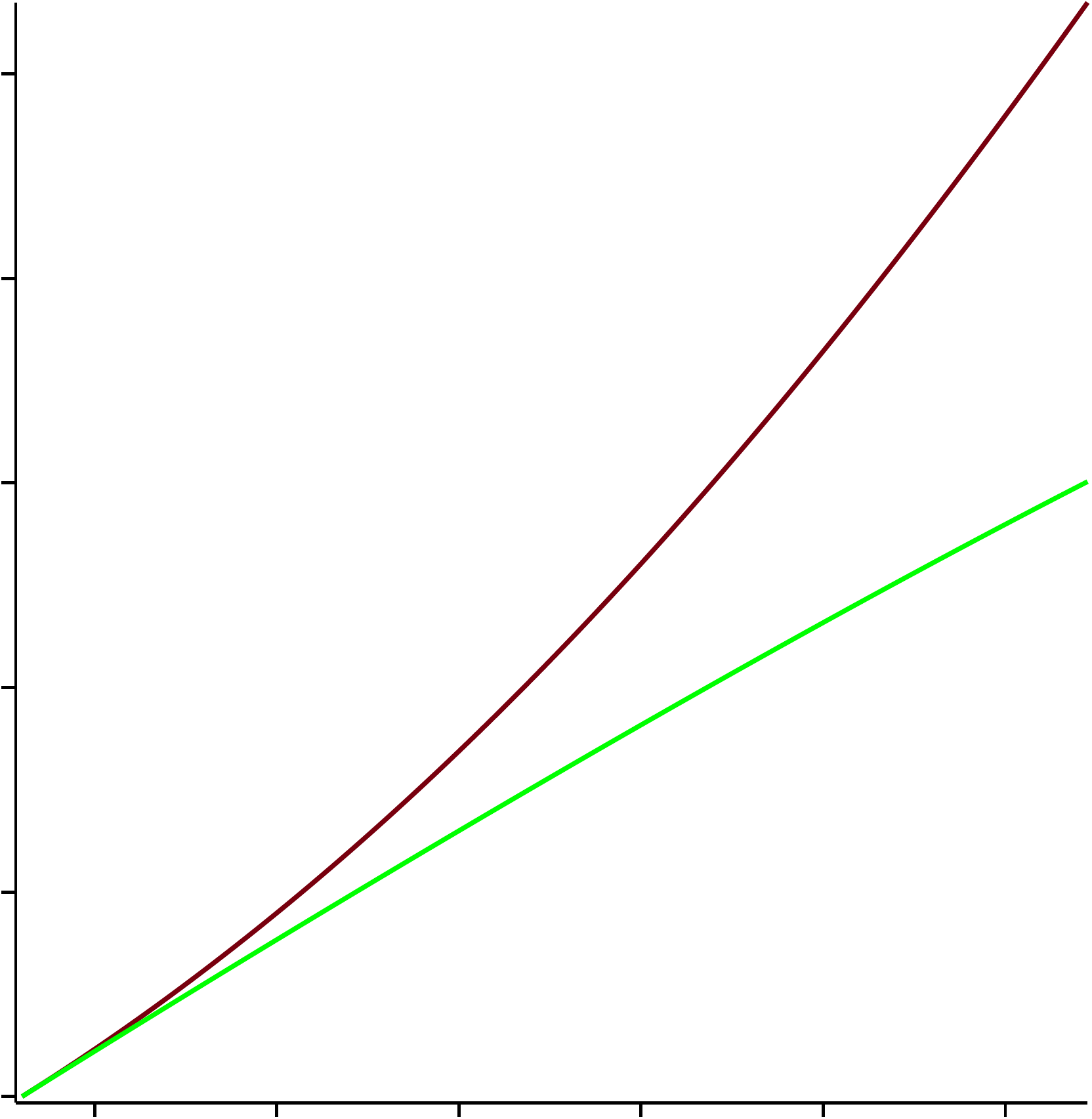}
\begin{small}
\put(-23,-9){b)}
\put(1,-9){$0.2$}
\put(44,-9){$c_-$}
\put(83,-9){$0.3$}
\put(-7,-2){$1$}
\put(-16,90){$1.1$}
\put(-9,48){$r_s$}
\end{small}         
  \end{overpic} 
		\hfill \quad
\begin{overpic}
    [height=33mm]{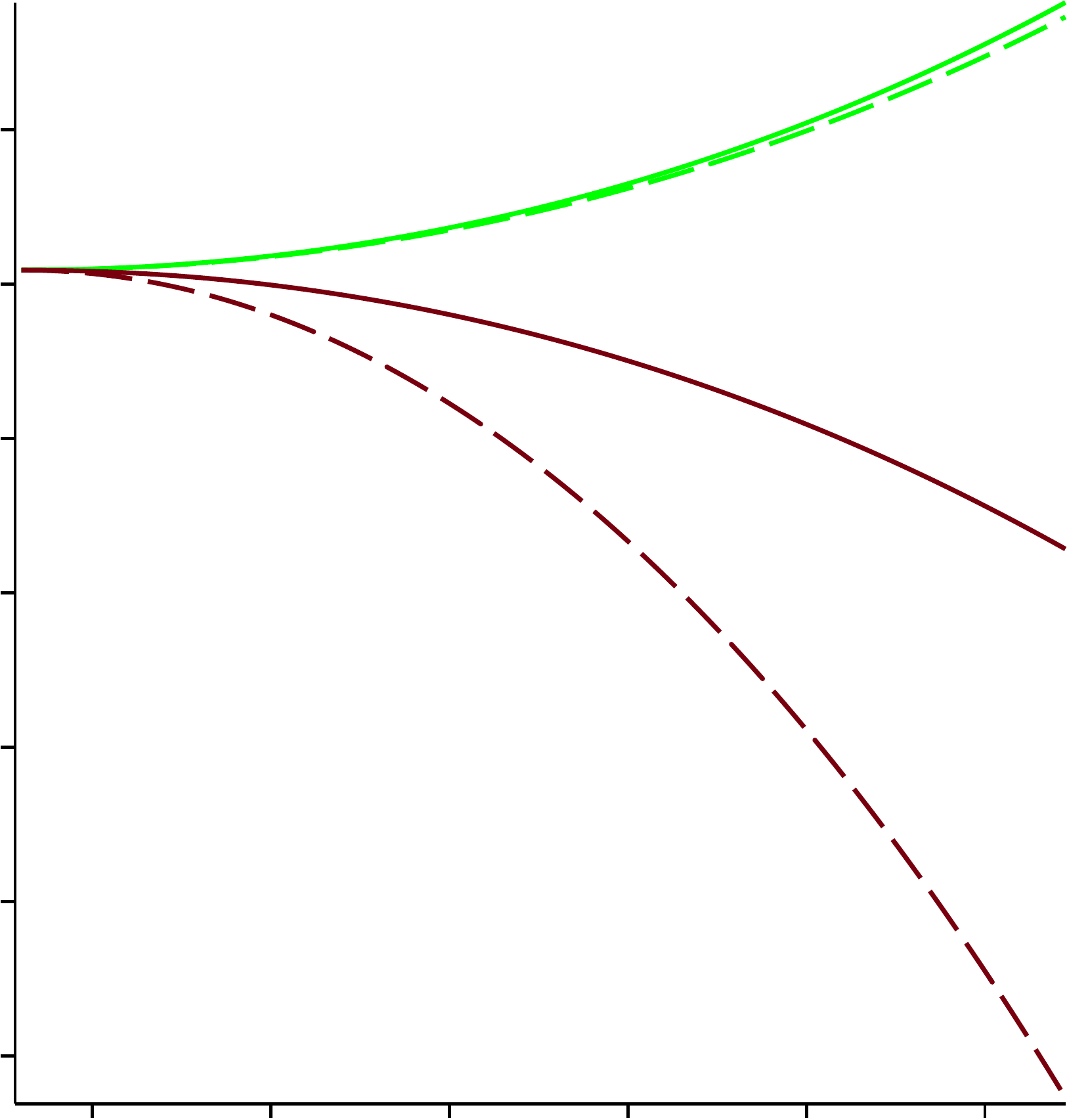}
\begin{small}
\put(-35,-9){c)}
\put(1,-9){$0.2$}
\put(43,-9){$c_-$}
\put(81,-9){$0.3$}
\put(-23,15){$0.79$}
\put(-23,72){$0.81$}
\end{small}         
  \end{overpic} 	
	\hfill 
	 \\
$\phm$
\caption{(a) The graph of $\varsigma$  with respect to the 
method of Section \ref{sec:roughspiral} (red curve) and Section \ref{sec:impspiral} (green curve), respectively.
The corresponding values of $r_s$ are also illustrated in (b). (c) The graphs of $\tfrac{p_s^{2}}{p_{\pm}}$ (solid) and 
$\tfrac{p_s^{4}}{p_{\pm}^3}$ (dashed) verify that the geometric series involved in the computation of $U_{total}$ converge 
as both values are always less than $1$.
}
\label{fig14}
\end{figure}

\subsubsection{Rough computation}\label{sec:roughspiral}

A first estimate (upper border) on the snappability of the cone structure can be obtained by
assuming that $\lambda_s=\lambda_{\pm}$ ($\Rightarrow$ $q_s=q_{\pm}$) holds. 
The consequences of this assumption are discussed on basis of  Example \ref{ex:advanced_snap}, which we have
already analyzed in detail.  

We consider the set of global self-intersection free snapping realizations, which  
corresponds to the curve segment of $\go h$ bounded by $H_-^{5,1}$ and $H_s^1$ (cf.\ Fig.\ \ref{fig9a}b). 
The assumption $\lambda_s=\lambda_{\pm}$ does not necessarily imply that the snap has to pass the singular configuration 
$\mathcal{R}_s^1$ but a scaled version of it, where $r_s$ is the scaling factor. 
The values of $c_s$ and $p_s$ equal those of $\mathcal{R}_s^1$ given in the table of Example \ref{ex:shaky}. 
Based on $p_s$ it can also be verified that the geometric series involved in the computation of $U_{total}$ converge (cf.\ Fig.\ \ref{fig14}c)

The graph of $\varsigma$ for $c_-\in]c_s; \tfrac{\sqrt{5}-1}{4}]$ is displayed in Fig.\ \ref{fig14}a and the  
graph of the corresponding $r_s$-values, which can be computed from $\tfrac{\partial U_{total}}{\partial r_s}=0$, is 
illustrated in  Fig.\ \ref{fig14}b.

%%%%%%%%%%%%%%%%%%%%%%%%%%%%%%%%%%%%%%%%%%%%

\subsubsection{Improved computation}\label{sec:impspiral}

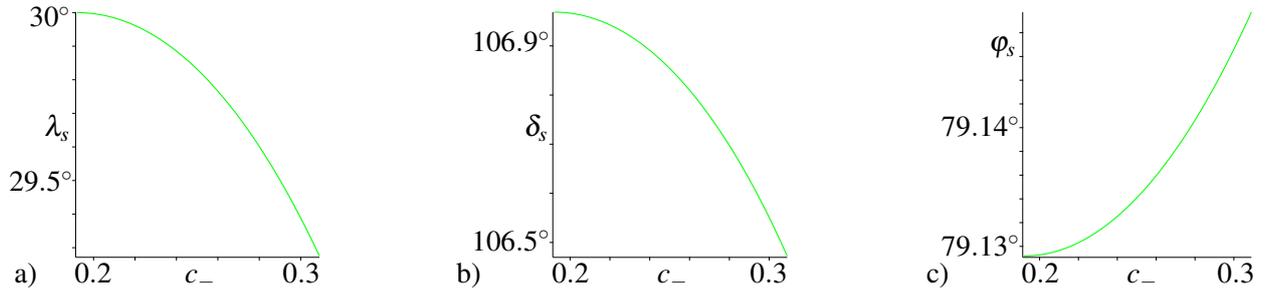
\begin{figure}[t]
\qquad
\begin{overpic}
    [height=33mm]{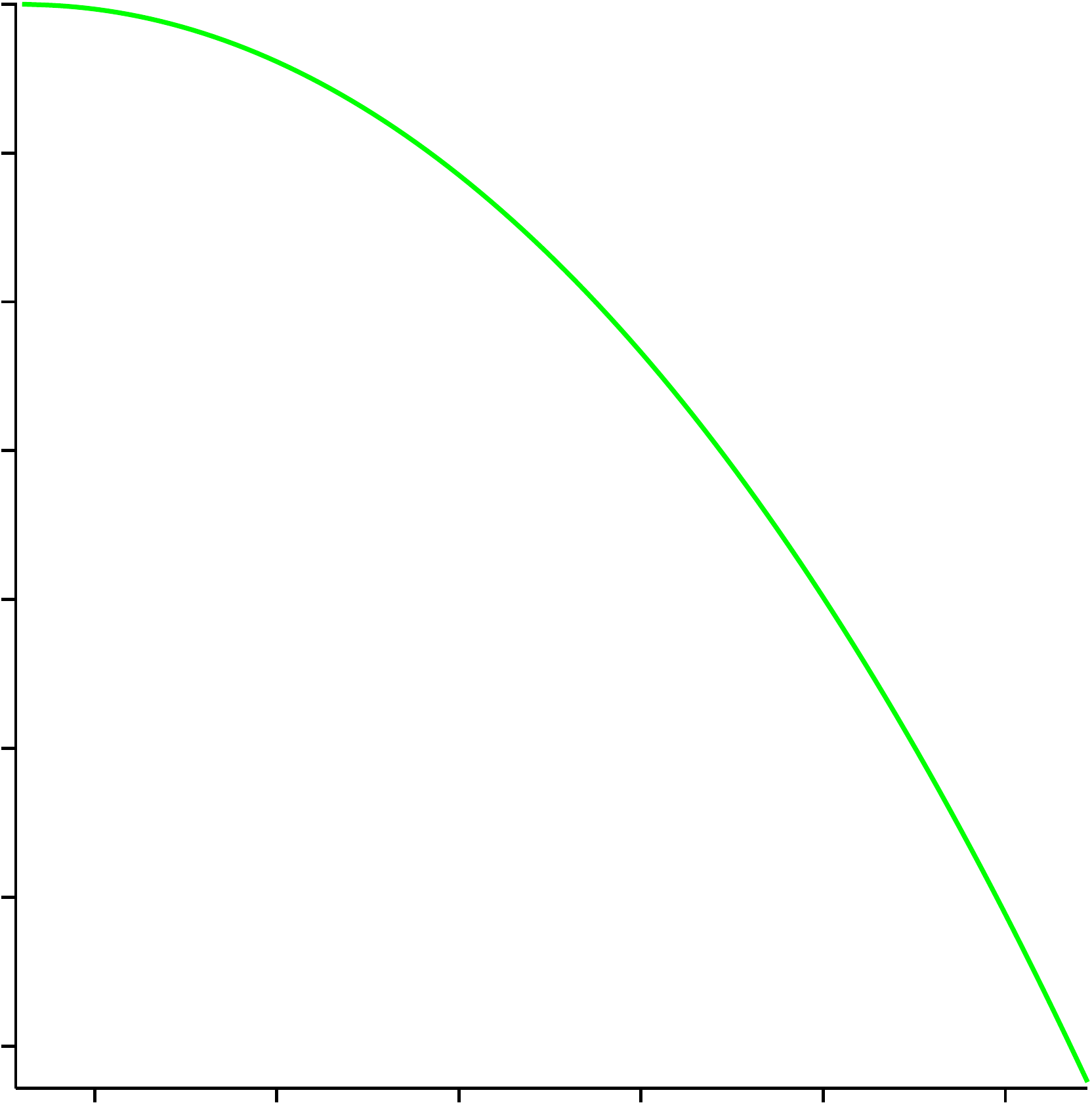}
\begin{small}
\put(-23,-9){a)}
\put(1,-9){$0.2$}
\put(45,-9){$c_-$}
\put(84,-9){$0.3$}
\put(-11,50){$\lambda_s$}
\put(-17,94){$30^{\circ}$}
\put(-25,28){$29.5^{\circ}$}
\end{small}         
  \end{overpic} 	
		\hfill 
\begin{overpic}
    [height=33mm]{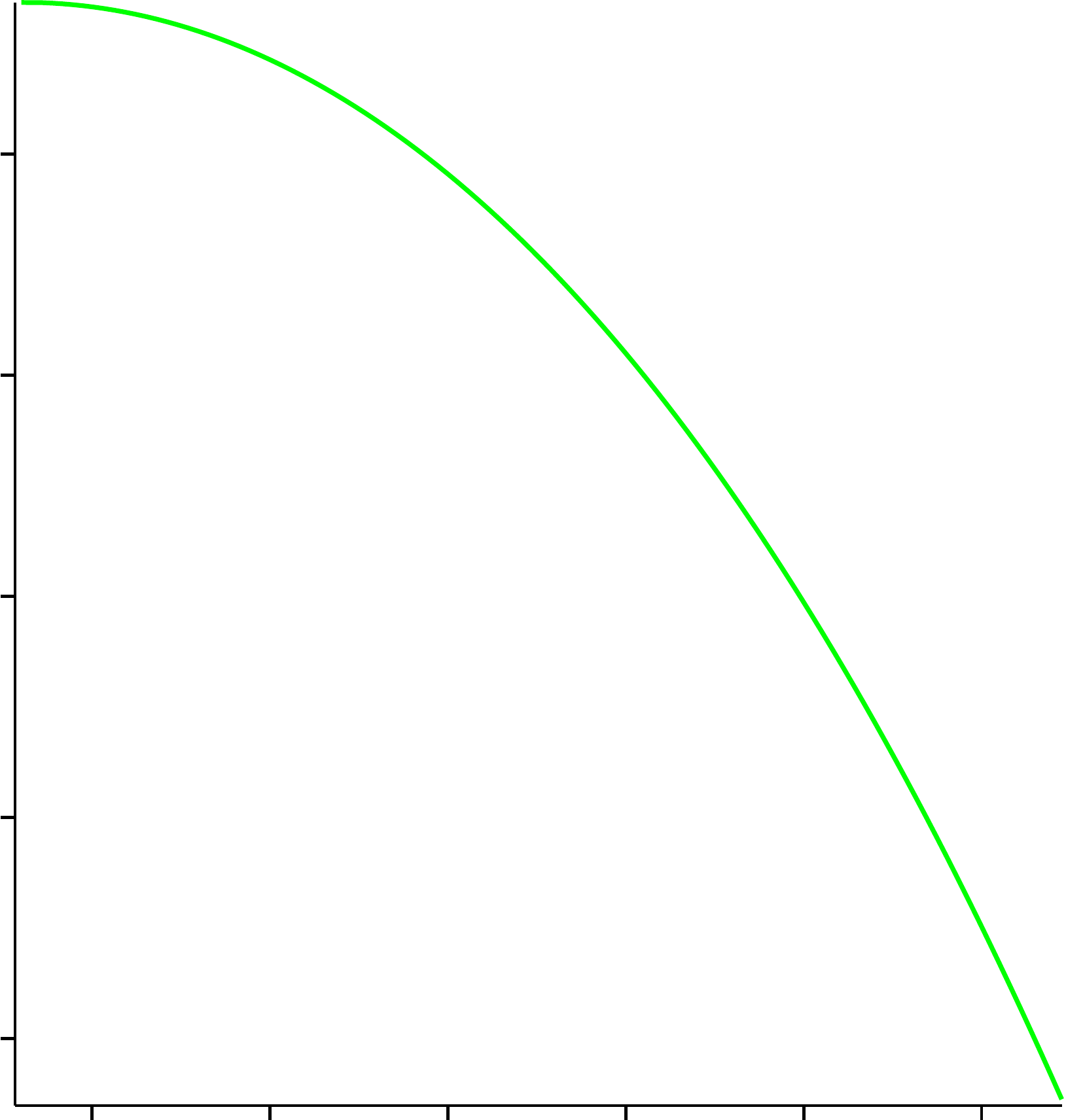}
\begin{small}
\put(-37,-9){b)}
\put(1,-9){$0.2$}
\put(43,-9){$c_-$}
\put(80,-9){$0.3$}
\put(-10,50){$\delta_s$}
\put(-31,4){$106.5^{\circ}$}
\put(-31,84){$106.9^{\circ}$}
\end{small}         
  \end{overpic} 	
		\hfill 
\begin{overpic}
    [height=33mm]{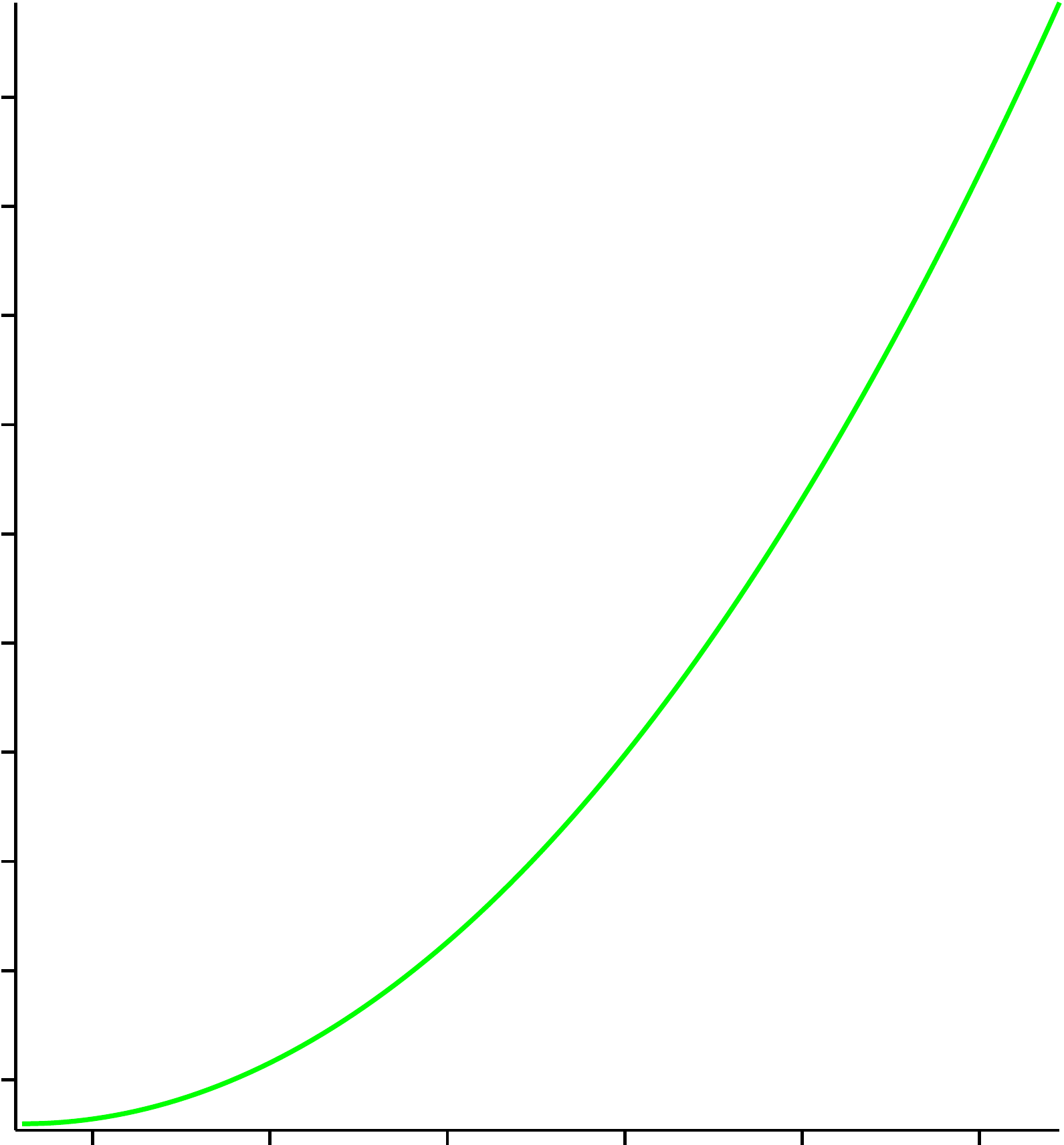}
\begin{small}
\put(-37,-9){c)}
\put(1,-9){$0.2$}
\put(43,-9){$c_-$}
\put(79,-9){$0.3$}
\put(-12,84){$\phi_s$}
\put(-31,2){$79.13^{\circ}$}
\put(-31,50){$79.14^{\circ}$}
\end{small}         
  \end{overpic} 	
	\hfill 
	 \\
$\phm$
\caption{The graphs of $\lambda_s$ (a), $\delta_s$ (b) and $\phi_s$ (c), respectively, for the 
results obtained by the method described in Section \ref{sec:impspiral}.
}
\label{fig15}
\end{figure}

Within this more accurate approach we skip the assumption on $\lambda_s$. 
As a consequence one has to solve the following system of equations:
\begin{equation}
\tfrac{\partial U_{total}}{\partial \lambda_s}=0,\quad
\tfrac{\partial U_{total}}{\partial p_s}=0,\quad
\tfrac{\partial U_{total}}{\partial c_s}=0,\quad
\tfrac{\partial U_{total}}{\partial r_s}=0
\end{equation}
for $\lambda_s$, $c_s$, $p_s$ and $r_s$, which is done numerically for the same dataset as used in Section \ref{sec:roughspiral}. 
The obtained values for 
$r_s$ and $\lambda_s$  are displayed in Figs.\ \ref{fig14}b and \ref{fig15}a, respectively. 
Note that the cone $\Lambda_s$ associated with $\mathcal{R}_s$ gets more acute for increasing $c_-$-values.
The values of $\phi_s$ and $\delta_s$ resulting from $c_s$ and $p_s$ are illustrated in Fig.\ \ref{fig15}b,c. 
Based on $p_s$ it can again be checked that the geometric series involved in the computation of $U_{total}$ 
converge (cf.\ Fig.\ \ref{fig14}c). The resulting $\varsigma$-values for the snappability are visualized in 
Fig.\ \ref{fig14}a, which differ a lot from the first estimate computed in Section \ref{sec:roughspiral}.

\section{Conclusion and open problems}
In this paper we generalized the circular and helical arrangement of the Kresling pattern on rotary cylinders to cones of revolution 
with emphasis on multi-stability. The presented approach enables to design origami structures, which snap between conical realizations 
with prescribed apex angles. We also analyzed these triangulated cones with respect to their capability to snap by means of the 
so-called snappability index.  
In addition we figured out shaky realizations, intervals for self-intersection free realizations and 
an interesting property related to the cross sectional area.

Finally, we  close the paper with a list of open problems:
\begin{enumerate}[$\bullet$]
\item
We conjecture that all the given theorems hold for arbitrary  $n>3$, but a geometric proof is still missing (cf.\ Remarks \ref{rem:moren} and \ref{rem:moren2}). 
\item 
Geometric characterization of infinitesimal flexibility for helical/spiral-motion based cylindrical/conical triangulated structures (cf.\ Remark \ref{rem:shaky2}).
\item
The question of existence of three self-intersection free realizations of a spiral-motion based triangulated structure on different cones (cf.\ Section \ref{sec:tristable}).
\item 
We conjecture that there are at most two realizations of a spiral-motion based triangulated structure on the same cone (cf.\ Remark \ref{rem:2oncone}).
\item
On can think of a further generalization of the paper at hand by studying 
polyhedral cone surfaces with planar quadrilateral faces 
in analogy to cylinders formed from  Kokotsakis' 
flexible tessellations \cite{stachel1,stachel2,wittenburg2}. 
In this context it should be noted, that a conical Miura-ori pattern, which can be seen as a generalization of the anti-frustum based triangulation of the cone, 
was presented recently in \cite{sharma2}.
\end{enumerate}

\bigskip

\noindent
{\bf Acknowledgments.} 
The author is supported by grant P\,30855-N32 of the Austrian Science Fund FWF 
as well as by project F77 (SFB ``Advanced Computational Design'', subproject SP7).

%%%%%%%%%%%%%%%%%%%%%%%%%%%%%%%%%%%%%%%%%%%%%%%%%%%%%%%%%%%%%%%%%%%%%%%%%%%%%%%%%%%%%%%%%%%%%%%%%%%%%%%%%%%%%%%
%%%%%%%%%%%%%%%%%%%%%%%%%%%%%%%%%%%%%%%%%%%%%%%%%%%%%%%%%%%%%%%%%%%%%%%%%%%%%%%%%%%%%%%%%%%%%%%%%%%%%%%%%%%%%%%
%%%%%%%%%%%%%%%%%%%%%%%%%%%%%%%%%%%%%%%%%%%%%%%%%%%%%%%%%%%%%%%%%%%%%%%%%%%%%%%%%%%%%%%%%%%%%%%%%%%%%%%%%%%%%%%

\appendix

\section{Trigonometric function value}\label{app:trig}

In the following we list the analytic expressions for some needed trigonometric function values:
\begin{align}
&\cos{\tfrac{2\pi}{3}}=-\tfrac{1}{2} \\
&\cos{\tfrac{\pi}{2}}=0 \\
&\cos{\tfrac{2\pi}{5}}=\tfrac{\sqrt{5}-1}{4} \\
&\cos{\tfrac{\pi}{3}}= -\cos{\tfrac{2\pi}{3}}=\tfrac{1}{2} \\
&\cos{\tfrac{2\pi}{7}}= \tfrac{w^{2/3} - 2w^{1/3} + 28}{12w^{1/3}} \quad\text{with} \quad w:=28 + 84I\sqrt{3}\\
&\cos{\tfrac{\pi}{4}}=-\cos{\tfrac{3\pi}{4}}=\tfrac{\sqrt{2}}{2} \\
&\cos{\tfrac{2\pi}{9}}=\tfrac{v^{2/3} + 4}{4v^{1/3}}   \quad\text{with} \quad v:=4I\sqrt{3}-4 \\
&\cos{\tfrac{4\pi}{5}}=-\tfrac{1+\sqrt{5}}{4} \\ 
&\cos{\tfrac{4\pi}{7}}= \tfrac{w_1^{1/3}+w_1^{1/3}I\sqrt{3}}{24} - \tfrac{5w_1^{2/3} - w_1^{2/3}I\sqrt{3}}{336} - \tfrac{1}{6} \\
&\cos{\tfrac{6\pi}{7}}=\tfrac{w_1^{2/3}}{336} + \tfrac{w_1^{2/3}I\sqrt{3}}{112} -\tfrac{w_1^{1/3}}{12}  - \tfrac{1}{6} \\
&\cos{\tfrac{4\pi}{9}}= \tfrac{v_1^{2/3} - v_1^{1/3} + v_1^{1/3}I\sqrt{3}}{8} \quad\text{with} \quad  v_1:=\overline{v} \\
&\cos{\tfrac{8\pi}{9}}= -\tfrac{v_2^{2/3} + v_2^{2/3}I\sqrt{3}}{16} - \tfrac{v_2^{1/3}}{4} \quad\text{with} \quad  v_2:=-v 
\end{align}

\section{Proof of Theorem \ref{thm:prop3}}\label{app:proof3}

As in the proof for Theorem \ref{thm:prop} we start with the 
computation of the algebraic condition $f=0$ indicating the coplanarity of $V_0,V_{1},V_{n-1},V_{n}$ with
\begin{equation}
g:=\det\begin{pmatrix} 1 & 1 & 1 & 1 \\
V_0 & V_1 & V_{n-1} & V_{n}
\end{pmatrix}.
\end{equation}
By using  Chebyshev polynomials $U_i(x)$ of the second kind
we can rewrite $g$ as $q\sin\phi_{+}g_*$ with
\begin{equation}\label{eq:f*neu}
\begin{split}
g_*=&\left(p^{2n}-p^{n}\right)\left(D_{n-3}(c_+)-D_{n-2}(c_+)+1\right) +\\
&\left(D_{n-1}(c_+)-D_{n-2}(c_+)-1\right)p^{2n-1} +  \left(D_{n-2}(c_+)-D_{n-1}(c_+)+1\right)p^{n+1}.  
\end{split}
\end{equation}
We proceed as in the  proof for Theorem \ref{thm:prop3} where we substitute the  
analytic expression for the trigonometric function value $\cos{\tfrac{2\pi}{n}}$ according to \ref{app:trig}
into  $d_*(0,n-1)$ and $d_*(0,n)$.
Again we compute the three possible resultants of the expressions $g_*$, $d_*(0,n-1)$ and $d_*(0,n)$ each with respect to $p$. 
The only factors of $Res(d_*(0,n-1),d_*(0,n))$ 
which do not appear in $Res(g_*,d_*(0,n-1))$ as well as $Res(g_*,d_*(0,n))$ are:
\begin{align}
&n=4: &\quad &(q^2+1) \\  
&n=5: &\quad &(2q^2 + \sqrt{5} - 1)(q^2+1) \\ 
&n=6: &\quad &\text{same as Eq.\ (\ref{thm:3})} \notag \\ 
&n=7: &\quad &\text{same as Eq.\ (\ref{thm:4}) and Eq.\ (\ref{thm:5}), respectively} \notag \\
&n=8: &\quad &\text{same as Eq.\ (\ref{thm:6})} \notag \\
&n=9: &\quad & (v^{2/3}I\sqrt{3} + v^{2/3} - 24q^2 - 4v^{1/3)} - 8) \\
&  &\quad & (v^{2/3}I\sqrt{3} - v^{1/3}I\sqrt{3} + 2v^{2/3} - 4q^2 - 5v^{1/3} + 8) \\
&  &\quad & (v^{2/3}I\sqrt{3} + 2v^{1/3}I\sqrt{3} - v^{2/3} - 8q^2 - 2v^{1/3} + 8)
\end{align}
with $v:=4I\sqrt{3}-4$.
All given factors do not have a real solution for $q$, which closes the proof.  \hfill $\BewEnde$

\section{Proof of Theorem \ref{thm:prop4}}\label{app:proof4}

The proof can be done in the same fashion as the one of Theorem \ref{thm:prop3}. We only have to replace for type  $\left\{\tfrac{n}{d} \right\}$ 
the variable $c_-$ by the analytic expression for the trigonometric function value $\cos{\tfrac{2d\pi}{n}}$ according to  \ref{app:trig}. 
In the following we list the factors of $Res(d_*(0,n-1),d_*(0,n))$ 
which do not appear in $Res(g_*,d_*(0,n-1))$ as well as $Res(g_*,d_*(0,n))$:
\begin{align}
&n=5, d=2: &\quad & (-2q^2 + \sqrt{5} + 1)(q^2+1)\\  
&n=6, d=2: &\quad & \text{same as Eq.\ (\ref{thm2:2})} \notag \\ 
&n=7, d=2: &\quad & \text{same as Eq.\ (\ref{thm2:3}) and Eq.\ (\ref{thm2:4}), respectively} \notag \\
&n=7, d=3: &\quad & \text{same as Eq.\ (\ref{thm2:5}) and Eq.\ (\ref{thm2:6}), respectively} \notag \\
&n=8, d=2: &\quad & \text{same as Eq.\ (\ref{thm2:7})} \notag\\
&n=8, d=3: &\quad & \text{same as Eq.\ (\ref{thm2:8})} \notag   \\
&n=9, d=2: &\quad &  (v_1^{2/3}I\sqrt{3} - v_1^{2/3} - 8q^2 + 4v_1^{1/3} + 8 )\\
& &\quad & 					 (v_1^{2/3}I\sqrt{3} + 6v_1^{1/3}I\sqrt{3} + 5v_1^{2/3} + 8q^2 - 2v_1^{1/3} - 16) \\
& &\quad & 					 (v_1^{1/3}I\sqrt{3} + v_1^{2/3} + 12q^2 - v_1^{1/3} + 4) \\
&n=9, d=3: &\quad &  (q^6 - 6q^4 + 3q^2 + 1)\\
&n=9, d=4: &\quad &  (v_2^{2/3}I\sqrt{3} - 2v_2^{1/3}I\sqrt{3} - v_2^{2/3} - 8q^2 + 2v_2^{1/3} + 8) \\
& &\quad & 					 (v_2^{2/3}I\sqrt{3} + v_2^{2/3} - 24q^2 + 4v_2^{1/3} - 8)\\
& &\quad & 					 (v_2^{2/3}I\sqrt{3} + v_2^{1/3}I\sqrt{3} + 2v_2^{2/3} - 4q^2 + 5v_2^{1/3} + 8)
\end{align}
with $v_1:=\overline{v}$ and $v_2:=-v$.
The case $n$ for $n=5,\ldots ,8$ is the same as the case $n-1$ of Theorem \ref{thm:prop2} for all possible $d$. 
Therefore we only have to discuss in more detail the following cases:
\begin{enumerate}[$\bullet$]
\item
Case $n=9, d=2$: Only the first factor implies a possible solution for $q$, which reads as
\begin{equation}
q = \tfrac{1}{4}\sqrt{16 + 2v_1^{2/3}I\sqrt{3} - 2v_1^{2/3} + 8v_1^{1/3}}.
\end{equation}
Back-substitution into $d_*(0,n-1)$ and $d_*(0,n)$ shows that their gcd equals
\begin{equation}
(v_1^{1/3}-2p)(v_1^{2/3}I\sqrt{3} - v_1^{2/3} - 8p)
\end{equation}
which has no real solution for $p$. 
\item
Case $n=9, d=3$: There are two possible solutions for $q$, namely
\begin{equation}
\begin{split}
q&=\tfrac{\sqrt{2v_3^{1/3}(v_3^{2/3} + 4v_3^{1/3} + 12)}}{2v_3^{1/3}},  \\
q&=\tfrac{\sqrt{v_3^{1/3}(12I\sqrt{3} -v_3^{2/3}I\sqrt{3} - v_3^{2/3} + 8v_3^{1/3} - 12)}}{2v_3^{1/3}},
\end{split}
\end{equation}
with $v_3:=12(3 + I\sqrt{3})$.
Back-substitution into $d_*(0,n-1)$ and $d_*(0,n)$ shows that their gcd equals
\begin{equation}
\begin{split}
&(v_3^{1/3}I\sqrt{3} + 3v_3^{1/3} - 12p)(v_3^{2/3}I\sqrt{3} - v_3^{2/3} + 24p), \\
&(v_3^{1/3}I\sqrt{3} + 6p)(v_3^{2/3}I\sqrt{3} + v_3^{2/3} - 24p).
\end{split}
\end{equation}
No factor implies a real solution for $p$.
\item
Case $n=9, d=4$: Each factor implies a possible solutions for $q$, namely
\begin{equation}
\begin{split}
q&=\tfrac{1}{4}\sqrt{16 + 2v_2^{2/3}I\sqrt{3} - 4v_2^{1/3}I\sqrt{3} - 2v_2^{2/3} + 4v_2^{1/3}}, \\
q&=\tfrac{1}{12}\sqrt{6v_2^{2/3}I\sqrt{3} + 6v_2^{2/3} + 24v_2^{1/3} -48}, \\
q&=\tfrac{1}{2}\sqrt{v_2^{2/3}I\sqrt{3} + v_2^{1/3}I\sqrt{3} + 2v_2^{2/3} + 5v_2^{1/3} + 8}.
\end{split}
\end{equation}
Back-substitution into $d_*(0,n-1)$ and $d_*(0,n)$ shows that their gcd equals
\begin{equation}
\begin{split}
&(v_2^{1/3}I\sqrt{3} - v_2^{1/3} + 4p)(v_2^{2/3}I\sqrt{3} - v_2^{2/3} - 8p) ,\\
&(p^2 + p + 1), \\
&(v_2^{1/3}I\sqrt{3} + v_2^{1/3} - 4p)(v_2^{2/3} - 4p).
\end{split}
\end{equation}
None of these factors imply a real solution for $p$.
\end{enumerate}
Beside the fact that we do not get real solutions for $p$, the vanishing of the related factors imply again the 
vanishing of the numerator of $r$ (a contradiction). \hfill $\BewEnde$

\section{Proof of Theorem \ref{thm:prop6}}\label{app:proof6}

As $\mathcal{R}_-$  is located on a right pyramid over a  regular $n$-gon the 
realization $\mathcal{R}_+$ of Theorem \ref{thm:prop} cannot be located on 
a right pyramid over a regular star polygon of type $\left\{\tfrac{n}{d} \right\}$ as in this case not 
 $V_k,V_{k+1},V_{k+n-1},V_{k+n}$ have to be coplanar but $V_k,V_{k+1},V_{k+n},V_{k+n+1}$. 
Therefore only the possibilities remain that  $\mathcal{R}_+$ is located on a right pyramid over a regular: 

\begin{enumerate}[$\bullet$]
\item
$(n-1)$-gon: This discussion equals the one in the first bullet point of the proof of Theorem \ref{thm:prop5}. 
\item
star polygon of type $\left\{\tfrac{n-1}{d} \right\}$: 
We compute the expressions $d_*(0,n-1)$ and $d_*(0,n)$ as in the general case of Section \ref{sec:same}. 
Then we substitute $c_-$ and  $c_+$ by the analytic expression for $\cos{\tfrac{2\pi}{n}}$ and $\cos{\tfrac{2d\pi}{n-1}}$, respectively, according to \ref{app:trig}. 
We have to check this for the pairs of $n$ and $d$ listed in the proof of Theorem \ref{thm:prop2}. 
We compute the resultant of $d_*(0,n-1)$ and $d_*(0,n)$ with respect to $p$. In all cases the resulting expression has $q=0$ as a real solution, but 
there exist further solutions. These solutions for the case $n,d$ are identical with those listed for the case $n-1,d$ in the proof of Theorem \ref{thm:prop5} for 
$n=6,\ldots,9$. 
Again for all solutions the gcd of $d_*(0,n-1)$ and $d_*(0,n)$ has no real solution for $p$, which finishes the proof. \hfill $\BewEnde$
\end{enumerate}

\end{document}